\documentclass[12pt,epsfig]{article}
\usepackage{graphicx,pstricks,fancyhdr,ulem,t1enc,alltt,color}
\usepackage[latin1]{inputenc}
\usepackage{framed}
\usepackage{epsfig}
\usepackage{amsmath}
\usepackage{amssymb}
\usepackage{bigints}
\usepackage{appendix}
\newcounter{exercise}
\usepackage{cancel}
\usepackage{color}
\usepackage{xcolor}
\usepackage{hyperref} 
\usepackage{tikz}
\usepackage{cite}
\usepackage[nobottomtitles*]{titlesec}

\newcommand{\comments}[1]{}

\newenvironment{packed_enum}{
\begin{itemize}
  \setlength{\itemsep}{0pt}
  \setlength{\parskip}{0pt}
  \setlength{\parsep}{0pt}
}{\end{itemize}}

\newfont{\itg}{cmbxti10 at 10pt}
\begin{document}

\centerline{
\textcolor{red}{\bf \Large Introduction to Langevin Stochastic Processes}
}

\vspace{0.75cm}


\vspace{1cm}

\centerline{\large \bf Leticia F. Cugliandolo}

\vspace{0.5cm}

\centerline{
Sorbonne Universit\'e}
\centerline{
Laboratoire de Physique Th\'eorique et Hautes Energies}
\centerline{
4 Place Jussieu, Tour 13, 5\`eme \'etage 
}
\centerline{
75252 Paris Cedex 05 France}

\vspace{1cm}

{\small 
These lecture notes provide an introduction to Langevin processes
and briefly discuss some interesting properties and simple applications. 
They compile material presented at the "School of Physics and Mathematics Without Frontiers" (ZigZag), 
held La Havana, Cuba, in March 2024, the School "Information, Noise and Physics of Life"
held at Ni\v{s}, Serbia, in June 2024, both sponsored by ICTP, the 
PEBBLE summer camp at Westlake University, China, in August 2024, the Barcelona school on "Non-equilibrium Statistical Physics", in 
April 2025, and  in the 2012-2016 course "Out of Equilibrium Dynamics of Complex Systems" for the Master 2 program "Physics of 
Complex Systems" in the Paris area.
}

\vspace{0.7cm}

\centerline{\large \today}

\vspace{8cm}

\comments{
\begin{minipage}{.5\textwidth}
\hspace{3cm}
\includegraphics[scale=0.25]{logoUPMC.ps}
\end{minipage}
\begin{minipage}{.5\textwidth}
\hspace{1cm}
\includegraphics[scale=0.25]{iuf.jpg}
\end{minipage}
}

\pagestyle{empty}

\newpage

\tableofcontents

\newpage 

\pagestyle{fancy}

$\;$

\vspace{1cm}

\textcolor{red}{
\section{Introduction}
\label{sec:introduction}
}

\pagenumbering{arabic}

Langevin equations permeate the description of systems in contact with environments.
These notes present an introduction to this kind of modelling and they 
discuss a number of basic properties. 

Langevin proposed  a stochastic equation, later named after him~\cite{Langevin}, as an alternative
description to Einstein's~\cite{Einstein} of Brownian motion~\cite{Brown}. One of the most 
spectacular applications of Brownian motion was
performed by Perrin~\cite{Perrin} who received the Nobel Prize in Physics in 1926 
for his work on the discontinuous structure of matter, particularly for his experimental 
verification of Einstein's and Langevin's theoretical explanation of Brownian motion, and 
the measurement of the Avogrado number.

Many textbooks and review articles describe the Langevin equation and some of  their uses~\cite{Chandrasekhar,Risken,Gardiner,vanKampen,Zwanzig-book,Langevin-Coffey,vanKampen-Ito,Oksendal,HanggiMarchesoni,weiss,Razi,Pomeau}. 
In this Lecture Notes, we offer a modern perspective by detailing some 
recent applications after first addressing the fundamental aspects. We focus on simple single particle 
problems and we  leave aside 
applications to many-body problems, for instance,
critical dynamics~\cite{critical-dyn,Tauber},
nucleation~\cite{review-nucleation,dilute-gas-instanton},
coarsening~\cite{review-coarsening,Puri}, interface motion~\cite{Barabasi}, and
glasses and spin-glasses~\cite{Cugliandolo,Kurchan}. 
We discuss some applications, for example to active matter~\cite{active-matter-reviews,Cates} and 
we refer the readers to the vast literature on the fields of stochastic thermodynamics~\cite{Sekimoto,Peliti,Seifert}
(including fluctuation theorems~\cite{Searles,Gallavotti,Jarzynski,Kurchan1,Searles-review,Ciliberto-FT,Crooks,Crooks00,Lebowitz}) where Langevin processes have been especially useful.




\textcolor{red}{
\section{Equilibrium}
}
\setcounter{equation}{0}
\renewcommand{\thesection}{\arabic{section}}
\renewcommand{\theequation}{\thesubsection.\arabic{equation}}
\label{sec:environemnts-dissipation}
\setcounter{equation}{0}
\setcounter{exercise}{1}

In this section we revisit certain aspects of equilibrium statistical physics that are not
commonly discussed in usual undergraduate courses~\cite{Pathria,Reif,Huang,Reichl,Pauli,Landau,Berlinsky}. 
We then briefly discuss different sources of noise.

\textcolor{red}{
\subsection{Canonical setting}
}
\label{sec:canonical-setting}

In these lectures we consider the dynamics of a system of interest coupled to an 
environment with which it can exchange energy (not particles) and that can be the source of 
fluctuations. The total energy of the full
system is conserved but the contributions from the system, bath and interaction 
between the two are not. See the sketch in Fig.~\ref{fig:sketch-syst-bath}.

\setcounter{figure}{0}
\begin{figure}[h]
\begin{center}
\includegraphics[scale=0.4]{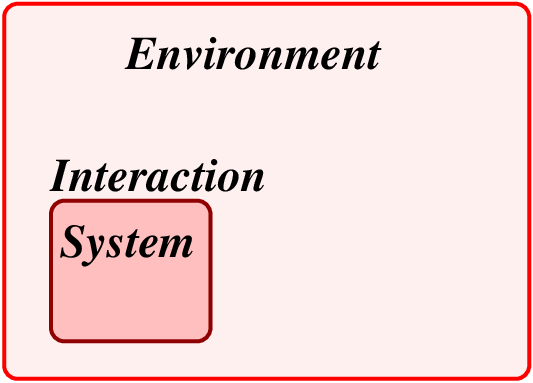}
\end{center}
\vspace{-0.5cm}
\caption{\small Sketch of a total systems consisting of the system of interest, the bath, and their coupling.}
\label{fig:sketch-syst-bath}
\end{figure}

The bath is usually considered to be macroscopic and not altered by the 
coupling to the system. This means that its statistical properties, which are 
assumed from the start,  and are typically those of equilibrium, do not change in time.
The most interesting problems are those in which the system of interest is itself many-body
with non-trivial collective effects.
 
\textcolor{red}{\subsection{Fluctuations}}
\label{sec:fluctuations}

There are several possible sources of fluctuations:

\begin{packed_enum}
\item[--]
\textcolor{blue}{Thermal}: the system is coupled to a classical environment 
that ensures fluctuations (noise) and dissipation (the fact that the energy of the system 
of interest is 
not conserved). Examples at the single particle level are passive colloidal particles immersed
in a fluid, like in Brownian motion, or active ones also immersed in a fluid, like Janus 
particles used to mimic active matter. At the macroscopic level, classical 
coarsening, glasses and spin-glasses are passive examples, while biological  matter in 
general are active ones.
\item[--]
\textcolor{blue}{Quantum}: the system is coupled to a 
quantum environment that ensures fluctuations (noise) and dissipation. The temperature 
of the bath can be zero or not.
Some examples are atom impurities confined by optical traps in contact with an ensemble of another atomic species; 
coarsening, glasses, and spin-glasses at very low temperatures where quantum effects are important.
\item[--]
\textcolor{blue}{Stochastic motors}: forces that act on the  particles 
stochastically. The energy injected in the sample is partially
dissipated to the bath and partially used as work. As the system is also coupled to 
a bath there are also thermal fluctuations in it. Active matter provides manifold realizations.
\end{packed_enum}

Classical and quantum environments are usually modeled as large 
assemblies of non-interacting variables (oscillators~\cite{Feve}, spins~\cite{Stamp}, fermions~\cite{Schaller}) 
with chosen distributions of coupling constants and energies.  

\textcolor{red}{
\subsection{The reduced partition function}
}
\label{subsec:classical-statics}

We analyze the statistical \textcolor{blue}{static} properties of a 
\textcolor{blue}{classical canonical system}
in equilibrium at inverse temperature $\beta$ and itself formed by two sub-parts,
one that will be treated as an environment (not necessarily of infinite size) 
and another one that will be the 
(sub-)system of interest. We study the \textcolor{blue}{partition function} or Gibbs functional, 
$Z_{\rm tot}$,
\begin{eqnarray}
Z_{\rm tot}[\eta]= 
\sum_{
{\tiny 
\begin{array}{c}
{\rm conf \; env}\\
{\rm conf \; syst}
\end{array}
} 
}
\; \exp(-\beta H_{\rm tot}-\beta \eta x)
\end{eqnarray}
where the sum represents an integration over the phase space of
the full system,  i.e.
the system's and the environmental ones. 
$\eta$ is a source. We take
\begin{equation}
H_{\rm tot} = H_{\rm syst} + H_{\rm env} + H_{\rm int}+H_{\rm counter}
= H_{\rm syst} + \tilde H_{\rm env}
\; .
\label{Htot}
\end{equation}
For simplicity we use as a system a single particle moving in $d=1$:
$H_{\rm syst}$ is the  Hamiltonian of the isolated 
particle, 
\begin{equation}
H_{\rm syst}= \frac{p^2}{2M} + V(x)
\; , 
\end{equation}
with $p$ and $x$ its momentum and position. 
$H_{\rm env}$ is the Hamiltonian of a `thermal bath' that, again for simplicity,
we take to be an ensemble of $a=1,\dots,N$ independent harmonic oscillators~\cite{Ford,Zwanzig,Feve}
with masses $m_a$ and frequencies $\omega_a$, 
\begin{equation} 
H_{\rm env}= \sum_{a=1}^{N} \frac{\pi_a^2}{2m_a} + 
\frac{m_a \omega_a^2}{2} q_a^2 
\label{osc-bath}
\end{equation}
with  $\pi_a$ and $q_a$ their momenta and positions.
This is indeed a very usual choice since it may represent phonons.
(These oscillators could be the normal modes of a generic Hamiltonian expanded to quadratic order around its absolute minimum,
written in terms of other pairs of conjugate variables; the bath could be, for instance, a chain of harmonic oscillators
with nearest-neighbor couplings.)
$H_{\rm int}$ is the coupling between system and environment.
We will restrict the following discussion to a linear interaction 
in the oscillator coordinates, $q_a$, and in 
the particle coordinate, 
\begin{equation}
H_{\rm int}=x\sum_{a=1}^{N} c_a q_a
\; , 
\end{equation} 
with $c_a$ the coupling constants.
The counter-term 
$H_{\rm counter}$ is added to avoid the generation of a negative harmonic 
potential on the particle due to the coupling to the oscillators (that may render
the dynamics unstable). We choose it to be
\begin{equation}
H_{\rm counter} = \frac{1}{2} \sum_{a=1}^{N} \frac{c^2_a}{m_a \omega_a^2}   x^2
\;.
\end{equation}

We note that the addition of the counter-term makes the combination of the 
environmental, interaction and counter-term Hamiltonians take a 
rather simple and natural  form
\begin{equation}
\tilde H_{\rm env} =  H_{\rm env} + H_{\rm int} + H_{\rm counter} = \sum\limits_{a=1}^N \frac{m_a\omega_a^2}{2} \left( q_a + \frac{c_a}{m_a \omega_a^2} x \right)^2
\; . 
\end{equation}

The generalization to more complex systems and/or to more 
complicated baths and higher dimensions 
is straightforward. The calculations can also 
be easily generalized to an interaction of the oscillator coordinate with a  
more complicated dependence on the system's coordinate, ${\cal V}(x)$,  
that may be dictated by the symmetries of the system at the expense of modifying the 
counter-term. Non-linear 
functions of the oscillator coordinates cannot be used since they render the 
problem unsolvable analytically.

Having chosen a quadratic bath and 
a linear coupling, the integration over the oscillators' coordinates 
and momenta can be easily
performed. This yields the \textcolor{blue}{reduced partition function} 
\begin{eqnarray}
Z_{\rm red}[\eta] \propto \sum_{{\small {\rm conf \; syst}}} 
\exp\left[-\beta \left(H_{\rm syst}+H_{\rm counter} +\eta x- 
\frac{1}{2}\sum_{a=1}^{N} \frac{c_a^2 }{m_a \omega_a^2} \; x^2 \right)\right]
\;. 
\end{eqnarray}
The `counter-term' $H_{\rm counter}$ is chosen to cancel the last term
in the exponential and it avoids the renormalization of the coefficient of the quadratic term in the potential
due to the coupling to the environment that could have even 
destabilized the potential by taking negative values. 
An alternative way of curing this problem would be to take a vanishingly 
small coupling to the bath in such a way that the last term must
vanish by itself (say, all $c_a\to 0$).  However, this might be
problematic when dealing with the stochastic dynamics since a very
weak coupling to the bath implies also a very slow relaxation.  It is
then conventional to include the counter-term to cancel the mass
renormalization.  One then finds
\begin{equation}
\textcolor{magenta}{
\fbox{$
Z_{\rm red}[\eta] \propto \sum\limits_{{\small {\rm conf \; syst}}} 
\exp\left[-\beta \left(H_{\rm syst}+\eta x\right)\right]
= Z_{\rm syst}[\eta] 
$}
}
\end{equation}
For a non-linear coupling, the counter-term has to be modified:
\begin{equation}
H_{\rm int}=\sum_{a=1}^N c_a q_a {\cal V}(x)
\qquad\qquad
H_{\rm counter} = \frac12 \sum_{a=1}^N
\frac{c_a^2 }{m_a \omega_a^2} [{\cal V}(x)]^2 
\; .
\end{equation}

\vspace{0.25cm}

\setcounter{exercise}{0}
\noindent
\refstepcounter{exercise}\textcolor{orange}{\bf Exercise \thesection.\theexercise}
Prove the last equation.
\addtocounter{exercise}{1}
\vspace{0.25cm}

The interaction with the reservoir does not modify the statistical 
properties of the particle since $Z_{\rm red} \propto Z_{\rm syst}$, independently 
of the choices of $c_a, \ m_a, \ \omega_a$ and $N$. 
 
If one is interested in the \textcolor{blue}{dynamics} of a coupled problem, the characteristics
of the sub-system that will be considered to be the bath have an influence on 
the reduced dynamic equations found for the system, that are of generic 
Langevin kind, as explained in Sect.~\ref{sec:brownian}.
 
\textcolor{blue}{Quantum mechanically} the reduced partition function depends explicitly
on the properties of the bath. The interaction with quantum 
harmonic oscillators introduces non-local interactions (along the Matsubara 
time direction) and there is no physical way to introduce a counter-term to correct for this
feature. 

The \textcolor{blue}{dynamics of quantum systems} has all these difficulties.

\textcolor{red}{\subsection{Ergodicity}}

Finally, let us discuss Boltzmann's and Gibb's interpretation 
of averages and the \textcolor{red}{\it ergodic hypothesis}.
Boltzmann interpreted macroscopic observations as time averages of the form\footnote{In practice, 
in an experiment or numerical simulation initiated at time $t=0$, 
averages are computed over a symmetric time interval around a measuring time 
$t$, in the form $\overline A
\equiv \lim_{t_0\ll \tau \leq t} 
\frac1{2\tau} \int_{t-\tau}^{t+\tau} dt' \; A(\vec x(t'),\vec p(t'))$
with the lower bound in the limit representing a microscopic time-scale. The result
should be independent of the measuring time $t$, that is why we did not write it in the expression of the main text.} 
\begin{equation}
\overline A
\equiv \lim_{\tau\to\infty} 
\frac1{\tau} \int_{0}^{\tau} dt \; A(\vec x(t),\vec p(t))
\label{eq:average-time}
\end{equation}
(focusing on observables $A$ that depend on the phase space variables collected in $\vec x$ and $\vec p$but are not explicitly time dependent).
The fact that this limit exists is the content of a 
Theorem in Classical Mechanics initially proven by Birkhoff and later by Kolmogorov~\cite{Khinchin,Castiglione-etal}.
Note that in classical mechanics the choice of the initial time is irrelevant.

With the introduction of the \textcolor{red}{\it concept of ensembles} Gibbs gave a
different interpretation (and an actual way of computing) macroscopic
observations.  For Gibbs, these averages are statistical ones 
over all elements of the statistical ensemble,
\begin{equation}
\langle  A  \rangle = \int d\vec x d\vec p \; \rho(H(\vec x,\vec p)) A(\vec x,\vec p)
\; , 
\label{eq:average}
\end{equation}
with $\rho$ the measure. 
 In the microcanonical ensemble this is an
average over micro-states on the constant energy surface taken with the
microcanonical distribution:
\begin{equation}
\langle  A  \rangle = \rho_0 \int d\vec x d\vec p \; \delta(H(\vec x,\vec p)-E) 
A(\vec x,\vec p)
\; ,  
\label{eq:average-micro}
\end{equation}
and the normalization constant $\rho_0^{-1}=\int d\vec x d\vec p \, 
\delta(H(\vec x,\vec p)-E)$. In the canonical ensemble the 
average is computed
with the Gibbs-Boltzmann weight:
\begin{equation}
\langle  A  \rangle = Z^{-1} \int 
d\vec x d\vec p \; e^{-\beta H(\vec x,\vec p)} A(\vec x,\vec p)
\; .  
\label{eq:average-canonical}
\end{equation}
$Z$ is the partition function $Z=\int d\vec x d\vec p
 \; e^{-\beta H(\vec x,\vec p)} $.

The \textcolor{red}{\it (weak) ergodic hypothesis} states that under the dynamic
evolution the representative point in phase space of a classical
system governed by Newton laws can get as close as desired to any
point on the constant energy surface.

\begin{framed}
The \textcolor{red}{\it ergodic hypothesis} states that time and ensemble averages, 
(\ref{eq:average-time}) and (\ref{eq:average}), coincide in 
equilibrium for {\it all reasonable observables}:
\begin{equation}
\overline A  = \langle A \rangle\; . 
\end{equation}
\end{framed}
\noindent
 This hypothesis cannot 
be proven in general but it has been verified in a large number of cases.
In general, the great success of Statistical Mechanics in predicting
quantitative results has given enough evidence to accept this hypothesis.

An important activity in modern Statistical Mechanics is devoted to
the study of macroscopic (non-integrable) systems that \textcolor{red}{\it do not satisfy the ergodic
hypothesis}.  A well-understood case is the one of phase transitions. Other cases are related to
the breakdown of equilibration. This can occur either because they
are externally driven or because they start from an initial condition
that is far from equilibrium and their interactions are such that they
do not manage to equilibrate. One may wonder whether certain concepts
of thermodynamics and equilibrium statistical mechanics can still be
applied to the latter problems. At least for cases in which the
macroscopic dynamics are slow one can hope to derive an extension of
equilibrium statistical mechanics concepts to describe their
behaviour.


\textcolor{red}{
\section{The Langevin equation}
\label{sec:brownian}
}
\setcounter{equation}{0}
\setcounter{exercise}{1}

Here we first introduce the 
Langevin equation phenomenologically and then with a strict calculation.
We also present several of its main characteristics and we mention some applications. 


\setcounter{equation}{0}
\textcolor{red}{
\subsection{Definition}
}

Examples of experimental and theoretical interest in condensed matter
and biophysics in which quantum fluctuation can be totally neglected
are manifold. In this context one usually concentrates on systems in
contact with an environment: one selects some relevant degrees of
freedom and treats the rest as a bath. It is a canonical view.  Among
these instances are colloidal suspensions which are particles
suspended in a liquid, typically salted water, a `soft condensed
matter' example; spins in ferromagnets coupled to lattice phonons, a
`hard condensed matter' case; and proteins in the cell, a `biophysics'
instance.  These problems are modelled as stochastic processes with
Langevin equations, 
the Kramers-Fokker-Planck formalism or master
equations depending on the continuous or discrete character of the
relevant variables and analytic convenience~\cite{Chandrasekhar,Risken,Gardiner,vanKampen,Langevin-Coffey,Zwanzig-book,vanKampen-Ito,Oksendal,HanggiMarchesoni,weiss,Razi,Pomeau}.

The Langevin equation was originally proposed to model the motion of a 
colloidal particle in a liquid but it was soon realised that generalisations of 
it can be used in a much wider context to describe, e.g. the motion of ions 
in water, the reorientation of dipolar molecules, or some collective variable of a 
macroscopic system. 
It is a stochastic differential equation
that describes phenomenologically a large variety of 
problems. Concretely, it models the time evolution of 
a set of slow variables coupled to a much larger set of 
fast variables that are usually (but not necessarily) 
assumed to be in thermal equilibrium 
at a given temperature. We first introduce it 
in the context of Brownian motion in Sect.~\ref{subsubsec:Lagevins} and we derive it
in more generality in Sect.~\ref{sec:Langevin-eq-gen}.

\textcolor{red}{
\subsubsection{Langevin's Langevin equation}
\label{subsubsec:Lagevins}
}

The Langevin equation~\cite{Langevin}
for a particle moving in one dimension in contact with a 
\textcolor{blue}{  white-noise} bath reads
\begin{equation}
\fbox{$ 
\ \ \ \underbrace{m\dot v \ - \ F}_{\rm Newton} = \underbrace{\overbrace{\ \ - \gamma_0 v \ \ }^{\rm friction} + \overbrace{\xi}^{\rm noise}}_{\rm bath}
\qquad \qquad\qquad 
v =\dot x  \ \ \
$}
\label{eq:langevin}
\end{equation}
with $x$ and $v$ the particle's position and velocity. 

The fluctuating force $\xi$ is supposed 
to come from occasional impacts of the Brownian particle with  
molecules of the surrounding medium. The force during an impact is  
assumed to vary with extreme rapidity over the time of any observation. 
The fluctuating force is then taken to be Gaussian, due to a time average
over an infinitesimal time interval, 
and its first and second moments are chosen to be $\langle \xi(t)\rangle =0$ 
and  $\langle \xi(t) \xi(t')\rangle = 2\gamma_0 k_B T \delta(t-t')$.
The delta  function in time indicates that there is no correlation between impacts in 
any distinct time intervals around $t$ and $t'$. The friction force
 $\gamma_0 v$ opposes the motion of the particle.  
The force $F$ designates all external deterministic forces and depends,
in the most common cases, on the position of the particle  $x$ only. In cases in which the 
force derives from a potential, $F= -d V/d x$.
The generalization to higher dimensions is straightforward.
Note that $\gamma_0$ is the parameter that controls the strength of the 
coupling to the bath (it appears in the friction term as well as in the 
noise term). In the case $\gamma_0=0$ one recovers Newton's equation of motion.
The relation between the friction term and the thermal correlation
is non-trivial.  Langevin fixed it by requiring
\begin{equation}
\langle v^2(t)\rangle \to \langle v^2\rangle_{eq} = \frac{k_BT}{m}
\; .
\end{equation} 
We will 
give a different argument for this choice in the next section.

\newpage

\textcolor{red}{
\subsubsection{Derivation of the Langevin equation}
\label{sec:Langevin-eq-gen}
}

Let us take a system in contact with an  environment. The interacting system+environment ensemble 
is  `closed' while the system is `open'.  The nature of the 
environment, {\it e.g.} whether it can be modeled by a  classical 
or a quantum formalism, depends on the problem under study. 
We focus here on the classical problem defined by $H_{\rm tot}$ in Eq.~(\ref{Htot}).
A derivation of a generalized Langevin equation with memory
is very simple starting from Newton dynamics of the full system~\cite{Zwanzig,Zwanzig-book,weiss}.
\comments{  
\begin{equation}
H_{\rm tot} = H_{\rm syst} + H_{\rm env} + H_{\rm int}+H_{\rm counter}
= H_{\rm syst} + \tilde H_{\rm env}
\; .
\label{Htot}
\end{equation}
For simplicity we use a single particle moving in $d=1$:
$H_{\rm syst}$ is the  Hamiltonian of the isolated 
particle, 
\begin{equation}
H_{\rm syst}= \frac{p^2}{2M} + V(x)
\; , 
\end{equation}
with $p$ and $x$ its momentum and position. 
$H_{\rm env}$ is the Hamiltonian of a thermal bath that, for simplicity,
we take to be an ensemble of $N$ independent Harmonic oscillators
with masses $m_a$ and frequencies $\omega_a$, 
$a=1,\dots,N$
\begin{equation} 
H_{\rm env}= \sum_{a=1}^{N} \frac{\pi_a^2}{2m_a} + 
\frac{m_a \omega_a^2}{2} q_a^2 
\label{osc-bath}
\end{equation}
with  $\pi_a$ and $q_a$ their momenta and positions.
This is indeed a very usual choice since it may represent phonons.
$H_{\rm int}$ is the coupling between system and environment.
We will restrict the following discussion to a linear interaction 
in the oscillator coordinates, $q_a$, and in 
the particle coordinate, 
\begin{equation}
H_{\rm int}=x\sum_{a=1}^{N} c_a q_a
\; , 
\end{equation} 
with $c_a$ the coupling constants.
The counter-term 
$H_{\rm counter}$ is added to avoid the generation of a negative harmonic 
potential on the particle -- due to the coupling to the oscillators --
 that may render
the dynamics unstable. We choose it to be
\begin{equation}
H_{\rm counter} = \frac{1}{2} \sum_{a=1}^{N} \frac{c^2_a}{m_a \omega_a^2}   x^2
\;.
\end{equation}
} 
The generalization to 
more complex systems and/or to more 
complicated baths and higher dimensions 
is straightforward. The calculations can also 
be easily generalized to an interaction of the oscillator coordinate with a  
more complicated dependence on the system's coordinate, ${\cal V}(x)$,  
that may be dictated by the symmetries of the system, see 
\textcolor{orange}{\bf Exercises~3.\ref{ex-Nu}-3.\ref{ex-Nu3}}.

We recall Hamilton's equation of motion
\begin{equation}
\dot x(t) = \dfrac{\partial H}{\partial p(t)}
\qquad\qquad
\dot x(t) = -\dfrac{\partial H}{\partial x(t)}
\; . 
\end{equation}
Hamilton's equations for the particle are
\begin{eqnarray}
\dot x(t) = \frac{p(t)}{m}
\; , \;\;\;\; && \;
\dot p(t) =  
-V'[x(t)]
- \sum_{a=1}^N c_a q_a(t)
- \sum_{a=1}^N \frac{c_a^2}{m_a\omega_a^2} x(t)
\label{eq-p0}
\end{eqnarray}
(the counter-term yields the last term)
while the dynamic equations for each member of the environment read
\begin{eqnarray}
\dot q_a(t) = \frac{\pi_a(t)}{m_a}
\; , \;\;\;\; & & \;\;\;\;
\dot \pi_a(t) =  -m_a\omega_a^2 q_a(t) - c_a x(t)  
\; ,
\end{eqnarray}
showing that they are all stable harmonic oscillators
\textcolor{blue}{forced by the chosen particle}.
These equations are readily solved by
\begin{equation}
q_a(t) = q_a(0) \cos(\omega_a t)  + \frac{\pi_a(0)}{m_a\omega_a}
\sin(\omega_a t)
-
\frac{c_a}{m_a\omega_a}
\int_0^t dt' \sin[\omega_a(t-t')] x(t')  
\end{equation} 
with $q_a(0)$ and $\pi_a(0)$ the initial coordinate
and position at time $t=0$ when the particle is set in contact
with the bath. It is convenient to integrate by parts the last term.
The replacement of the resulting expression
in the last term in the {\rm rhs} of Eq.~(\ref{eq-p0})
yields
\begin{eqnarray}
\fbox{$
\;\; \dot p(t) =
\underbrace{-V'[x(t)]}_{\rm deterministic \;\; force}+ \underbrace{\xi(t) - \int_0^t dt' \
\Gamma(t-t') \dot x(t')}_{\rm coupling \;\; to \;\; the \;\; bath}
\;\; $}
\label{eq:langevin_generalized}
\end{eqnarray}
with the \textcolor{blue}{symmetric and stationary kernel} $\Gamma$ given by 
\begin{equation}
\fbox{$
\ \ \
\Gamma(t-t') = \sum\limits_{a=1}^N \dfrac{c_a^2}{m_a\omega^2_a} \cos[\omega_a(t-t')]
\ \ \ 
$}
\label{eq:Gamma}
\end{equation}
$\Gamma(t-t')=\Gamma(t'-t)$, 
and the \textcolor{blue}{time-dependent force} $\xi$ given by
\begin{eqnarray}
\fbox{$ \ \ \
\xi(t) = -\sum\limits_{a=1}^N 
c_a \left[ 
\dfrac{\pi_a(0)}{m_a\omega_a} 
 \sin(\omega_a t)
+
\left(q_a(0)+\dfrac{c_a x(0)}{m_a\omega_a^2 }\right)
  \cos(\omega_a t)
\right]
\ \ \ $}
\label{eq:xi}
\end{eqnarray} 
This is the equation of motion of the \textcolor{blue}{ reduced} system.
It is still \textcolor{blue}{deterministic}.

$\xi(t)$ is a sum of oscillating functions of time.
The third term on the {\rm rhs} of
Eq.~(\ref{eq:langevin_generalized}) represents 
a rather complicated \textcolor{blue}{friction force}. Its value at time $t$ depends
explicitly on the history of the particle at times $0\leq t'\leq t$
and makes the equation \textcolor{blue}{non-Markovian}.
One can rewrite it as an integral running up to a total time ${\cal T}>\max(t,t')$
introducing the \textcolor{blue}{  retarded friction}:
\begin{equation}
\fbox{$\ \ \ 
\gamma(t-t')=\Gamma(t-t') \theta(t-t')
\ \ \ $
}
\end{equation}

Until this point the dynamics of the system remain deterministic and
are completely determined by its initial conditions as well as those of
the reservoir variables. Two important points can be make here. 
On the one hand, one can check, by simple numerical generation, that 
the function $\xi$ at, say, fixed $t$  resembles more and more a random 
variable as the number of oscillators increases (for incommensurate frequencies 
$\omega_a$). The initial conditions for the oscillators are the seeds of the random number generator.
This is similar to what happens with random number generators in the sense that 
these are, ultimately, periodic functions with a finite recurrence time that, 
however, can be made sufficiently long for computational purposes. 
Therefore, $\xi$ is a \textcolor{blue}{pseudo random number}.
On the other hand, one can directly introduce 
the \textcolor{blue}{statistical element}  into play 
when one realizes that it is impossible to know the initial configuration 
of the large number of oscillators with great precision and
one proposes  that the initial coordinates and 
momenta of the oscillators have a canonical distribution at an \textcolor{blue}{inverse
temperature $\beta$}. Note that one needs to assume that the oscillators interacted in the 
past to establish ergodicity and reach this probability density function (pdf), though they do not do any longer.
Then, one chooses $\{\pi_a(0),q_a(0)\}$ to be 
initially distributed according to a canonical phase space distribution:
\begin{equation}
\fbox{$
\ \ \ P(\{\pi_a(0),q_a(0)\}, x(0)) = 
\dfrac{1}{\tilde{\cal Z}_{\rm env} [x(0)]} \ e^{-\beta \tilde H_{\rm env}[\{\pi_a(0),q_a(0)\}, x(0)]}
\ \ \ $}
\label{eq:P-initial-osc}
\end{equation}
with  
$\tilde H_{\rm env}=H_{\rm env}+H_{\rm int}+H_{\rm counter}$, that can be rewritten as
\begin{equation}
\tilde H_{\rm env} = \sum\limits_{a=1}^N \left[ 
\frac{m_a\omega_a^2}{2} 
\left( q_a(0) + \frac{c_a x(0) }{m_a\omega_a^2} \right)^2
+ \frac{\pi_a^2(0)}{2m_a} \right]
\; .
\end{equation}
Again, the presence of $H_{\rm counter}$ here is for convenience.
The randomness in the initial conditions implies
\begin{eqnarray}
&& \langle \pi_a \rangle =0
\qquad\qquad
\left\langle q_a + \frac{c_a x(0)}{m_a\omega_a^2}  \right\rangle =0
\qquad\qquad
\left\langle \pi_a \left( q_a + \frac{c_a x(0)}{m_a\omega_a^2}  \right)  \right\rangle =0
\nonumber\\
&&
\langle \pi^2_a \rangle = k_BT m_a
\qquad\qquad
\left\langle \left( q_a + \frac{c_a x(0) }{m_a\omega_a^2} \right)^2  \right\rangle = \frac{k_BT}{m_a\omega_a^2}  
\end{eqnarray}
where the angular brackets denote an average over the measure  (\ref{eq:P-initial-osc}).
It gives rise to a random force $\xi$
acting on the reduced system.  Indeed, $\xi$ is now a
\textcolor{blue}{Gaussian random variable}, that is to say a noise, with
\begin{eqnarray}
\langle \xi(t) \rangle =  
0 \, , \qquad \qquad 
\langle \xi(t) \xi(t') 
\rangle = 
k_B T \; \Gamma(t-t')
\end{eqnarray}
and $\Gamma(t-t')$ given in Eq.~(\ref{eq:Gamma}).
One can easily check that higher-order correlations vanish for an odd
number of $\xi$ factors and factorize as products of two time
correlations for an even number of $\xi$ factors. In consequence $\xi$ has
Gaussian statistics. Defining the inverse of $\Gamma$ over the
interval $[0,t]$, $\int_0^t dt'' \ \Gamma(t-t'') \Gamma^{-1}(t''-t') =
\delta(t-t')$, one has the Gaussian pdf:
\begin{equation}
P[\xi] = Z^{-1} e^{-\frac{1}{2k_B T} \int_0^t dt \int_0^t  dt' \ \xi(t) \Gamma^{-1}(t-t') \xi(t')}
\; . 
 \end{equation}
 $Z$ is the normalization.  A random force with non-vanishing
 correlations on a finite support is usually called a
 \textcolor{blue}{ coloured
   noise}. Equation~(\ref{eq:langevin_generalized}) is now a genuine
 Langevin equation. A \textcolor{blue}{multiplicative retarded noise} arises from a
 model in which one couples the coordinates of the oscillators to a
 generic function of the coordinates of the system, see \textcolor{orange}{\bf Exercise~3.\ref{ex-Nu}} and 
Eq.~(\ref{eq:multiplicative-noise}).

The use of an \textcolor{blue}{equilibrium measure} 
for the oscillators implies  the relation between the friction kernel
and  the noise-noise correlation, which
are proportional, with a constant of proportionality of value $k_B T$.
This is a generalized form of the
\textcolor{blue}{fluctuation-dissipation relation}, and it applies to
the environment.

\vspace{0.25cm}

\noindent
\textcolor{red}{
{\it About the bath}
}

\vspace{0.25cm}

One last comment is in order here. A closed ensemble of harmonic oscillators is an integrable system
 that does not  equilibrate in strict sense. Still, we are using it as a model for a thermal bath. One can argue that the 
oscillators were in interaction in the past, before being connected to the system, and that this allowed them 
to reach the Boltzmann probability distribution that is used for their initial states in this calculation. 
Or else, one can follow the calculations by Mazur \& Montroll and see that in the limit of a large number of 
degrees of freedom the \textcolor{blue}{Poincar\'e recurrence time} for the system of non-interacting harmonic oscillators 
becomes so large that it lies beyond any 
relevant time for the relaxation of the system that is coupled to the oscillator bath~\cite{Mazur}.

\pagebreak

\noindent
\textcolor{red}{\it About the counter-term}

\vspace{0.2cm}

Had we not added the counter term the equation would read
\begin{eqnarray}
\dot p(t) =
-V'[x(t)]+\sum_{a=1}^{N} \frac{c_a^2}{m_a \omega_a^2} x(t) +  \xi(t) - \int_0^t dt' \
\Gamma(t-t') \dot x(t')
\; ,
\label{eq:langevin_generalized_without_counterterm}
\end{eqnarray}
that is like Eq.~(\ref{eq:langevin_generalized})  
for $V\mapsto V - \frac{1}{2} \sum_{a=1}^N \frac{c_a^2}{m_a \omega_a^2} x^2$, as we found with the 
analysis of the partition sum. Note that, as we will take $c_a = \tilde c_a/\sqrt{N}$, with $\tilde c_a$ of $O(1)$, 
 the constant resulting from the sum over $a$ is $O(1)$. 
For the distribution of the initial values we can still use $\widetilde H_{\rm env}$ or we can choose a Maxwell-Boltzmann
distribution with $H_{\rm env}+H_{\rm int}$ alone. The result will be the same, as the supplementary 
term goes into the normalisation constant
for $P_{\rm env}(0)$. 

\vspace{0.25cm}

\noindent
\textcolor{red}{\it The bath kernel $\Gamma$}

\vspace{0.25cm}

Different choices of the environment are possible by selecting different ensembles 
of harmonic oscillators~(see~\cite{Ford} for a detailed analysis). The simplest one, that leads to an 
approximate Markovian equation, is to consider
that the oscillators are coupled 
to the particle via coupling constants $c_a = \tilde c_a/\sqrt{N}$
with $\tilde c_a$ of order one. One defines
\begin{equation}
\fbox{$ \ \ 
S(\omega) \equiv 
\dfrac{1}{N}
\sum\limits_{a=1}^N  \dfrac{{\tilde c}_a^{\, 2}}{m_a \omega_a} \ \delta(\omega-\omega_a)
\ \ $}
\end{equation}
a function of $\omega$, of order one with respect to $N$, 
and rewrites the kernel $\Gamma$ as
\begin{equation}
\fbox{$ \ \ 
\Gamma(t-t') =
\displaystyle{\int_0^\infty} d\omega \, \frac{S(\omega)}{\omega} \, \cos[\omega (t-t')]
\ \ $}
\end{equation} 

The spectral density $S(\omega)$ is a weighted 
sum over the frequencies of the oscillators in the bath. For all finite $N$ it is then just a 
discrete sum of delta functions. In the limit $N\to\infty$ it can become, instead,  a regular 
function of  $\omega$. Assuming this limit is taken, several proposals for the resulting 
function $S(\omega)$ are made. A common choice~is
\begin{eqnarray}
\fbox{$ \ \ 
\displaystyle{
\frac{S(\omega)}{\omega}
= \frac{2 \gamma_0}{\pi}
\left( \frac{|\omega|}{\tilde \omega}\right)^{\alpha-1} \, 
f_c\left( \frac{|\omega|}{\Lambda}\right)
}
\ \ $}
\label{eq:spectral-density-bath}
\end{eqnarray}
The function $f_c(x)$ is a high-frequency cut-off of typical width 
$\Lambda$ and is usually chosen to be an exponential. The frequency
$\tilde \omega\ll \Lambda$ is a reference frequency that allows one 
to have a coupling strength $\gamma_0$ with the dimensions of 
viscosity. If $\alpha=1$, the friction is said to be \textcolor{blue}{  Ohmic}, 
$S(\omega)/\omega$ is constant when $|\omega| \ll
\Lambda$ as for a white noise. This name is motivated by the electric circuit analog 
exposed in Sec.~\ref{subsec:phenomenological-Langevin-eqs}.
When $\alpha > 1 $ ($\alpha<1$) the bath is \textcolor{blue}{  superOhmic} (\textcolor{blue}{  subOhmic}).
The exponent $\alpha$ is taken 
to be $>0$ to avoid divergencies at low frequency.
For the exponential cut-off the integral over $\omega$ can be computed
for $\alpha=1$ and $\alpha\neq 1$. In the former Ohmic case one finds
\begin{equation}
\Gamma(t) = \frac{2\gamma_0}{\pi}  \; \frac{\Lambda}{[1+(\Lambda t)^2]}
\qquad\quad
\alpha=1
\; .
\end{equation}
In the $\Lambda\to\infty$ limit one approaches the Stratonovich limit and $\Gamma(t)$ becomes a delta-function, 
$\Gamma(t)\to 2\gamma_0 \delta(t)$ such that $\int_0^t dt' \ \Gamma(t-t') = 2\gamma_0 \mbox{arctan}(\Lambda t) 
\to \gamma_0$ for $\Lambda \to \infty$.
In the latter non-Ohmic case the integral over $\omega$  yields
\begin{eqnarray}
\Gamma(t) = \frac{2\gamma_0 \tilde \omega^{-\alpha+1}}{\pi}  \ \Gamma_{\rm E}(\alpha) \ \Lambda^\alpha 
\;  
\frac{\cos[\alpha\arctan(\Lambda t)]}{[1+(\Lambda t)^2]^{\alpha/2}} 
\qquad\quad
\alpha \neq 1
\label{eq:kernel-Gamma}
\end{eqnarray}
with $\Gamma_{\rm E}(\alpha)$ the Euler Gamma-function.
At long times, for any $\alpha >0$ and $\alpha\neq 1$, one has
\begin{eqnarray}
\lim_{\Lambda t \to \infty} \Gamma(t)
&=&
\frac{2\gamma_0 \tilde \omega}{\pi}   \
\cos(\alpha\pi/2) \Gamma_{\rm E} (\alpha) \;  (\tilde\omega t)^{-\alpha}
\; , 
\end{eqnarray}
a \textcolor{blue}{power law decay}.

\vspace{0.25cm}

\noindent
\textcolor{red}{
{\it Dimensional analysis}
}

\vspace{0.25cm}

The noise $\xi$ is a force and it should have dimensions $[\xi] = ML/T^2$ with $[m_a]=M$,
$[q_a]=L$ and the frequencies  $[\omega_a] = 1/T$. From their
definition one finds $[c_a]=M/T^2$, $[\Gamma]=[S(\omega)]=M/T^2$ and $[\gamma_0]=M/T$.

\vspace{0.25cm}

\noindent
\refstepcounter{exercise}\label{ex-Nu}\textcolor{orange}{\bf Exercise \thesection.\theexercise}
Prove that for a non-linear coupling $H_{\rm int}= {\cal V}[x] \sum_{a=1}^N c_a q_a $
there is a choice of counter-term for which the Langevin equation reads
\begin{equation}
\fbox{
$
\dot p(t)=-V'[x(t)]+
   \xi(t) {\cal V}'[x(t)] - 
{\cal V}'[x(t)]
\bigintsss_{\, 0}^t dt' \ \Gamma(t-t') {\cal V}'[x(t')] \dot x(t')
$}
\label{eq:multiplicative-noise}
\end{equation}
 with the same $\Gamma$ as in
   Eq.~(\ref{eq:Gamma}) and $\xi(t)$ given by Eq.~(\ref{eq:xi}) with
   $x(0) \to {\cal V}[x(0)]$. The noise appears now 
\textcolor{blue}{multiplying} a function of the particles' coordinate.
Applications of this kind of equations are manifold. For instance, the 
random motion of a colloid in a confined medium is mimicked with a 
Langevin equation in which the friction coefficient depends on the 
position notably close to the walls of the container.



\vspace{0.25cm}

\noindent
\refstepcounter{exercise}\label{ex-Nu1}\textcolor{orange}{\bf Exercise \thesection.\theexercise}
Take now a system made of $i=1,\dots, n$ variables collected in two $n$-component 
vectors $\vec p, \vec x$. Use $H_{\rm int}= \sum_{i=1}^n \sum_{a=1}^N c_{ai} q_{ai} x_i$ as the coupling 
between system and bath and an ensemble of $n$ independent sets of harmonic 
oscillators for the bath. Prove that the stochastic equation is
\begin{equation}
\dot p_i(t)=-\frac{\partial V[x(t)]}{\partial x_i(t)}
+
   \xi_i(t) - 
\int_0^t dt' \ \Gamma_i(t-t') \dot x_i(t')
\qquad\qquad i=1,\dots,n
\label{eq:multiplicative-noise-neqs}
\end{equation}
where there is no sum over repeated indices and 
 $\Gamma_i$  and $\xi_i$ are given by
 \begin{eqnarray}
&&
\displaystyle{
\Gamma_i(t-t') = \sum\limits_{a=1}^N \frac{c_{ai}^2}{m_{ai}\omega^2_{ai}} \cos[\omega_{ai}(t-t')]
\; ,
}
\label{eq:Gamma_i}
\\
&&
\displaystyle{
\xi_i(t) = -\sum\limits_{a=1}^N 
c_a \left[ 
\frac{\pi_{ai}(0)}{m_{ai}\omega_{ai}} 
 \sin(\omega_{ai} t)
+
\left(q_{ai}(0)+\frac{c_{ai} x_i(0)}{m_{ai}\omega_{ai}^2 }\right)
  \cos(\omega_{ai} t)
\right]
\; .
\qquad\quad
}
\label{eq:xi_i}
\end{eqnarray} 
The $i$ dependence in $\Gamma_{i}$ can be ignored if the ensembles of 
oscillators are equivalent (i.e. same distribution of parameters). 
Characterise next the mean $\langle \xi_i(t)\rangle$ and the 
correlation $\langle \xi_i(t) \xi_j(t')\rangle$
and see under which conditions $\langle \xi_i(t) \xi_j(t')\rangle
= \delta_{ij} \Gamma(t-t')$.


\vspace{0.25cm}

\noindent
\refstepcounter{exercise}\label{ex-Nu2}\textcolor{orange}{\bf Exercise \thesection.\theexercise}
Take now a system made of $i=1,\dots, n$ variables collected in $\vec p, \vec x$.
Use $H_{\rm int}= {\cal V}[x] \sum_{a=1}^N c_a q_a $ as the coupling 
between system and bath, where $x$ is the modulus of the vector $\vec x$. 
Prove that the stochastic equation is
\begin{eqnarray}
&& 
\dot p_i(t)=-\frac{\delta V[x(t)]}{\delta x_i(t)}
+
\xi(t) \frac{\delta {\cal V}[x(t)]}{\delta x_i(t)} 
\nonumber\\
&&
\qquad\qquad\qquad
- 
\frac{\delta {\cal V}[x(t)]}{\delta x_i(t)}
\int_0^t dt' \ \Gamma(t-t') \sum_{j=1}^n \frac{\delta {\cal V}[x(t')]}{\delta x_j(t')} \; \dot x_j(t')
\label{eq:multiplicative-noise}
\end{eqnarray}
 with the same $\Gamma$ and $\xi$ as in
   Eqs.~(\ref{eq:Gamma}) and~(\ref{eq:xi}) with
   $x(0) \to {\cal V}[x(0)]$. There is only one noise component and it appears 
\textcolor{blue}{multiplying} a function of the particles' coordinate.


\vspace{0.25cm}

\noindent
\refstepcounter{exercise}\label{ex-Nu3}\textcolor{orange}{\bf Exercise \thesection.\theexercise}
The classical phonon Hamiltonian in one dimension is
\begin{equation}
H=\sum_{\ell=-N/2+1}^{N/2} \frac{p^2_\ell}{2} + \sum_{\ell = -N/2+1}^{N/2} \frac{(q_\ell- q_{\ell-1})^2}{2} 
\end{equation}
where $\ell$ is the lattice site index and, for simplicity, we took $m_\ell = \omega_\ell=1$. The last term is a neares-negihbour 
coupling. We assume periodic boundary conditions on the chain.
Let us also consider a linear coupling of the form $H_I=-q_0 x$ between the central 
site and the system's coordinate $x$. 
Show that going to the Fourier modes $Q_k = N^{-1/2} \sum_{\ell = -N/2+1}^{N/2} e^{ik\ell} q_\ell$
all phonons decouple into independent harmonic oscillators with frequencies $\omega_k^2 = 2(1-\cos ka)=4 \sin^2 (ka/2)$, with 
$a$ the lattice spacing, and the coupling constants between the variable and all Fourier modes are $c_k = N^{-1/2}$. 
The periodicity $q_{N}=q_0$ imposes the quantisation of the wave-vectors, $k=2\pi n/(Na)$ with $n\in {\mbox Z}$.
Each $k$ describes a normal mode of vibration  and within the first Brillouin zone there are $N$ modes.

This calculation can be extended to higher dimensions.

\addtocounter{exercise}{1}

\pagebreak

\textcolor{red}{\subsubsection{Energy conserving baths}}

In some situations one wishes to study a system coupled to a bath 
and  conserve the energy of the system.   This can be done with a 
(noiseless) `Gaussian' thermostat~\cite{Nose,Hoover}, extensively used in
the context of molecular dynamics simulations and, on the analytical side,
the study of entropy production and the Gallavotti-Cohen theorem.

\vspace{0.25cm}

\noindent
\textcolor{red}{\it Deterministic constant termperature thermostat}

\vspace{0.25cm}

The so-called Gaussian thermostat~\cite{Hoover,Nose} conserves the kinetic energy 
of a particle system and sets it to be given by an equipartition law, and thus
introduces temperature
\begin{equation}
K(\vec p) = \sum\limits_{i=1}^N \dfrac{p_i^2}{2m} = g(d,N) \; \frac{k_BT}{2}
\end{equation}
with $g(d,N)$ is the number of degrees of freedom in the problem, which will be 
fixed to be $g(d,N)=dN-1$, where the subtraction of one is due to the fact that 
there is a global constraint. The proposed equations of motion read
\begin{eqnarray}
&& \dot x_i = \dfrac{p_i}{m}
\label{eq:Nose1}
\\
&& \dot p_i = - \dfrac{\partial V}{\partial x_i} - \zeta p_i
\label{eq:Nose2}
\end{eqnarray}
with $\zeta$ a Lagrange multiplier set to conserve the kinetic energy. Its explicit form in 
terms of the phase space variables is obtained by multiplying the second equation 
by $p_i/(2m)$, summing over $i$ and requiring that the time-derivative of the kinetic energy 
built in the left-hand-side vanishes. Then, 
\begin{equation}
0=\dfrac{dK(\vec p)}{dt} = -\sum\limits_{i=1}^N \dfrac{\partial V}{\partial x_i} \dfrac{p_i}{2m} - \zeta K(\vec p)
\quad
\implies
\quad
\zeta = - \dfrac{1}{g(d,N) k_BT} \; \sum\limits_{i=1}^N \dfrac{\partial V}{\partial x_i} \dfrac{p_i}{m} 
\label{eq:zeta}
\end{equation}
Note that here we used a conservative force, with respect to unconstrained Newtonian dynamics, 
but a similar form for the Lagrange
multiplier can be derived for a more generic non-potential force with
$-\partial V/\partial x_i \mapsto F_i$. 

Importantly enough, equations~(\ref{eq:Nose1})-(\ref{eq:Nose2}) do not conserve the total energy. This 
can be simply verified by multiplying Eq.~(\ref{eq:Nose1}) by $\partial V/\partial x_i$, Eq.~(\ref{eq:Nose1}) by
$p_i/m$ and summing over $i$. One obtains, for $H(\vec x, \vec p)=K(\vec p) + V(\vec x)$, 
\begin{equation}
\frac{d H(\vec x,\vec p)}{dt} = \sum\limits_{i=1}^N \frac{p_i}{m} \dfrac{\partial V}{\partial x_i} 
-\sum\limits_{i=1}^N \dfrac{\partial V}{\partial x_i}  \frac{p_i}{m}  
- \zeta K(\vec p) = - \zeta K(\vec p) \neq 0 
\; . 
\end{equation}

The equations (\ref{eq:Nose1})-(\ref{eq:Nose2}) with $\zeta$ given in (\ref{eq:zeta}) 
are fully deterministic. A Liouville equation then rules the 
evolution of the probability density $f(\vec x, \vec p; t)$ in phase space. The analysis of this equation 
shows that, asymptotically, the phase space density approaches the stationary form
\begin{equation}
f(\vec x, \vec p; t) \to \delta(K(p)-g(d,N) k_BT) \; \frac{e^{-\beta V(\vec x)}}{{\mathcal Z}}
\end{equation}
with $\beta = 1/(k_BT)$. This is the canonical distribution for the position variables, while the momentum ones 
are constrained to be constant by the delta prefactor.

Equations~(\ref{eq:Nose1})-(\ref{eq:Nose2}) can be derived from the coupling of the system to an extra degree
of freedom that represents a full bath, with which the system can exchange energy. That setting of the 
problem is called the Nos\'e-Hoover bath~\cite{Nose}.

\vspace{0.25cm}

\noindent
\textcolor{red}{\it Generalization}

\vspace{0.25cm}

The procedure can be extended to conserve the total energy, or to include an energy-preserving noise.
Both such conditions are satisfied by the following equations:
\begin{eqnarray}
&& \dot x_i = \dfrac{p_i}{m} 
\label{Gau1}
\\
[15pt]
&& \dot p_i = 
 -\dfrac{\partial V[\vec x] }{\partial x_i} 
 + 
 \underbrace{ \ \ g_{ij}(\vec p) \xi_j \ \ }_{\stackrel{\rm \Large conservative}{\rm noise}} 
 - 
 \underbrace{\ \ f_i({\vec x})\ \ }_{\rm forcing} 
 + \underbrace{\ \ \gamma(t) p_i\ \ }_{\rm bath}
\label{Gau2}
\end{eqnarray}
\begin{packed_enum}
\item[--]
$\xi_j$ are white, independent noises of variance $\epsilon$, unrelated 
to temperature,
since the energy is fixed.
\item[--] $g_{ij}=\delta_{ij}-p_i p_j/p^2$ is the projector
onto the space tangential to the energy surface.
\item[--]  Multiplying Eq.~(\ref{Gau1}) by 
$\partial V({\vec x })/\partial x_i$, and then Eq.~(\ref{Gau2}) 
by $p_i/m$, adding, and using usual rules of calculus, 
one concludes that the total energy is conserved provided 
\begin{equation}
\gamma(t) = m({\vec f}\cdot {\vec p})/p^2
\; .
\end{equation}
Note that in the absence of non-potential forces $\gamma(t)=0$. Under non-potential forces 
the dissipative term compensates the work done by these forces.
\end{packed_enum}

 The product $g_{ij}({\vec p})\xi_j$ is rather ill-defined because both
$g_{ij}$  and $\xi_j$  are discontinuous functions of time. The ambiguity is
raised by discretising time~\cite{Risken}, or by specifying the evolution
of the probability density, as in the Kramers or Fokker-Planck approaches.

\setcounter{equation}{0}
\textcolor{red}{
\subsection{Properties}
}

In this section we discuss a number of important properties of the Langevin processes.

\textcolor{red}{\subsubsection{Novikov theorem}}

A rather trivial property of Gaussian white noises goes under the name of Novikov theorem and is 
a consequence of a simple integration by parts. Take the following expression
\begin{equation}
\left\langle \dfrac{\delta f_\xi[x(t),p(t)]}{\delta \xi(t')} \right\rangle
= {\mathcal N}^{-1} \int {\mathcal D} \xi \; e^{-\frac{1}{4 k_BT\gamma_0} \int_0^t dt''\,  \xi^2(t'') } \; \dfrac{\delta f_\xi[x(t),p(t)]}{\delta \xi(t')}
\end{equation}
where $f_\xi[x(t),p(t)]$ is any functional of the stochastic process ruled by a Langevin equation with white 
noise $\xi$ and ${\mathcal N}$ is the normalization constant. ${\mathcal D}\xi$ is a short-hand-notation 
for the functional differential over all time-dependent noises (which can be made precise using a time discretization).
With a simple integration by parts,
and noticing that there is no contribution from the borders, 
\begin{eqnarray}
\left\langle \dfrac{\delta f_\xi[x(t),p(t)]}{\delta \xi(t')} \right\rangle
\!\! & \!\! = \!\! & \!\!  - {\mathcal N}^{-1} \int {\mathcal D} \xi \; \frac{\delta e^{-\frac{1}{4 k_BT\gamma_0} \int_0^t dt'' \,\xi^2(t'') } }{\delta \xi(t')} 
 f_\xi[x(t),p(t)]
 \nonumber\\
 \!\! & \!\! = \!\! & \!\!  
 {\mathcal N}^{-1} \int {\mathcal D} \xi \; \frac{1}{2 k_BT\gamma_0} \xi(t') \; e^{-\frac{1}{4 k_BT\gamma_0} \int_0^t dt'' \,\xi^2(t'') } 
 \;  f_\xi[x(t),p(t)]
 \nonumber\\
 [5pt]
 \!\! & \!\! = \!\! & \!\!  
 \frac{1}{2k_BT\gamma_0}
 \left\langle  f_\xi[x(t),p(t)] \xi(t') \right\rangle
 \label{eq:Novikov}
\end{eqnarray}
The generalization to the coloured noise case is 
\begin{eqnarray}
\left\langle \dfrac{\delta f_\xi[x(t),p(t)]}{\delta \xi(t')} \right\rangle
 \!\! & \!\! = \!\! & \!\!  
 \frac{1}{k_BT}
 \int_0^t dt'' \; \Gamma^{-1}(t'-t'') 
 \left\langle  f_\xi[x(t),p(t)] \xi(t'') \right\rangle
 \label{eq:Novikov-coloured}
\end{eqnarray}
which reduces to (\ref{eq:Novikov}) for white noise.

\vspace{0.25cm}

\noindent
\refstepcounter{exercise}\label{ex-Nu3}\textcolor{orange}{\bf Exercise \thesection.\theexercise}
Extend the relation (\ref{eq:Novikov}) to the many variable case and prove that the generalization to 
the case with  coloured noise is correct.

\addtocounter{exercise}{1}

\vspace{0.25cm}

\textcolor{red}{
\subsubsection{The emergence of irreversibility}
}

\textcolor{red}{\it Time reversibility}  is the statement that 
$\{ x(-t), - p(-t) \}$ at time $-t$ satisfy the same equations as
$\{ x(t), p(t) \}$ at time $t$. 

In other words, define a reversed time $t_R=-t$, and a new variable $x_R(t_R)=x(-t)$, and 
ask whether the equation respected is the same. It is well known that Newton dynamics 
for conservative systems with time-independent potentials 
are time reversal. One was to see this is the following. Use the dynamics written in terms
of the position variable only written in the reversed frame:
\begin{equation}
0 = m \frac{d^2 x_R(t_R)}{dt_R^{\, 2}}  + V'(x_R(t_R)) = 0 \qquad\qquad \mbox{Newton}
\end{equation}
Rewrite now $t_R$ in terms of $t$ and $x_R(t_R)$ in terms of $x(-t)$:
\begin{equation}
0 = m \frac{d^2 x(-t) }{d(-t)^{2}} + V'(x(-t)) = m \frac{d^2 x(-t)}{dt^2}  + V'(x(-t))
\end{equation}
and the equation acting on $x(-t)$ is the same as the one acting on $x(t)$. 
This argument can be equivalently done introducing the velocity and recognizing that 
$v_R(t_R) = - v(-t)$.
Another way of writing the transformation is 
\begin{eqnarray}
{\mathcal T}_R:
\left\{
\begin{array}{rcl}
t & \mapsto & -t 
\\
[5pt]
\left. x(t') \right|_{t'=t} & \mapsto & \left. x(t') \right|_{t'= -t}
\\
[5pt]
\left. v(t')\right|_{t'=t} & \mapsto & \left. - v(t')  \right|_{t'=-t}
\end{array} 
\right.
\label{eq:time-reversal}
\end{eqnarray}


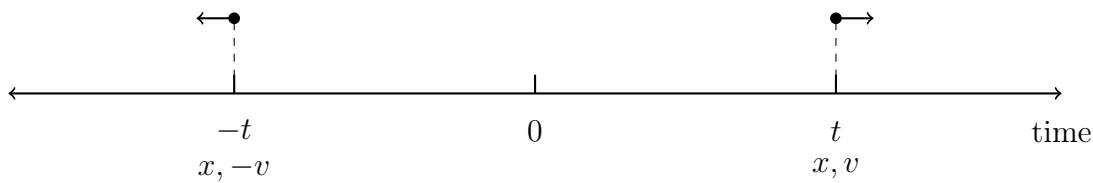
\begin{figure}[h!]
\centerline{
\begin{tikzpicture}
\draw[black, thick, <->] (-7,2) -- (7,2);
\draw[black, thick, -] (0,2) -- (0,2.25);
\draw[black, thick, -] (4,2) -- (4,2.25);
\draw[black, thick, -] (-4,2) -- (-4,2.25);
\draw[black, dashed, -] (4,2) -- (4,3);
\draw[black, dashed, -] (-4,2) -- (-4,3);
\draw (7,1.5) node {time};
\draw (4,1.5) node {$t$};
\draw (4,1) node {$x,v$};
\draw (-4,1.5) node {$-t$};
\draw (-4,1) node {$x,-v$};
\draw (0,1.5) node {$0$};
\filldraw[black] (4,3) circle (2pt);
\filldraw[black] (-4,3) circle (2pt);
\draw[black, thick, ->] (4,3) -- (4.5,3);
\draw[black, thick, ->] (-4,3) -- (-4.5,3);
\end{tikzpicture}
}
\caption{
Sketch of the time reversal process at the level of the Newton or Langevin equation
for the system's variables.}
\end{figure}

Now, consider the effect of the friction term in the Langevin equation in its simplest 
Markov form. In reversed variables it reads
\begin{equation}
-\gamma_0 v_R(t_R) = -\gamma_0 \frac{dx_R(t_R)}{dt_R} =  -\gamma_0 \frac{dx(-t)}{d(-t)}
= \gamma_0 \frac{dx(-t)}{dt} = \gamma_0 v(-t) 
\end{equation}
and the sign has changed. Therefore, this term is not invariant under time reversal. 

\comments{ 
Consider now the case with delayed friction. Start from the original one and 
transform according to (\ref{eq:time-reversal}):
\begin{equation}
-\int_0^t dt' \; \Gamma(t-t') \, v(t')
\quad
\mapsto
\quad
-\int_0^{-t} dt' \; \Gamma(-t-t') \, [-v(-t')]
\end{equation}
Now, the expression in the right, after a change of interaction variable from $t'$ to $-t'$, is also
\begin{equation}
\int_0^{t} d(-t') \; \Gamma(-t+t') \, v(t') = 
-\int_0^{t} dt' \; \Gamma(t-t') \, v(t')
\end{equation}
}

However, time reversibility is a property that has to be
respected by any set of \textcolor{blue}{microscopic Newtonian} dynamic equations.  Newton
equations describing the whole system, the particle and all the
molecules of the fluid, {\it are} time reversal invariant. 
Indeed, the dynamics of the selected variable is still reversible, 
in the sense that the pair $\{ x(-t), -p(-t) \}$ satisfies the same equation as 
$\{ x(t), p(t) \}$, as long as $N$ is kept finite {\it and} the oscillator positions and 
momenta are taken as $\{ q_a(0), -\pi_a(0)\}$ (instead of $\{ q_a(0), \pi_a(0)\}$). In order to prove this statement, 
we can start from the equation evaluated at $-t$. 
The tricky terms, stemming from the integration over the oscillators, evaluated at 
$-t$ are:
\begin{eqnarray}
&&
 \displaystyle{
\underbrace{- \frac{1}{m}  \int_0^{-t} dt' \ 
 \frac{c_{a}^2}{m_{a}\omega^2_{a}} \cos[\omega_{a}(-t-t')]
\, \overline p(t')}_{\rm dissipation}
}
\nonumber\\
&& \displaystyle{
\qquad  -
\underbrace{c_a \left[ 
\frac{\overline{\pi}_{a}(0)}{m_{a}\omega_{a}} 
 \sin(-\omega_{a}t)
+
\left(\overline{q}_{a}(0)+\frac{c_{a} \overline{x}(0)}{m_{a}\omega_{a}^2 }\right)
  \cos(-\omega_{a}t)
\right]
}_{\rm noise}
\; 
}
\nonumber\\
&&
= \displaystyle{
- \frac{1}{m} \int_0^{-t} dt' \ 
 \frac{c_{a}^2}{m_{a}\omega^2_{a}} \cos[\omega_{a}(t+t')]
\, \overline{p}(t')
}
\nonumber\\
&& \displaystyle{
\qquad  -
c_a \left[ 
-\frac{\overline{\pi}_{a}(0)}{m_{a}\omega_{a}} 
 \sin(\omega_{a}t)
+
\left(\overline{q}_{a}(0)+\frac{c_{a} \overline{x}(0)}{m_{a}\omega_{a}^2 }\right)
  \cos(\omega_{a}t)
\right]
\; 
}
\nonumber\\
&&
= \displaystyle{
+ \frac{1}{m} \int_0^{t} dt' \ 
 \frac{c_{a}^2}{m_{a}\omega^2_{a}} \cos[\omega_{a}(t-t')]
\,  \overline{p}(-t')
}
\nonumber\\
&& \displaystyle{
\qquad  -
c_a \left[ 
-\frac{\overline{\pi}_{a}(0)}{m_{a}\omega_{a}} 
 \sin(\omega_{a}t)
+
\left(\overline{q}_{a}(0)+\frac{c_{a} \overline{x}(0)}{m_{a}\omega_{a}^2 }\right)
  \cos(\omega_{a}t)
\right]
\; 
}
\label{eq:rev}
\end{eqnarray} 
where we did not write the sum over $a$ to alleviate the notation.  The overlined variables are 
yet not known, they are the ones that we will later choose to get the same form as for the 
original terms originating in the bath coupling. It is now clear that this aim is 
attained if one uses $\overline x(t) = x(-t)$, $\overline p(-t) = -p(-t)$, 
$\overline q_a(t) = q_a(-t)$, $\overline \pi_a(-t) = -\pi_a(-t)$, and the last condition implies
that $-\pi_a(0)$ has to be used in the last line.

\vspace{1cm}

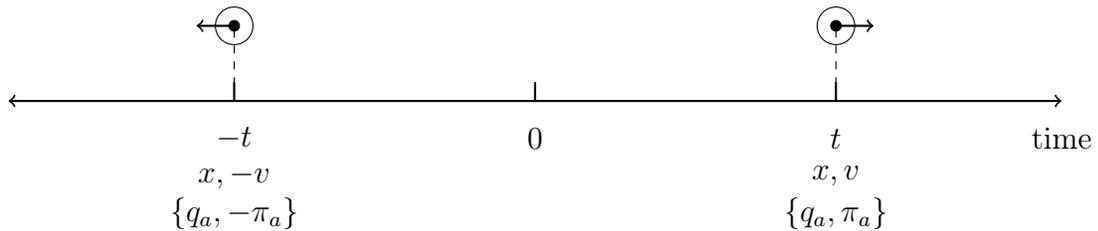
\begin{figure}[h!]
\centerline{
\begin{tikzpicture}
\draw[black, thick, <->] (-7,2) -- (7,2);
\draw[black, thick, -] (0,2) -- (0,2.25);
\draw[black, thick, -] (4,2) -- (4,2.25);
\draw[black, thick, -] (-4,2) -- (-4,2.25);
\draw[black, dashed, -] (4,2) -- (4,3);
\draw[black, dashed, -] (-4,2) -- (-4,3);
\draw (7,1.5) node {time};
\draw (4,1.5) node {$t$};
\draw (4,1) node {$x,v$};
\draw (4,0.5) node {$\{q_a,\pi_a\}$};
\draw (-4,1.5) node {$-t$};
\draw (-4,1) node {$x,-v$};
\draw (-4,0.5) node {$\{q_a,-\pi_a\}$};
\draw (0,1.5) node {$0$};
\filldraw[black] (4,3) circle (2pt);
\filldraw[black] (-4,3) circle (2pt);
\draw[fill=none] (4,3) circle (0.25);
\draw[fill=none] (-4,3) circle (0.25);
\draw[black, thick, ->] (4,3) -- (4.5,3);
\draw[black, thick, ->] (-4,3) -- (-4.5,3);
\end{tikzpicture}
}
\caption{
Sketch of the time reversal process at the level of the Newton's equations of the system and 
environmental variables.}
\end{figure}

However, \textcolor{blue}{time-reversal is broken in the reduced} equation
in which the fluid is treated in an effective statistical form. Indeed, the friction force $-\gamma_0 v$ in Eq.~(\ref{eq:langevin}) -- or 
its retarded extension in the non-Markovian case -- explicitly breaks
time-reversal ($t\to -t$) invariance. 

\textcolor{red}{
\subsubsection{Dissipation}
}

Even in the case in which all forces derive from a potential, 
$F=-dV/dx$, the energy of the particle, $H_{\rm syst}=mv^2/2+V$, is not conserved by Eq.~(\ref{eq:langevin}). 
This can be easily seen
by taking $dH_{\rm syst}/dt=mv \dot v + V' v = v (-\gamma_0 v +\xi)$, say, in the case of additive white noise.
(Note that we used the standard rules of calculus here, the Stratonovich convention. We will say more 
about discretization subtleties later.)
The second member does not vanish in general. Its sign is not determined  
unless at zero-temperature, when it is negative semi-definite, $-\gamma_0 v^2$, indicating that the 
dynamics are of \textcolor{blue}{gradient descent}.
At zero temperature and in the absence of non-potential energy injecting forces,  one finds that 
on average the energy flows to the bath leading to \textcolor{blue}{dissipation}. At very long times,
however, the particle may reach a stationary regime in which  
the particle gives and receives energy from the bath at equal rate,
and then $\langle H_{\rm syst}\rangle \to {\rm ct}$ saturates to a constant value. 
We will see this mechanism at work in some examples.

\vspace{0.25cm}

\noindent
\textcolor{orange}{\bf Exercise \thesection.\theexercise}
Prove the time-irreversibility of the Langevin equation and the fact that 
the symmetry is restored if $\gamma_0=0$.

\addtocounter{exercise}{1}

\vspace{0.25cm}

\noindent
\textcolor{orange}{\bf Exercise \thesection.\theexercise}
Show that $d\langle H_{\rm syst} \rangle/dt\neq 0$ 
when $\gamma_0 \neq 0$. Prove that for a single variable, the non-zero terms can be found from the 
analysis of  $t'\to t^-$ and $t'\to t^+$ limits of
$\gamma_0 \langle \dot x(t) \dot x(t') \rangle - 2\gamma_0 T \langle \xi(t) \dot x(t')\rangle$. Relate the last
term to the linear response function.
Discuss the equilibrium case, in which $m \langle {\dot x}^2(t) \rangle = k_B T$ (equipartition).

\addtocounter{exercise}{1}

\textcolor{red}{
\subsubsection{Smoluchowski (over-damped) limit}
\label{sec:smoluchowski}
}

In many situations in which friction is very large, the characteristic 
time for the relaxation of the velocity degrees of freedom to their 
Maxwell distribution, $t_r^v$, is very short (see the examples in Sect. 2.5). In consequence, 
observation times are very soon longer than this time-scale, 
the inertia term $m\dot v$ can be dropped, and the Langevin equation becomes 
\begin{equation}
\gamma_0 \dot x =F +\xi
\label{eq:smoluchowski}
\end{equation}
(for simplicity we wrote the white-noise case).
Indeed, this \textcolor{blue}{ over-damped} limit is acceptable
whenever the observation times are much longer than the characteristic
time for the velocity relaxation. Inversely, the cases in which the
friction coefficient $\gamma_0$ is small are called \textcolor{blue}{
  under-damped}.

In the over-damped limit with white-noise the friction
coefficient $\gamma_0$ can be absorbed in a rescaling of time.
One defines the new time $\tau$ 
\begin{equation}
t=\gamma_0\tau 
\; , 
\end{equation}
the new position,  
${\tilde  x}(\tau) = x(\gamma_0\tau)$,
and the new noise $\eta(\tau) =
\xi(\gamma_0\tau)$. In the new variables the Langevin equation reads
$\dot {\tilde x}(\tau) = F(\tilde x,\tau) + \eta(\tau)$ with $\langle
\eta(\tau) \eta(\tau') \rangle = 2 k_B T \delta (\tau-\tau')$.

\vspace{0.25cm}

\noindent
\textcolor{orange}{\bf Exercise \thesection.\theexercise }
Redefine time as $\tau=k_BT \gamma_0^{-1} \ t$ and $\tilde x(\tau) = x(\gamma_0 \beta t)$,
and similarly for the other function of time, to transform the Langevin equation into
\begin{equation}
\dot {\tilde x} = \beta \tilde F + \tilde \xi
\; ,
\qquad\qquad
\langle \tilde \xi(\tau) \tilde \xi(\tau') \rangle = 2 \delta(\tau-\tau')
\; . 
\end{equation}
This equation is not well adapted to take the $T\to 0$ limit.

\addtocounter{exercise}{1}

\vspace{0.25cm}

\textcolor{red}{
\subsubsection{Markov character and generation of memory}
}

In the case of a white noise (delta correlated) 
the full set of equations defines a \textcolor{blue}{ 
  Markov process}, that is a stochastic process that depends on its
history only through its very last step. 

The Langevin equation (\ref{eq:langevin}) is actually a set of two
first order differential equations. Notice, however, that the
pair of first-order differential equations could also be described by
a single second-order differential equation:
\begin{equation}
m \ddot x + \gamma_0 \dot x = F + \xi
\; . 
\label{eq:langevin-second}
\end{equation}
Having replaced the velocity by its definition in terms of the
position $x(t)$ depends now on $x(t-\delta)$ and $x(t-2\delta)$. This is a very
general feature: by integrating away some degrees of freedom (the
velocity in this case) one generates memory in the evolution.
Generalizations of the Langevin equation, such as the one that we have
just presented with colored noise, and the ones that will be generated to describe the
slow evolution of super-cooled liquids and glasses in terms of
correlations and linear responses, do have memory~\cite{Cugliandolo}.

A simple solvable example that illustrates this feature is the under-damped white-noise 
Langevin equation for a harmonic oscillator:
\begin{eqnarray}
m\dot x = p \; , \qquad\qquad
\dot p = - m \omega^2 x - \frac{\gamma_0}{m}  \ p + \xi \; , 
\end{eqnarray}
with initial condition $p(0) =0$ and white noise such that 
$\langle \xi(t) \xi(t') \rangle = 2\gamma k_BT \delta(t-t')$. The second equation can be solved exactly 
\begin{equation}
p(t) = \int_{0}^t dt' \ e^{-\frac{\gamma_0}{m}  (t-t')} \ [-m\omega^2 x(t') + \xi(t') ] 
\end{equation}
and when this solution is put back into the equation for $\dot x$ one finds an 
equation with memory,
\begin{equation}
\dot x(t) = - \int_0^t dt' \ K(t-t') x(t') + \xi_x(t) 
\; , 
\end{equation}
with 
\begin{eqnarray}
K(t) = \omega^2 \ e^{-\frac{\gamma_0}{m}  t}  
\qquad \mbox{and} \qquad 
\xi_x(t) = \frac{1}{m} \int_0^\infty dt' \ e^{-\frac{\gamma_0}{m} (t-t')} \ \xi(t')
\; . 
\end{eqnarray}
The statistical properties  of the new noise can be derived from the ones  of the 
original noise $\xi$. It keeps the zero average and its correlations are
\begin{equation}
\langle \xi_x (t) \xi_x(t') \rangle = \frac{k_BT}{m\omega^2} \ K(|t-t'|) \; . 
\end{equation}
The solution of this problem will be developed in Sect.~\ref{sec:harmonic oscillator}.

Whenever a variable is integrated out from a Markov set of equations a non-Markov equation is obtain. 
Proceeding in the reverse sense one can transform a non-Markov equation with exponentially correlated
noise into a set of Markov equations that are easier to integrate numerically.

\vspace{0.25cm}
\textcolor{red}{
\subsubsection{Distinction between relaxation and equilibration}
}

A system coupled to an environment can relax to a non-equilibrium steady state, usually called a {\rm NESS}, 
that is not the one of thermal equilibrium, $P(v,x; t) \to P_{\rm NESS}(v,x)\neq P_{\rm GB}(v,x)$, that is, 
that is different from the Gibbs-Boltzmann measure.

\textcolor{red}{
\subsubsection{Discretization of stochastic differential equations}
}
\label{subsubsec:discretization}

The way in which the stochastic differential equation (\ref{eq:smoluchowski})
(with no inertia and with white noise)
is to be discretized is a subtle matter. Two schemes are the most popular ones, 
called the It\^o and Stratonovich calculus,
and are rather well documented in the literature~\cite{vanKampen-Ito,Oksendal}.

\begin{figure}[h!]
\centerline{
\includegraphics[scale=0.75]{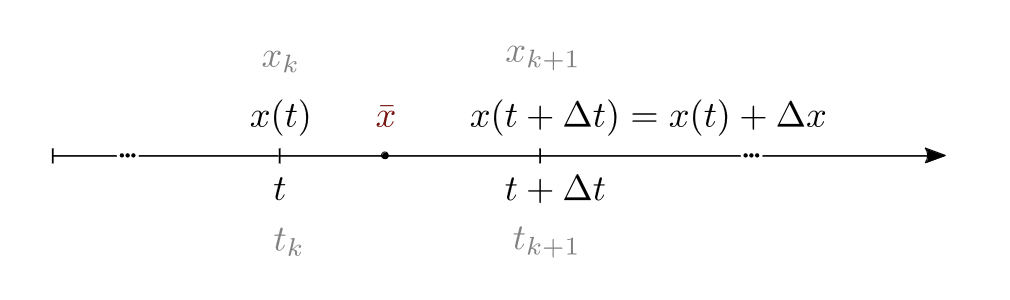}
}
\caption{\small A sketch of the time axis, discretized with step $\Delta t$.}
\end{figure}

\comments{
In short, we will use a prescription in which the pair velocity-position of the
particle at time $t+\delta$, with $\delta$ an infinitesimal time-step,
depends on the pair velocity-position at time $t$ and the value of the
noise at time $t$. 
}

Let us try to explain, in a simple way, the origin of the subtleties and how they are 
controlled. We discretize time according to $t_n=n \Delta t + t_0$ with $n$ an 
integer running as $n=0, \dots, {\cal N}$. The continuous time limit corresponds
to $\Delta t\to 0$, ${\cal N} \to \infty$ while keeping ${\cal N} \Delta t = {\cal T}-t_0$ fixed and 
$t_{\cal N} = {\cal T}$. We now 
take a single real variable $x$ the dynamics of which is governed by 
the following stochastic equation
\begin{equation}
\fbox{$d_t x(t) = f(x) + g(x) \xi(t) 
$}
 \label{eq:x-eom}
\end{equation}
with multiplicative white noise distributed according to a Gaussian pdf 
with zero mean and 
variance $\langle \xi(t) \xi(t') \rangle = 2 k_BT \delta(t-t')$. 
This stochastic differential equation makes sense only when a discretization rule 
is explicitly given to define it. 

We use the short-hand notation $x_n = x(t_n)$ and $\xi_n = \xi(t_n)$ The white noise 
statistics correspond to $\langle \xi_n\rangle =0$ and $\langle \xi_n \xi_m \rangle = (2k_BT/\Delta t) \ \delta_{nm}$ that implies $\xi_n^2 \simeq 2k_BT/\Delta t$ and 
 $\xi_n \simeq {\cal O}(\Delta t^{-1/2})$ (we use here the step realisation of the Dirac delta function as being identical to 0 away from the 
interval $[-\Delta t/2, \Delta t/2]$ and equal to $1/\Delta t$ within this interval).
We will use the generic $\alpha$ prescription~\cite{Gardiner}
\begin{equation}
\boxed{
x_{n+1}-x_n = f(\overline x_n) \Delta t + g(\overline x_n) \xi_n \Delta t
}
\label{eq:used-for-calc}
\end{equation}
with 
\begin{equation}
\boxed{
\overline x_n \equiv \alpha x_{n+1} + (1-\alpha) x_n 
}
\end{equation}
and $0 \leq \alpha \leq 1$ in the following. $\alpha=0$ is the It\^o prescription while $\alpha=1/2$ is the Stratonovich one.

The definition of $\overline x_n$ implies
\begin{eqnarray}
\overline x_n &=& x_n + \alpha ( x_{n+1} - x_n) 
\\
\overline x_n &=& x_{n+1} - (1-\alpha) (x_{n+1}-x_n)
\end{eqnarray}
and
\begin{eqnarray}
x_n &=& \overline x_n - \alpha (x_{n+1} - x_n) 
\\
x_{n+1} &=& \overline x_{n} + (1-\alpha) (x_{n+1} - x_n) 
\end{eqnarray}
which allow one to rewrite the $\alpha$-prescription equation as
\begin{eqnarray}
x_{n+1} - x_n &=& 
[f(x_n) + f'(x_n) \alpha (x_{n+1}-x_n) +{\cal O}((x_{n+1}-x_n)^2)] \ \Delta t \nonumber\\
&& + [g(x_n) + \alpha g'(x_n) (x_{n+1}-x_n) +{\cal O}((x_{n+1}-x_n)^2)] \ \xi_n \Delta t
\nonumber\\
&\simeq &
f(x_n) \Delta t + g(x_n) \xi_n \Delta t + \alpha g'(x_n) (x_{n+1}-x_n)  \xi_n \Delta t
\end{eqnarray}
where we used $\xi_n ={\cal O}( \Delta t^{-1/2})$ to estimate the relevant contributions up to ${\cal O}(\Delta t)$
and we dropped terms ${\cal O}(\Delta t^{3/2})$. Replacing $x_{n+1}-x_n$ in the last term by the 
outcome of the same equation to  order $\Delta t^{1/2}$:
\begin{eqnarray}
\Delta x_n \equiv x_{n+1} - x_n \simeq 
f(x_n) \Delta t + g(x_n) \xi_n \Delta t + \alpha g'(x_n) g(x_n) (\xi_n \Delta t)^2 
\; . 
\end{eqnarray}
As $(\xi_n \Delta t)^2 \simeq 2k_BT \Delta t$, all terms in the rhs are of order $\Delta t$.
We will use this expression to derive the Fokker-Planck equation.

The force term in the stochastic equation is also sometimes written as 
\begin{eqnarray*}
f(\overline x_n) &=& 
\alpha f(\overline x_n) + (1-\alpha) f(\overline x_n) 
\nonumber\\
&=&
\alpha f(x_n + \alpha \Delta x_n) + (1-\alpha) f(x_{n+1} - (1-\alpha) \Delta x_n)
\nonumber\\
&=& 
\alpha [f(x_n) + \alpha f'(x_n) \Delta x_n +\dots ] 
\nonumber\\
&& \qquad  + (1-\alpha) [f(x_{n+1} - (1-\alpha) f'(x_{n+1}) \Delta x_n + \dots]
\nonumber\\
&=&
\alpha f(x_n) + (1-\alpha) f(x_{n+1}) + {\mathcal O}(\Delta x_n)
\end{eqnarray*}
and dropping the ${\mathcal O}(\Delta x_n)$ terms that appear multiplied by $\Delta t$ and give rise to negligible
terms of ${\mathcal O}(\Delta t^{3/2})$, one has 
\begin{eqnarray}
f(\overline x_n) = 
\alpha f(x_n) + (1-\alpha) f(x_{n+1})
\end{eqnarray}
This is ultimately equivalent to writing the force term as $f(x_n) \Delta t$ in the Langevin equation.

\vspace{0.25cm}
\noindent 
\textcolor{red}{\it The chain rule}
\vspace{0.25cm}

As explained in~\cite{Gardiner}, the chain-rule for the time-derivative of a function $V$ of the 
variable $x$ depends on the stochastic equation  governing the evolution of $x$; we call it the $x$-chain rule and 
for Eq.~(\ref{eq:x-eom}) it reads
\begin{equation} 
\label{eq:x-chain0}
\boxed{d_t V = \dot x \, V' + (1-2\alpha) k_BT  g^2  V''
}
\end{equation}
where $\dot x=d_t x = dx/dt$, $v' = \partial_x V$ and $V''= \partial^2_x V$.
Note that the chain rule is independent of $f(x)$ (that is to say, it will take the same form for a Langevin equation 
with the drift term, Eq.~(\ref{eq:x-eom-drifted}), to be discussed below). Somehow surprisingly, the second term 
is still present for $g=1$, the additive noise case. It only disappears and one recovers normal calculus
for $\alpha=1/2$.

We now prove Eq.~(\ref{eq:x-chain0}).
Let us write the difference between a generic function $V$ evaluated at $x$ at two subsequent times $n+1$ and $n$. 
We expand $x_n$ around the generic $\alpha$ point $\bar x_n$ we get
\begin{eqnarray}
& V(x_{n+1}) - V(x_n) = V(\bar x_n + (1-\alpha)(x_{n+1} - x_n)) - V(\bar x_n - \alpha(x_{n+1} - x_n)) \nonumber \\
& 
\qquad\quad
= (x_{n+1} - x_n) V'(\bar x_n) + \frac{1}{2}(1-2\alpha) (x_{n+1} - x_n)^2 V''(\bar x_n) + \mathcal{O}(\Delta t^{3/2})
\nonumber
\end{eqnarray}
where $\Delta x_n =x_{n+1}-x_n$.
Using \fbox{$(x_{n+1} - x_n)^2  = 2 k_BT g(\bar x_n)^2 \Delta t + \mathcal{O}(\Delta t^{3/2})$} 
from Eq.~(\ref{eq:used-for-calc}) after setting 
$(\xi_n \Delta t)^2 = 2k_BT \Delta t$, where the crucial fact is that this square is of
order $\Delta t$ (instead to $\Delta t^2$) 
we obtain
\begin{align}
V(x_{n+1}) - V(x_n) &= (x_{n+1} - x_n) V'(\bar x_n) + (1-2\alpha)k_BT g(\bar x_n)^2 V''(\bar x_n) \Delta t + \mathcal{O}(\Delta x^3)
\nonumber
\end{align}
and dropping terms of order $\Delta t^{1/2} $ or higher,
\begin{equation}
\boxed{
\frac{V(x_{n+1}) - V(x_n)}{\Delta t}= \frac{x_{n+1} - x_n}{\Delta t} V'(\bar x_n) + (1-2\alpha)k_BT g(\bar x_n)^2 V''(\bar x_n)} 
\label{eq:chain-ruledisc} 
\end{equation} 
which is the chain-rule.
As above, at this order one can replace the 
$\overline x_n$ in $g$, $V'$ and $V''$ by any $x$ in the interval.
This expression is next written as in Eq.~(\ref{eq:x-chain0}).


\pagebreak

\noindent
\textcolor{red}{\it Influence of the discretisation on the trajectories}

\vspace{0.25cm}

One can estimate the importance of the discretisation on the actual trajectories found by computing the 
difference between the two terms in the right-hand-side of the Langevin equation obtained for a 
discretisation $\alpha$ and another discretisation $\overline \alpha$.

Let us start with the first term, the one equal to $f$.
\begin{eqnarray}
&& f(\overline x_\alpha(t_k)) - f(\overline x_{\overline\alpha}(t_k)) 
\nonumber\\
&& 
\qquad
= 
f(x(t_k)) + f'(x(t_k)) \alpha \Delta x_k - f(x(t_k)) - f'(x(t_k)) \overline\alpha \Delta x_k + {\cal O}(\Delta x_k^2)
\nonumber\\
&& 
\qquad
= 
f'(x(t_k)) (\alpha - \overline \alpha) \Delta x_k + {\cal O}(\Delta x_k^2) \to 0 \qquad \mbox{for} \qquad \Delta t \to 0
\end{eqnarray}
the last results being due to the fact that $\Delta x_k = {\cal O}(\Delta t^{1/2})$.

Now, we will find that the difference between the two noise-dependent terms evaluated at different 
discretisation parameters does not vanish in the same limit:
\begin{eqnarray}
&& 
g(\overline x_\alpha(t_k)) \xi(t_k) - g(\overline x_{\overline\alpha}(t_k)) \xi(t_k) 
\nonumber\\ 
&&
\qquad
= [g(x(t_k)) + g'(x(t_k)) \alpha \Delta x_k - g(x(t_k)) - g'(x(t_k)) \overline\alpha \Delta x_k + {\cal O}(\Delta x_k^2)] \xi(t_k)
\nonumber\\
&& 
\qquad
= g'(x(t_k)) (\alpha - \overline \alpha) \xi(t_k) \Delta x_k + {\cal O}(\Delta x_k^2) \xi(t_k) = {\cal O}(\Delta t^0)  
\end{eqnarray}

In consequence, there is a difference of order one in the trajectories found with one and the another 
discretisation scheme.

\vspace{0.25cm}

\noindent
\textcolor{red}{\it Numerical integration of the Langevin equation}

\vspace{0.25cm}

The numerical integration of the Langevin equation requires the discretisation of time, $t_k=k \Delta t$ where 
$k$ is an integer and $\Delta t$ the time-step. The choice of the optimal value of $\Delta t$ has to be gauged 
depending on the accuracy of the numerical integration desired (the smallest the $\Delta t$ the best) 
and the length of the time-interval wished to be analysed (one cannot take it to be so small because otherwise
only too short time-scales are explored). Several algorithms are described in~\cite{Kloeden}.

In the over-damped limit with additive white noise the most common algorithm used 
is just the simple iteration of the \textcolor{blue}{Ito relation}
\begin{equation}
x(t_k) = x(t_{k-1}) - \Delta t \, V'(x(t_{k-1})) + \Delta t \, \xi(t_{k-1})
\; . 
\end{equation}
The only practical issue to stress here is that one needs to consider the time-discretised version of the 
delta-correlated white noise $\xi$:
\begin{equation}
\langle \xi(t_k) \xi(t_n) \rangle = \frac{2k_BT}{\Delta t} \qquad \mbox{if} \qquad |t_k-t_n| = \Delta t |k-n| < dt/2
\end{equation}
that implies $\xi(t_k) = \sqrt{2k_BT/\Delta t} \ \eta_k$ with $\langle \eta_k \eta_n\rangle = \delta_{kn}$.
 
\pagebreak

\noindent
\textcolor{red}{\it The inertial Langevin equation}

\vspace{0.25cm}

Take now the generic stochastic system
\begin{eqnarray}
&& d_tx = \frac{\partial H}{\partial p} = \frac{p}{m}
\label{eq:x-eq}
\\
&& d_t p = -\frac{\partial H}{\partial x} - \gamma d_t x + g(x,p) \xi(t)
\label{eq:p-eq}
\end{eqnarray}
with $\xi(t)$ a white noise with zero average and correlations $\langle \xi(t) \xi(t') \rangle = 2\gamma k_B T \delta(t-t')$
which imply $\xi_n = {\mathcal O}(\Delta t^{-1/2})$.
Write it in discretized form
\begin{eqnarray}
&& x_{n+1} - x_n = h(\overline x_n, \overline p_n) \, \Delta t
\\
&&  p_{n+1} - p_n = f(\overline x_n, \overline p_n) \, \Delta t - \gamma (x_{n+1} - x_n)
 + g(\overline x_n, \overline p_n) \xi_n \Delta t
\end{eqnarray}
where we introduced the names $h$ and $f$ for the two first terms in the right-hand-sides 
for notation convenience and to make contact with what we discussed in the case 
without inertia.
Take $\overline x_n = x_n + \alpha \Delta x_n$ and $\overline p_n = p_n + \alpha' \Delta  p_n$. One can repeat 
the calculations done so far to show that the chain rule becomes
\begin{equation}
\frac{dV(x,p)}{dt} = \frac{\partial V(x,p)}{\partial x} \, \dot x + \frac{\partial V(x,p)}{\partial p} \, \dot p
+  (1-2\alpha') \frac{\partial^2 V(x,p)}{\partial p^2} k_BT g(x,p)^2
\end{equation}
What about the trajectories? Do they depend on $\alpha$ and/or $\alpha'$?
Let us compare the expressions of $h$ and $f$ for two processes, one with $\alpha, \alpha'$ and 
the other with  $\overline\alpha, \overline\alpha'$:
\begin{eqnarray}
&& 
h(\overline x_\alpha(t_k), \overline p_{\alpha'}(t_k))-h(\overline x_{\overline\alpha}(t_k), \overline p_{\overline\alpha'}(t_k))
\nonumber\\
&& 
\quad\;\;\;
\sim
h(x(t_k),  p(t_k))
+ \frac{\partial h( x(t_k),  p(t_k))}{\partial x(t_k)} \alpha \Delta x_k 
+ \frac{\partial h( x(t_k),  p(t_k))}{\partial p(t_k)} \alpha' \Delta p_k
\nonumber\\
&& 
\quad\;\;\;\;\;\;
- h(x(t_k),  p(t_k))
- \frac{\partial h( x(t_k),  p(t_k))}{\partial x(t_k)} \overline\alpha \Delta x_k 
- \frac{\partial h( x(t_k),  p(t_k))}{\partial p(t_k)} \overline\alpha' \Delta p_k
\nonumber\\
&& 
\quad\;\;\;
=
(\alpha- \overline\alpha)
 \frac{\partial h( x(t_k),  p(t_k))}{\partial x(t_k)} \Delta x_k 
 +
(\alpha'- \overline\alpha') \frac{\partial h( x(t_k),  p(t_k))}{\partial p(t_k)} \overline\alpha' \Delta p_k
\nonumber
\end{eqnarray}
The first term is ${\mathcal O}(\Delta t)$ and the second ${\mathcal O}(\Delta t^{1/2})$. Therefore, for 
$\Delta t\to 0$, the full difference vanishes. This also applies to the first term in the right hand 
side of Eq.~(\ref{eq:p-eq}), the $f$ term in the discretized version. Care should be taken 
with the noise term, and a similar analysis yields
\begin{eqnarray}
&& 
g(\overline x_\alpha(t_k), \overline p_{\alpha'}(t_k)) \xi(t_k) dt
\sim
\frac{\partial g(x(t_k), p(t_k))}{\partial p(t_k)}  (\alpha'-\overline\alpha') \xi(t_k) \Delta p_k
\sim {\mathcal O}(\Delta t^0)
\nonumber
\end{eqnarray}
Thus, for processes with a multiplicative noise such that the function $g$ depends on 
the momentum the trajectories will diverge. Otherwise they will not.

\vspace{0.25cm}

\noindent
\textcolor{orange}{\bf Exercise \thesection.\theexercise} Write an algorithm that integrates the Langevin equation and reproduce and check the claims made about the trajectories
in this Section.

\addtocounter{exercise}{1}

\textcolor{red}{
\subsection{Observables}
}

As usual in statistical and quantum mechanics 
meaningful quantities are averaged observables. For an 
equilibrated system, due to ergodicity, one can either take an ensemble 
average or an average over a sufficiently long time-window. Out of equilibrium 
these do not coincide in general. 

The interaction with the environment induces fluctuations and 
the Langevin equation is solved in a probabilistic sense, 
\begin{equation}
x^{\rm sol}_{\xi_n}(t) = {\cal F}[(\xi_n),x_0,t] 
\; .
\label{sol_stoch}
\end{equation}
The index $n$ labels different realizations of the thermal history, 
{\it i.e.} different realizations of the noise at each instant. 
$x_0$ is the initial condition $x_0=q(0)$ and, for each `run' of the 
dynamics, $(\xi_n)$ encodes all
noise values in the interval $[0,t]$.  
Equation (\ref{sol_stoch}) means that there is a different solution
for each noise history. 


\textcolor{red}{
\subsubsection{One-time averages}
}

 Any {\it one-time} functional of $x$, 
$A[x](t)$, must be averaged over all histories of the thermal 
noise to obtain a deterministic result
\begin{equation}
\langle  A[x](t) \rangle =
\lim_{{\cal N}\to \infty} 
\sum_{n=1}^{{\cal N}} A[x^{\rm sol}_{\xi_n}](t) P(\xi_n) 
\; .
\label{calG}
\end{equation}
${\cal N}$ is here the number of noise realizations.
$P(\xi_n)$ is the probability distribution of the $n$-th thermal history.
For a Gaussian noise
\begin{equation}
P(\xi_n) \propto \exp\left[-\frac{1}{2k_BT} \sum_{ab} \xi_n(t_a) 
\Gamma^{-1}(t_a-t_b) \xi_n(t_b) \right]
\; .
\label{gaussian-noise}
\end{equation}
One condenses these expressions in a continuous notation such that 
\begin{equation}
\langle  A[x](t) \rangle =
\int {\cal D}\xi \, P[\xi] \, A[x^{\rm sol}_{\xi}](t)
\; .
\label{calG}
\end{equation}
The measure of this functional integral is just 
${\cal D}\xi \equiv \prod_{k} d\xi(t_k)$ with $k$ now 
labelling the discretized times.

\textcolor{red}{
\subsubsection{Two-time correlations}
}

{\it Two-time functions} characterize the out of 
equilibrium dynamics in a more detailed way and they are 
defined as 
\begin{equation}
C_{AB}(t,t') \equiv 
\langle A[x](t) B[x](t') \rangle
= \int {\cal D}\xi \, P[\xi] \, A[x^{\rm sol}_{\xi}](t) B[x^{\rm sol}_{\xi}](t')
\; .
\end{equation}
The observable $B[x]$ is measured
at time $t'$, the observable $A[x]$ is measured at time $t$
for each noise realization and the average is taken 
afterwards. 

\vspace{0.25cm}

\begin{figure}[h!]
\centerline{
\begin{tikzpicture}
\draw (-7,1.75) node {(a)};
\draw (-7,1) node {$h$};
\draw[gray, dashed, -] (-6.5,1) -- (-5,1);
\draw[black, thick, ->] (-7,0) -- (-2,0);
\draw[red, thick, -] (-5,0) -- (-5,1);
\draw[red, thick, -] (-5,1) -- (-4,1);
\draw[red, thick, -] (-4,1) -- (-4,0);
\draw (-2,-0.5) node {$t$};
\draw[black, thick, -] (-4.5,0) -- (-4.5,0.125);
\draw (-4.5,-0.5) node {$t'$};
\draw (-5.75,-0.5) node {\tiny $t'- \frac{\Delta t}{2}$};
\draw (-3.25,-0.5) node {\tiny $t'+\frac{\Delta t}{2}$};
\draw (0,1) node {$h$};
\draw[gray, dashed, -] (0.5,1) -- (2,1);
\draw (0,1.75) node {(b)};
\draw[black, thick, ->] (0,0) -- (5,0);
\draw[red, thick, -] (2,0) -- (2,1);
\draw[red, thick, -] (2,1) -- (5,1);
\draw (5,-0.5) node {$t$};
\end{tikzpicture}
}
\caption{(a) An instantaneous perturbation of strength $h$ applied at $t'$, 
of duration $\Delta t \to 0$.
(b) A step perturbation of strength $h$ applied at $t'$ and held constant for
all subsequent times.}
\label{pert-resp}
\end{figure}
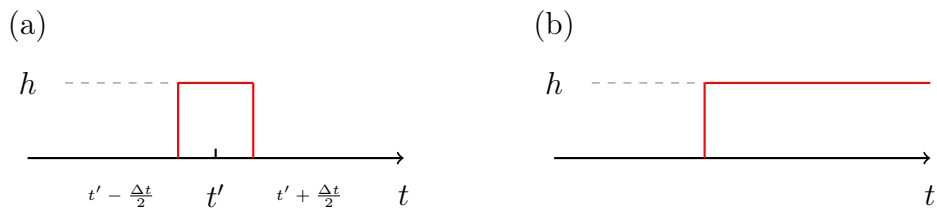

\textcolor{red}{
\subsubsection{The averaged linear response}
}

The instantaneous linear response is also a two-time function.
Imagine that $x$ represents the position of a Brownian 
particle that one {\it kicks}
with a 
weak perturbing force at time $t'$ (see Fig.~\ref{pert-resp}). 
The subsequent position
of the particle is modified by the perturbation. The linear
response is given  by the comparison of the
perturbed dynamics with the unperturbed one, in which no force has been 
applied, up to linear order in the perturbation:
\begin{equation}
R_{AB}(t,t') \equiv \left. \frac{\delta \langle A[x]\rangle(t)}
{\delta h_B(t')} \right|_{h_B=0}
\; . 
\label{linear-response}
\end{equation}
The subindex $B$ indicates that the perturbation applied at $t'$ 
is conjugated to 
the observable $B[x]$ when
the Hamiltonian is modified as
$H \to H - h_B B[x]$. 
The subindex $A$ indicates that we examine how the observable $A[x]$ 
at time $t$ 
reacts to the perturbation. At the end of the calculation 
we set $h_B=0$ to extract the linear response. Keeping $h_B\neq 0$ 
yields information about the nonlinear terms in the response function.
For {\it causal} systems the response function vanishes if $t'>t$.

One is also interested in the integrated linear response
rather than the instantaneous one. This quantity 
represents the linear response of the system to a {\it step}-like 
perturbation of duration $t-t'$ that starts at $t'$, as represented on 
the right panel of Fig.~\ref{pert-resp}:
\begin{equation}
\chi_{AB}(t,t') \equiv \int_{t'}^t dt'' \; R_{AB}(t,t'')
\; .
\label{int-linear-response}
\end{equation}

Rather often results are 
presented in the frequency domain. One defines the Fourier 
transform and its inverse
\begin{eqnarray}
\tilde A(\omega) = \int_{-\infty}^\infty  dt \, 
\exp\left(-i\omega t\right) \, A(t)
\; ,
\label{fourier}
\;\;\;\;\;\;\;\;
A(t) = \int_{-\infty}^\infty  \frac{d\omega}{2\pi} \, 
\exp\left(i\omega t\right) \, \tilde A(\omega)
\; .
\label{inv-fourier}
\end{eqnarray}
For a stationary process,  
the linear susceptibility, $\tilde\chi(\omega)$,
is simply given by
the Fourier transform of the linear response (\ref{linear-response}). 
However
$R_{AB}(t,t')$ is not necessarily stationary out of equilibrium.
Hence, we define two generalized linear susceptibilities,
\begin{eqnarray}
\tilde \chi^{(1)}_{AB}(\omega,t') &\equiv& \int_0^\infty d\tau \; 
\exp\left(-i\omega\tau \right) R_{AB}(t'+\tau,t')
\; ,
\label{chi1}
\\
\tilde \chi^{(2)}_{AB}(\omega,t) &\equiv& \int_0^\infty d\tau \; 
\exp\left(-i\omega\tau \right) R_{AB}(t,t-\tau)
\; ,
\label{chi2}
\end{eqnarray}
that reduce to the  well-known expression for $\tilde\chi_{AB}(\omega)$
in a stationary system. Note that in the first line we kept the
shorter time ($t'$) fixed while in the second line we kept the 
longer time ($t$) fixed.
These expressions have a real and an imaginary part that yield
the in-phase ($\tilde\chi'$) 
and the out-of-phase ($\tilde\chi''$) susceptibilities, respectively. 
The integrations run over positive values of 
$\tau$ only due to causality. 


\vspace{0.25cm}

\noindent
\textcolor{red}
{\it The linear response as a correlation with the noise}

\vspace{0.25cm}

Thanks to the Novikov theorem, one can derive a useful expression for the 
linear response in the absence of an applied field. Take the simple case in 
which one is interested in the linear response of a generic function of the 
stochastic variable to a linear perturbation applied to the stochastic variable:
\begin{equation}
R(t,t') = \left. \frac{\delta \langle x(t)\rangle}{\delta h(t')}\right|_{h=0}
\end{equation}
for the Langevin process with $F \mapsto F + h$
\begin{equation}
m\dot x = p
\qquad\qquad
 \dot p + \frac{\gamma_0}{m}  p = F + h + \xi
\end{equation}
Then,  it is clear that 
\begin{eqnarray}
\left\langle \dfrac{\delta f_\xi[x(t),p(t)]}{\delta h(t')} \right\rangle
=
\left\langle \dfrac{\delta f_\xi[x(t),p(t)]}{\delta \xi(t')} \right\rangle
\end{eqnarray}
for any $f_\xi[x(t),p(t)]$ and using Novikov's theorem
\begin{eqnarray}
2k_BT R_{fx}(t,t') =  \left\langle  f_\xi[x(t),p(t)] \xi(t') \right\rangle
 \label{eq:NovikovR}
\end{eqnarray}
for white noise, and 
\begin{eqnarray}
k_BT R_{fx}(t,t') = \int_0^t dt'' \; \Gamma^{-1}(t'-t'') \left\langle  f_\xi[x(t),p(t)] \xi(t'') \right\rangle
 \label{eq:NovikovR-coloured}
\end{eqnarray}
for coloured noise.

To compute the linear response from its definition can be hard. One needs to 
apply a sufficiently strong field to have a signal but also take it to zero to 
ensure the linear response regime. For these reasons several expressions 
of the linear response as correlation of observables in the absence of perturbation 
have been derived and used. We refer to them in Sec.~\ref{subsubsec:FDT}.
 
\textcolor{red}{
\subsubsection{Higher-order correlations}
}

Up to now  we have only discussed one-point 
and two-point functions. In general problems, higher order
functions are not trivially related to the previous ones and 
bear richer information. These are
{\it four-point functions}, 
$\langle A(t) B(t') C(t'') D(t'') \rangle$, 
or any other form of a more general type. In most
of the solvable models we shall discuss below, and in most of the 
approximations used to analyze realistic models, higher-order 
functions do not appear. The reasons for their disappearance
are manifold. In simplified models one can 
simply prove that higher order functions are exactly given 
in terms of one and two-point functions. In more realistic cases, 
higher order functions are 
approximated with expressions that depend on one and 
two-point functions only. This is done, for instance, in 
Gaussian approximations and mode-coupling theory~\cite{Cugliandolo}.
However, a complete solution to a finite dimensional model should 
be able to predict the behavior of such higher order correlations. 

\textcolor{red}{
\subsubsection{The fluctuation dissipation theorem}
\label{subsubsec:FDT}
}

The fluctuation dissipation theorem (FDT) is a model independent relation between the linear response and 
the correlation function. It states that in equilibrium the induced and spontaneous fluctuations have the same 
origin~\cite{Kubo}. Out of equilibrium the FDT does not need to hold. Modifications of this relation with an interesting
structure have been found in the slow relaxation of spin-glass models~\cite{Cuku1} and later in a variety of 
other probelms~\cite{Crri,Calabrese-Gambassi,Corberi1}. These modifications turn out of have 
a therodynamic interpretation with temperature replaced by an effective temperature~\cite{Cukupe,Teff}, see the 
reviews~\cite{Teff,Leuzzi,Vulpiani}.

The FDT and its violation can be partially
understood from the following considerations. For simplicity, 
let us look at a single particle system in one dimension by described a variable $x(t)$ which satisfies the
Langevin equation
\begin{equation}
m\frac{d^2}{d t^2} x(t) + \gamma_0 \frac{d}{d t} x(t) =  F[x](t) + \xi(t)
\label{lang0}
\end{equation}
where  $\xi$ is a Gaussian random noise with zero mean and correlation
\begin{equation}
\langle \xi(t) \, \xi(t') \rangle = 2 \gamma_0 k_BT \, \delta (t-t')
\; ,
\label{eta}
\end{equation}
$T$ being the temperature.
We now take $t>t'$ for definiteness and we proceed as follows:
\begin{packed_enum}
\item[-]
We write the Langevin equation at time $t$, we multiply it by $x(t')$ and we take the noise average. 
\item[-]
We write the Langevin equation at time $t'$, we multiply it by $x(t)$ and we take the noise average.
\item[-]
We subtract these equations.
\item[-]
We define $C(t,t') = \langle x(t) x(t') \rangle$ and we use the fact that it is symmetric with 
respect to the exchange of times $t$ and $t'$.
\item[-]
We use the Novikov's theorem, $\langle x(t) \, \xi(t') \rangle = 2\gamma_0 k_BT \, R(t,t')$, 
 to make the linear response appear and we use causality, that is, $R(t',t)=0$.
\end{packed_enum}
\vspace{-0.25cm}
Then we have
\begin{equation}
m\left({\partial^2 \over \partial t^2} - {\partial^2 \over \partial {t'}^2} \right) \, C(t,t')
+
\gamma_0
\left({\partial \over \partial t} - {\partial \over \partial t'} \right) \, C(t,t')
= - 2 \gamma_0 k_BT R(t,t') + A(t,t')
\label{giorgio1}
\end{equation}
where we defined
\begin{equation}
A(t,t') \equiv \langle F[x](t) \, x(t') \rangle - \langle F[x](t') \, x(t)\rangle
\; .
\end{equation}
The position-position correlation function is obviously symmetric with respect to the exchange of times.
Moreover, in equilibrium, it must be a function of time difference only. This implies that 
\begin{equation}
C(t,t') = {\mathcal C}(|t-t'|) \qquad
 \Rightarrow
\qquad
\left(
\dfrac{\partial}{\partial t'} +\dfrac{\partial}{\partial t}
\right) \,
C(t,t') = 0
\end{equation}
Therefore, the first terms, with the second time derivatives cancel while the 
next terms with the first time derivatives are equal and of opposite sign.

It remains to deal with the term $A(t,t')$. For it one can argue that  in equilibrium 
any correlation functions must satisfy
$ \langle B(t) D(t') \rangle = \langle B(t') D(t) \rangle$, if $B(t)$ and
$D(t')$ are any two functions of $x(t)$. This is a generalization of the
Onsager reciprocity relations. Hence the asymmetry $A$ vanishes and
the fluctuation-dissipation theorem may be recovered 
\begin{equation}
R(t,t') = \beta \, \frac{\partial C(t,t')}{\partial t'} \qquad t> t'
\; .
\label{eq:FDT-gen}
\end{equation}

A system can be out of equilibrium still respecting time-translational invariance
but keeping the asymmetry $A$
as in a NESS,  or it can also break this time-translational invariance. In both cases
 Eq.~(\ref{eq:FDT-gen})
is not valid  and a generalization has to be considered. 

\comments{ 
The argument above was applied to the Markov Langevin equation. 
One difference is that the subtraction of the two friction terms yields
\begin{equation}
\int_0^t dt'' \; \Gamma(t-t'') \langle x(t') \dot x(t'')\rangle -
\int_0^{t'} dt'' \; \Gamma(t'-t'') \langle x(t) \dot x(t'')\rangle
\end{equation}
Another one is that 
 the linear responses appear after inverting the relation in Novikovs's:
 \begin{equation}
 \langle x(t) \xi(t''') \rangle = 
 k_BT 
 \int_0^{t'''} dt' \; \Gamma(t'''-t') R(t,t')
 \end{equation}
 Putting these two ingredients together
 \begin{eqnarray}
 && 
 \int_0^t dt'' \; \Gamma(t-t'') \langle x(t') \dot x(t'')\rangle -
\int_0^{t'} dt'' \; \Gamma(t'-t'') \langle x(t) \dot x(t'')\rangle
\nonumber\\
&&
\qquad\qquad
= k_BT 
 \int_0^{t'} dt'' \; \Gamma(t'-t'') R(t,t'')
 \end{eqnarray}
The FDT in the very same form (\ref{eq:FDT-gen})
holds also for more complex processes with coloured noise 
provided they take the system to equilibrium.
}

In mean-field spin-glass models the auto-correlation and response functions
are defined as $C(t,t')$ $=$ $(1/N)\sum_{i=1}^N \langle s_i(t) s_i(t') \rangle$ and
$R(t,t')$ $=$ $(1/N) \sum_{i=1}^N \delta \langle s_i(t) \rangle / \delta h_i(t') $,
respectively.
In the analysis of the asymptotic dynamics, that $t , t' \rightarrow \infty$ after
$N \rightarrow\infty$, 
it has been proposed that,
for long enough times and  small time differences,
$t,t' \rightarrow \infty$ and $(t-t')/t \ll 1$,
 FDT is satisfied, while
for long enough times and long time differences,
$t,t' \rightarrow \infty$ and $(t-t')/t \sim O(1)$,
\begin{equation}
R(t,t') = \beta \theta(t-t') \, X[ C(t,t') ] \,
\frac{\partial C(t,t')}{\partial t'}
\; .
\label{fdtasym}
\end{equation}
A self-consistent  asymptotic solution for the  mean-field out of equilibrium
dynamics within this assumption has been found both for the $p$-spin
spherical
and the Sherrington-Kirkpatrick models~\cite{Teff,Leuzzi}. 

Equation~(\ref{giorgio1}) - and its generalizations for other types of stochastic 
processes~\cite{Chatelain} - has been used to measure the linear response function without the 
application of a perturbation~\cite{Ricci-Tersenghi,Lippiello,Baiesi,Szamel,DalCengio}. 

In the following section we will see some simple examples where FDT holds and 
is violated. Interestingly enough, the FDT serves as a very stringent test of equilibration. 
It is much more sensitive than simply looking at the convergence of one-time observables,
or the stationarity of correlation functions. Indeed, one can have non-equilibrium steady 
states in which FDT is violated. Experiments in which FDT has been tested in biological
systems are used to test the dead/alive state.

\textcolor{red}{
\subsubsection{Fluctuations}
}

All quantities defined in this Section were averaged over the noise. One can also be interested
in studying their  distribution functions over the realizations of the noise. For example, 
\begin{equation}
P(A[x_\xi^{\rm sol}](t))
\qquad\qquad
P(A[x_\xi^{\rm sol}](t) B[x_\xi^{\rm sol}](t'))
\qquad\qquad
P(A[x_\xi^{\rm sol}](t) \xi(t'))
\; . 
\end{equation}
This is the route followed in~\cite{ChamonCugliandolo} and references therein
in the study of dynamic  fluctuations in glassy systems.

\setcounter{equation}{0}
\textcolor{red}{
\subsection{The basic processes}
\label{subsec:basic-processes}
}

\begin{figure}[h]
\begin{center}
\includegraphics[scale=0.5]{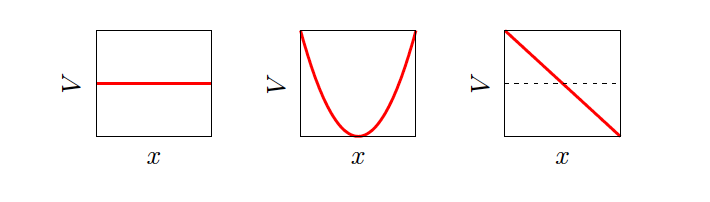}
\end{center}
\vspace{-1cm}
\caption{\small Three representative one-dimensional potentials.}
\label{fig:sketch-1da}
\end{figure}

\begin{figure}[h]
\begin{center}
\includegraphics[width=9cm]{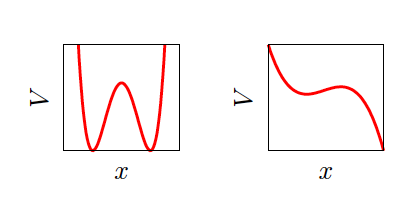}
\raisebox{0.75cm}{
\includegraphics[width=5cm]{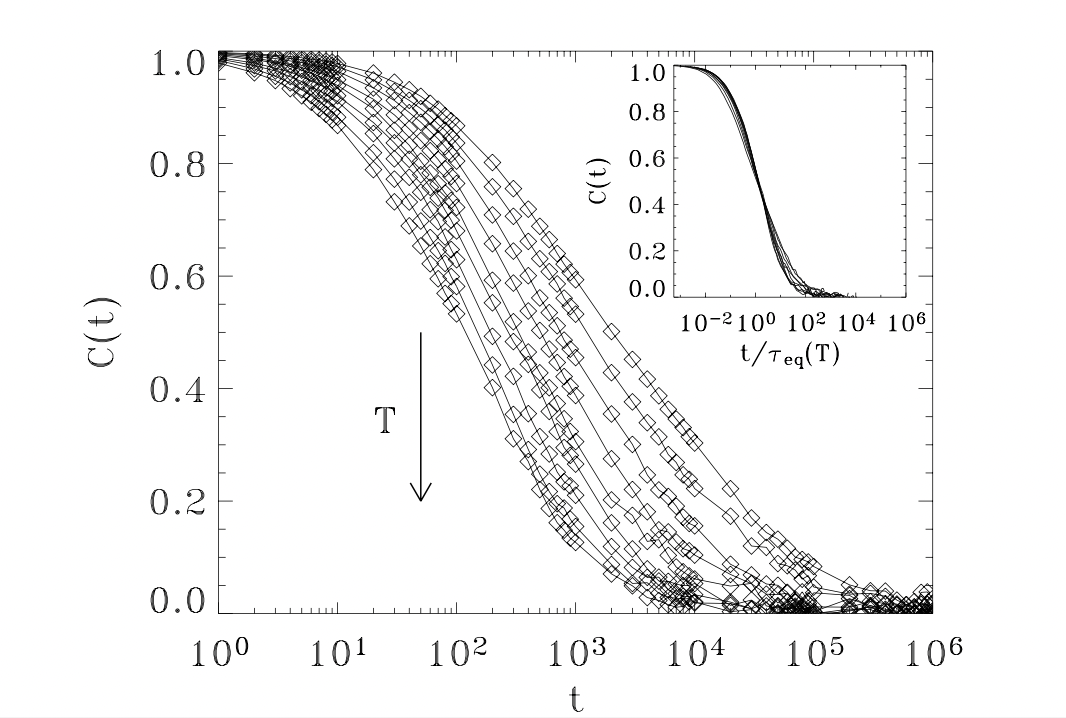}
}
\end{center}
\vspace{-1cm}
\caption{\small Left: sketch of a double-well potential. Center: sketch of a
  potential with a local minimum. Right: correlation function
  decay in a classical model of the $3d$ Coulomb glass at
  nine temperatures ranging from $T=0.1$ to $T=0.05$ in steps of
  $0.05$ and all above $T_g$. In the inset the scaling plot 
   $C(t)\sim f(t/t_A)$ with a characteristic time-scale,
  $t_A$, that follows the Arrhenius activated law, $t_A\simeq 0.45/T$.
  Figure due to Kolton, Dom\'{\i}nguez and Grempel~\cite{Kolton-Dominguez-Grempel}.}
\label{fig:kolton-coulomb-arrhenius}
\end{figure}

We will discuss the motion of the particle in some $1d$ representative
potentials: 
under a constant force, in
a harmonic potential, in the flat limit of these two (Fig.~\ref{fig:sketch-1da}) and the escape from
a metastable state and the motion  in a double well potential (Fig.~\ref{fig:kolton-coulomb-arrhenius}).

\textcolor{red}{
\subsubsection{A constant force}
}

Let us first consider the case of a \textcolor{blue}{constant
  force}, $F$. The first thing to notice is that the 
  Maxwell-Boltzmann measure
  \begin{equation}
  P_{\rm GB}(v,x) \propto e^{-\beta\left(\frac{v^2}{2m} + V(x)\right)} 
  \end{equation}
  is not normalizable 
 if the size of the line is infinite, 
due to the $\exp[-\beta V(x)] = \exp(\beta Fx)$ term. Let us then study
the evolution of the particle's velocity and
position to show how these variables behave and the fact that 
they do very differently.

The problem to solve is a set of two coupled stochastic first order
 differential 
equations on $\{v(t), x(t)\}$, one needs two initial conditions
$v_0$ and $x_0$. 

\vspace{0.25cm}
\noindent{\textcolor{red}{\it The velocity}}
\vspace{0.25cm}

The time-dependent velocity follows from the
integration of Eq.~(\ref{eq:langevin}) over time 
\begin{displaymath}
v(t) = v_0 \ e^{-\frac{\gamma_0}{m} t} + \frac{1}{m} 
\int_0^t dt' \; e ^{-\frac{\gamma_0}{m}(t-t')} \; [\, F+\xi(t') 
\,] 
\; , 
\qquad \qquad v_0 \equiv v(t=0)
\; .
\end{displaymath}
The velocity is a \textcolor{blue}{Gaussian variable} 
that inherits its average and 
correlations from the ones of $\xi$. 
Using the fact that the noise has zero average 
\begin{displaymath}
\langle v(t) \rangle = v_0 \ e^{-\frac{\gamma_0}{m} t}
+ \frac{F}{\gamma_0} \left(1- e^{-\frac{\gamma_0}{m} t}\right)
\; . 
\end{displaymath}
In the short time limit, $t\ll t_r^v=m/\gamma_0$, this expression 
approaches the Newtonian result ($\gamma_0 =0$) in which the velocity grows
linearly in time $v(t) \approx v_0 (1-\gamma_0/m \ t)+ F/m \ t = v_0 +  (F\gamma_0^{-1}-v_0) \ \gamma_0 m^{-1} \ t$. In the opposite 
long time limit, $t\gg t_r^v=m/\gamma_0$,
for all initial conditions $v_0$ the averaged velocity 
decays exponentially to the constant value $F/\gamma_0$.  
The saturation when the bath is 
active $(\gamma_0\neq 0$) is due to the friction term. 
\begin{eqnarray}
\langle v(t) \rangle =
\left\{
\begin{array}{ll}
v_0 + \left( \frac{F}{m} - \frac{v_0 \gamma_0}{m} \right) \  t
& \qquad t \ll t_r^v 
\nonumber\\
\frac{F}{\gamma_0}
& \qquad t \gg t_r^v 
\nonumber
\end{array}
\right.
\end{eqnarray}
The \textcolor{blue}{  relaxation time} 
 separating the two regimes
is
\begin{equation}
\fbox{$ 
t_r^v = \dfrac{m}{\gamma_0} 
$}
\label{eq:velocity-relaxation-time}
\end{equation} 
The average of the squared velocity is 
\begin{eqnarray}
\langle v^2(t) \rangle &=&
v_0^2 e^{-2 \frac{\gamma_0}{m} t} 
+ 2 v_0  \, \frac{F}{\gamma_0}  \, e^{- \frac{\gamma_0}{m} t} \left(e^{\frac{\gamma_0}{m} t} -1 \right)
\nonumber\\
&&
+ \frac{F^2}{\gamma_0^2} \, e^{-2\frac{\gamma_0}{m} t} \left(e^{\frac{\gamma_0}{m} t} -1\right)^2
+ \frac{k_BT}{m} \left( 1- e^{-2 \frac{\gamma_0}{m} t}\right)
\end{eqnarray}
The velocity mean-square displacement is
\begin{equation}
\sigma^2_v (t) \equiv \langle (v(t) - \langle v(t)\rangle)^2 \rangle = 
\frac{k_BT}{m} \left( 1-e^{-2\frac{\gamma_0}{m} t} \right) 
\label{eq:equipartition}
\end{equation}
independently of $F$.  This is an example of the \textcolor{blue}{regression theorem}
according to which the equilibrium fluctuations decay in time following the same law as the 
average value. The short and long time limits yield
\begin{eqnarray}
&& \sigma_v^2(t) \equiv \langle (v(t) - \langle v(t)\rangle)^2 \rangle  \simeq
\frac{k_BT}{m} \left\{
\begin{array}{ll}
\dfrac{2\gamma_0}{m} \ t & \qquad t\ll t^v_r
\\
[15pt]
1 & \qquad t\gg t^v_r
\end{array}
\right.
\end{eqnarray}
and the two expressions match at $t\simeq t_r^v/2$. 
The asymptotic limit is the result expected
from equipartition of the velocity mean-square displacement, 
$\langle (v(t) -\langle v(t)\rangle)^2\rangle \to \langle (v(t)-\langle
v\rangle_{\rm stat})^2\rangle_{\rm stat}$ that implies for the `kinetic energy'
$\langle K\rangle_{\rm stat} = k_BT/2$ only if the velocity is measured
with respect to its average.  In the heuristic derivation of the
Langevin equation for $F=0$ the amplitude of the noise-noise correlation, say
$A$, is not fixed.  The simplest way to determine this parameter is to
require that equipartition for the kinetic energy holds $A/(\gamma_0
m) = k_BT/m $ and hence $A=\gamma_0 k_BT$.  This relation is known under the
name of \textcolor{blue}{ fluctuation--dissipa\-tion theorem} ({\rm
FDT}) \textcolor{blue}{ of the second kind} in Kubo's nomenclature.
It is important to note that this {\rm FDT} characterizes the
surrounding fluid and not the particle, since it relates the
noise-noise correlation to the friction coefficient.  In the case of
the Brownian particle this relation ensures that after a transient of
the order of $t^v_r$, the bath maintains the fluctuations of the velocity, 
$\sigma_v^2$,  constant and equal to its equilibrium value.  

\begin{figure}[h]
\begin{center}
\includegraphics[scale=0.5]{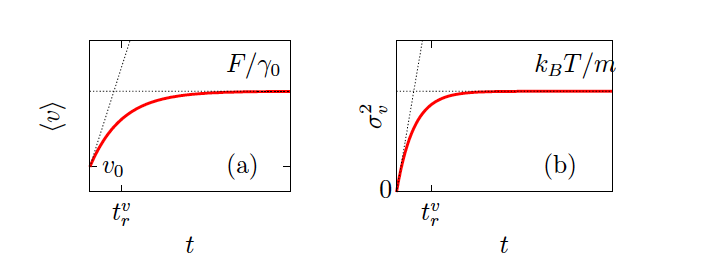}
\end{center}
\vspace{-1cm}
\caption{\small Results for the constant force problem.
(a) Mean velocity as a function of time.  (b) Velocity 
mean-square displacement as a function of time. In both cases the linear 
behavior at short times, $t\ll t_r^v$, and the saturation values are shown.
The slopes are $F/m$ and $k_BT/m \times 2\gamma_0/m$, respectively.}
\label{fig:sketch-results-1da}
\end{figure}

The velocity \textcolor{blue}{ two-time connected correlation} reads
\begin{displaymath}
\langle [v(t)-\langle v(t)\rangle] [v(t')-\langle v(t')\rangle] \rangle = 
\frac{k_B T}{m} 
\left[ 
e^{-\frac{\gamma_0}{m} |t-t'|}-
e^{-\frac{\gamma_0}{m} (t+t')}
\right] 
\; . 
\end{displaymath}
This is sometimes called the \textcolor{blue}{Dirichlet correlator}.
This and all other higher-order velocity correlation functions 
approach a \textcolor{blue}{stationary limit} when the shortest time 
involved is longer than $t_r^v$. At $t=t'$ one recovers the 
mean-square displacement computed in Eq.~(\ref{eq:equipartition}). 
When both times are short compared to $t_r^v$ the two-time correlator
behaves as $\sim 2k_BT\gamma_0/m^2 \ \max(t,t')$. When at least
one of the two times is much longer than $t_r^v$ the second term vanishes
and one is left with an exponential decay as a function of time delay:
\begin{equation}
C^c_{vv}(t,t') \equiv
\langle [v(t)-\langle v(t)\rangle] [v(t')-\langle v(t')\rangle \rangle 
\to \frac{k_BT}{m} e^{-\frac{\gamma_0}{m} |t-t'|}
\qquad t,t'\gg t_r^v
\; . 
\end{equation}
The two-time connected correlation falls off to, say, $1/e$ 
in a \textcolor{blue}{decay time} 
\begin{equation}
\fbox{
$t_d^v=\dfrac{m}{\gamma_0}$
}
\end{equation}
 In this simple case $t_r^v=t_d^v$ but this does not 
necessarily happen in more complex cases.

 More generally
one can show that for times $t_1 \geq t_2 \geq \dots \geq t_n \geq t_r^v$:
\begin{equation}
\fbox{$
\langle \delta v(t_1+\Delta) \dots \delta v(t_n+\Delta) \rangle = 
\langle \delta v(t_1) \dots \delta v(t_n) \rangle
\;\;\; $ (TTI)}
\label{eq:TTI1}
\end{equation}
with $\delta v(t) = v(t)- \langle v \rangle(t)$, 
for all delays $\Delta$. 
\textcolor{blue}{Time-translation invariance (TTI)} or \textcolor{blue}{ 
  stationarity} is one generic property of \textcolor{blue}{  equilibrium dynamics}. Another way 
  of stating (\ref{eq:TTI1}) is 
  \begin{equation}
  \langle v(t_1) \dots v(t_n) \rangle = f(t_1-t_2, \dots, t_{n-1}-t_n)
  \; . 
  \end{equation}

Another interesting object is the linear response of the averaged velocity 
to a small perturbation applied to the system in the form of $V\to V-fx$, 
i.e. a change in the slope of the potential in this particular case.
One finds
\begin{eqnarray}
&& R_{vx}(t,t') \equiv 
\left.
\frac{\delta \langle v(t)\rangle_f}{\delta f(t')}
\right|_{f=0}
=
\frac{1}{m} \  
e^{-\frac{\gamma_0}{m} (t-t') } \ \theta(t-t')
\noindent\\
&&\qquad\quad\quad
\simeq
\frac{1}{k_BT} 
\langle [v(t)-\langle v(t)\rangle] [v(t')-\langle v(t')\rangle] \rangle 
\ \theta(t-t')
\end{eqnarray}
the last identity being valid in the limit $t$ or $t'$ $\gg t_r^v$.
This is an {\rm fdt} relation between a linear response,
$R_{vx}(t,t')$, and a connected correlation, $C^c_{vv}(t,t')$, that
holds for one of the particle variables, its velocity, when this one reaches the 
stationary state.
\begin{equation}
\fbox{$
k_BT \ R_{vx}(t,t') =   C_{vv}^c(t,t')\ \theta(t-t') 
\qquad\qquad (\mbox{FDT})
\; . 
$}
\label{eq:FDT-vv}
\end{equation}

In conclusion, the velocity is a Gaussian variable that after a
characteristic time $t_r^v$ verifies `equilibrium'-like properties:
its average converges to a constant (determined by $F$), its
multi-time correlation functions become stationary and a fluctuation-dissipation
theorem links its linear response to the connected correlation at two
times. 

Finally, we can calculate the instantaneous correlation between the velocity and the noise, 
which appears in the variation of the total energy. One has
\begin{eqnarray}
\langle v(t) \xi(t) \rangle &=& \frac{1}{m} \int_0^t  dt' \, e^{-\frac{\gamma_0}{m} (t-t') }
\langle \xi(t') \xi(t) \rangle 
\nonumber\\
&=&  \frac{2\gamma_0 k_BT}{m} \int_0^t  dt' \, e^{-\frac{\gamma_0}{m} (t-t') }
\delta(t-t')
= \frac{\gamma_0 k_BT}{m} 
\end{eqnarray}
Back into the expression for the energy variation, at finite times, 
\begin{eqnarray}
\langle \frac{d H_{\rm syst}}{dt} \rangle &=& 
- \, \gamma_0 \langle v^2 \rangle 
 + \langle \xi(t) v(t)\rangle
\nonumber\\
&=&
- \, \gamma_0 v_0^2 e^{-2 \frac{\gamma_0}{m} t} 
- 2 v_0  \, F  \, e^{- \frac{\gamma_0}{m} t} \left(e^{\frac{\gamma_0}{m} t} -1 \right)
- \frac{F^2}{\gamma_0} \, e^{-2\frac{\gamma_0}{m} t} \left(e^{\frac{\gamma_0}{m} t} -1\right)^2
\nonumber\\
&&
- \frac{\gamma_0k_BT}{m} \left( 1- e^{-2 \frac{\gamma_0}{m} t}\right)
+ \frac{\gamma_0 k_BT}{m}
\nonumber\\
&=&
- \, \gamma_0 v_0^2 
e^{-2 \frac{\gamma_0}{m} t} 
- 2 v_0  \, F  \, e^{- \frac{\gamma_0}{m} t} \left(e^{\frac{\gamma_0}{m} t} -1 \right)
- \frac{F^2}{\gamma_0} \, e^{-2\frac{\gamma_0}{m} t} \left(e^{\frac{\gamma_0}{m} t} -1\right)^2
\nonumber\\
&&
+ \frac{\gamma_0k_BT}{m} e^{-2 \frac{\gamma_0}{m} t}
\end{eqnarray}
Not all these terms are exponentially decreasing. Indeed, 
\begin{equation}
\lim\limits_{t\gg \frac{m}{\gamma_0}} \langle \frac{d H_{\rm syst}}{dt} \rangle =
 -2 v_0 F - \frac{F^2}{\gamma_0}
\end{equation}
Naturally, under the force, the total energy of the system diverges to minus infinity and its negative 
velocity of change is proportional to $F$. If, instead, $F=0$ the asymptotic value is finite.
Note that at the initial time $t=0$ and close to it, 
\begin{eqnarray}
&& 
\langle \frac{d H_{\rm syst}}{dt} \rangle 
\sim
- \, \gamma_0 v_0^2 \left( 1-  \frac{2\gamma_0}{m} t  +  \frac{2\gamma_0^2}{m^2} t^2 \right)
- 2 v_0  \, F  \, \left(1 - \frac{\gamma_0}{m} t \right)  \frac{\gamma_0}{m} t 
- \frac{F^2}{\gamma_0} \,  
\frac{\gamma^2_0}{m^2} t^2 
\qquad\qquad\qquad
\nonumber\\
&& 
\quad\quad =
- \, \gamma_0 v_0^2
+ 2v_0\left( \gamma_0 v_0 -  F  \right) \, \dfrac{t}{t^{(v)}_r}
-
2\gamma_0 \left(  v_0^2 +  v_0 \frac{F}{\gamma_0} + \frac{F^2}{2\gamma^2_0}   \right) \left( \dfrac{t}{t_r^{(v)}} \right)^2
\end{eqnarray}
Consider the case $F=0$ 
\begin{eqnarray}
&& 
\langle \frac{d H_{\rm syst}}{dt} \rangle 
\sim
- \, \gamma_0 v_0^2 \left[ 1 -  2 \frac{t}{t_r^{(v)}} + 2 \left( \dfrac{t}{t_r^{(v)}} \right)^2 \right]
\end{eqnarray}
This is always negative.

\vspace{0.25cm}
\noindent{\textcolor{red}{\it The position}}
\vspace{0.25cm}

The particle's position, $x(t) = x_0+ \int_0^t dt' v(t')$ is 
still a Gaussian random variable:
\begin{eqnarray}
x(t) &=& x_0 +v_0 \ t^v_r  + \frac{F}{\gamma_0} (t-t^v_r)
+t_r^v \left(\frac{F}{\gamma_0}-v_0 \right) e^{-\frac{\gamma_0}{m} t}
\nonumber\\
&&
+ \frac{1}{m} \int_0^t dt' \int_0^{t'} dt'' \ e^{-\frac{\gamma_0}{m} (t'-t'')} \ \xi(t'')
\; . 
\label{eq:particle-position}
\end{eqnarray}
Its noise-average behaves as the Newtonian result,
\textcolor{blue}{ballistic motion}, 
\begin{equation}
\fbox{$
\langle x(t) \rangle \simeq x_0 +
v_0 t + \dfrac{\gamma_0}{2m} 
\left(\dfrac{F}{\gamma_0}- v_0 \right) \  t^2
\qquad\mbox{for} \qquad t\ll t_r^v
$}
\end{equation} 
at short times and it crossover to
\begin{equation}
\fbox{$
\langle x(t) \rangle \to  x_0 +v_0 \ t^v_r  + \dfrac{F}{\gamma_0} (t-t^v_r)
\qquad \mbox{for} \qquad t\gg t_r^v
$}
\end{equation}
at long times. Note the  reduction with respect to 
ballistic motion ($x\propto F t^2$) due to the friction drag
and the fact that this one-time observable does not saturate to a constant.

An interesting result, that we will use later, is the fact that the coordinate and the noise 
have vanishing correlation at equal times: $\langle x(t) \xi(t) \rangle =0 $. This 
can be easily proven by multiplying the expression for $x(t)$ by $\xi(t)$ and 
taking the average.

The position mean-square displacement approaches
\begin{equation}
\fbox{$
\sigma_x^2(t) \equiv \langle (x(t)-\langle x(t)\rangle)^2\rangle \to 2D_x  t 
\;\;\; \mbox{with} \;\;\; D_x \equiv \dfrac{k_BT}{\gamma_0} \;\;\;$ (Diffusion)}
\label{eq:diffusion}
\end{equation}
in the usual $t\gg t^v_r$ limit, that is to say \textcolor{blue}{  normal
  diffusion} with the \textcolor{blue}{diffusion constant} 
$D_x$. This expression 
can be computed using $x(t)-\langle x(t)\rangle$ as obtained from the 
$v(t)-\langle v(t)\rangle$ above 
(and it is quite a messy calculation) or one can simply go to
the Smoluchowski limit, taking advantage of the knowledge of what we have just
discussed on the behaviour of velocities, and obtain diffusion in 
two lines. 

\vspace{0.25cm}
\noindent
\textcolor{orange}{\bf Exercise \thesection.\theexercise} Do the calculation sketched above.

\addtocounter{exercise}{1}

\vspace{0.25cm}

When the friction coefficient $\gamma_0$ is given by the Stokes law,
$\gamma_0 = 6\pi\eta a$ for a spherical particle with radius $a$ in a liquid with 
dynamic viscosity $\eta$, the diffusion constant is given by the \textcolor{blue}{Stokes-Einstein}
relation $D_x= k_BT/(6\pi\eta a)$.

\begin{figure}[h]
\centerline{
\includegraphics[scale=0.25]{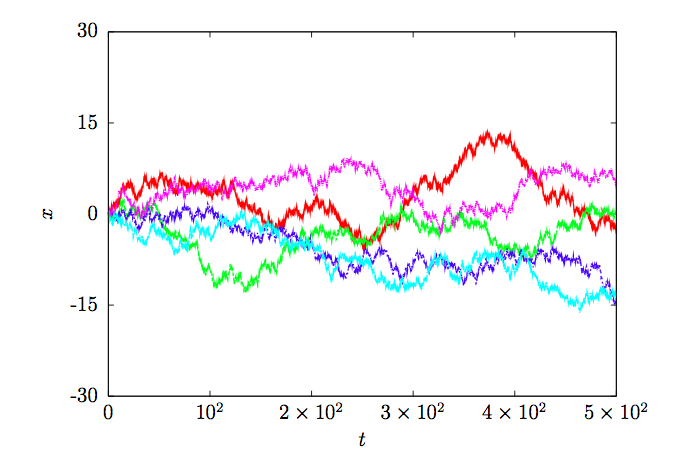}
\includegraphics[scale=0.25]{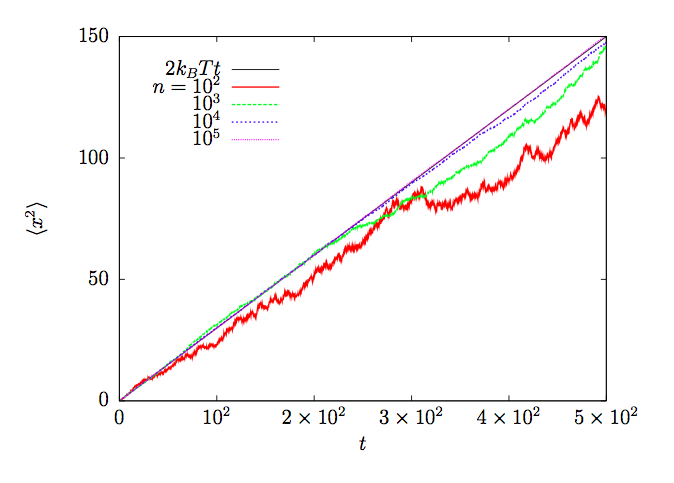}
}
\vspace{-0.5cm}
\caption{\small 
Left panel: five runs of the Langevin equation in the over-damped limit with no external force and 
a Gaussian white noise
at temperature $T$ and $\gamma_0=1$. Right panel: the average $\langle x^2\rangle$ computed with 
$n=10^2, \ 10^3, \ 10^4, \ 10^5$ runs. The straight line represents the normal diffusion
$\langle x^2 \rangle \simeq 2 k_BT t$.
}
\end{figure}

The searched result can also be found as follows. Multiply the Langevin equation 
evaluated at $t$ 
by $x$ evaluated at the same instant and use an obvious identity to find
\begin{equation}
m x \dot v = m \left( \frac{d}{dt} (xv) - v^2 \right) = -\gamma_0 v x + x F + x \xi 
\end{equation}
Take now the noise average. Use the fact that the average of $x\xi$, when the two 
factors are evaluated at the same time, vanishes identically,
and exchange time-derivative and noise-average 
(assuming this operation is permitted).  The resulting equation is 
\begin{equation}
\frac{d}{dt} \langle xv\rangle   = -\frac{\gamma_0}{m} \langle v x\rangle + \frac{F}{m}  \langle x\rangle   + \langle v^2\rangle
\; . 
\label{eq:xv-corr}
\end{equation}
The last two terms in the right-hand-side are a known time-dependent function, $A(t)$:
\begin{eqnarray}
A &\equiv&  \frac{F}{m} \langle x\rangle   + \langle v^2\rangle
\; , 
\\
\frac{F}{m}\langle x\rangle &=&
\frac{F}{m} \left[ x_0 + v_0 t_r^v + \frac{F}{\gamma_0} \left(t-t_r^v\right) + 
t_r^v \left( \frac{F}{\gamma_0}-v_0\right) e^{-\frac{\gamma_0}{m} t}\right]
\; , 
\\
\langle v^2\rangle
&=& 
\frac{k_BT}{m} \left( 1- e^{-2\frac{\gamma_0}{m} t}\right) + 
\left[ v_0 e^{-\frac{\gamma_0}{m} t} + \frac{F}{\gamma_0} \left( 1- e^{-\frac{\gamma_0}{m} t}\right)
\right]^2
\; .
\end{eqnarray}
One can now integrate Eq.~(\ref{eq:xv-corr}) over time 
\begin{equation}
\langle xv\rangle = 
x_0 v_0 \ e^{-\frac{\gamma_0}{m} t} + \int_0^t dt' \; e^{-\frac{\gamma_0}{m} (t-t')} A(t')
\end{equation}
to find  a rather lengthy expression.
In the long time limit, $t\gg t_r^v$, we drop all exponentially decaying terms to 
obtain 
\begin{eqnarray}
\langle xv\rangle 
&\to& 
\frac{k_BT}{\gamma_0} +
\frac{F}{\gamma_0} \left( t_r^v v_0 + x_0 \right) + \frac{F^2}{\gamma_0^2} \left( t-t_r^v \right)
\end{eqnarray}
Now, using 
$\langle xv\rangle = \frac{1}{2} \frac{d}{dt} \langle x^2\rangle $ one finally finds
\begin{equation}
\langle x^2\rangle \to 
2\frac{k_BT}{\gamma_0} t + 2 \frac{F}{\gamma_0} \left( t_r^v v_0 + x_0 \right) t + \frac{F^2}{\gamma_0^2} 
\left[ \left( t-t_r^v \right)^2 - t_r^2 \right]
\end{equation}
The last two terms are $\langle x\rangle^2$ in the same regime of times. Therefore,
Eq.~(\ref{eq:diffusion}) is recovered.

Another way to measure the diffusion coefficient directly from the 
velocity that is commonly used in the literature is 
\begin{equation}
\fbox{$
D_x = \lim_{\tau\to\infty} \lim_{t'\to\infty} 
\bigintsss_{\; 0}^{\, \tau} dt' \ \langle \delta v(\tau+t') \delta v(t') \rangle
\; . 
$}
\end{equation}
One can check that it gives the same result.

In contrast to the velocity mean-square displacement this 
quantity  does not saturate at any finite value.
Similarly, the particle
displacement between two different times $t$ and $t'$ is 
\begin{equation}
\Delta_{xx}(t,t')\equiv \langle [x(t)-x(t')]^2 \rangle \to 
2D_x |t-t'|
\; . 
\label{eq:diffusion-coeff}
\end{equation} 
It is interesting to note that the force 
dictates the mean position but it does not modify the fluctuations about 
it (similarly to what it did to the velocity). $\Delta_{xx}$ is stationary
for time lags longer than $t_r^v$. 

The two-time position-position connected correlation reads
\begin{equation}
C_{xx}^c(t,t') = \langle (x(t)-\langle x(t)\rangle) (x(t')-\langle x(t')\rangle) \rangle
= \dots
\end{equation}

\vspace{0.2cm}

\noindent
\textcolor{orange}{\bf Exercise \thesection.\theexercise} compute this correlation function.

\addtocounter{exercise}{1}

\begin{figure}[h]
\begin{center}
\includegraphics[scale=0.5]{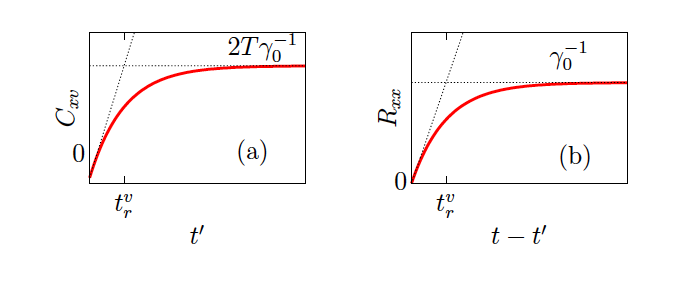}
\end{center}
\vspace{-1cm}
\caption{\small Results for the constant force problem.
(a) The correlation between the position and the velocity of the particle
measured at different times. 
(b) The linear response of the position to a kick applied linearly to itself
at a previous time. In both cases the linear 
behavior at short times, $t\ll t_r^v$ and the saturation values are shown.}
\label{fig:sketch-results-1db}
\end{figure}

The  linear response of the particle's position to a kick 
linearly applied to itself at a previous time, in the form $V\to V-f x$ at $t'<t$,  is
\begin{equation}
R_{xx}(t,t') \equiv \left. \frac{\delta \langle x(t)\rangle_f}{\delta f(t')} \right|_{f=0}
= \frac{1}{\gamma_0} \ [1-e^{-\frac{\gamma_0}{m}(t-t')}] \, \theta(t-t')
\; , 
\end{equation}
with the limits
\begin{eqnarray}
R_{xx}(t,t') \to 
\left\{
\begin{array}{ll}
m^{-1} \  (t-t') \ \theta(t-t') & \qquad t-t' \ll t_r^v
\; , 
\\
\gamma_0^{-1} \ \theta(t-t')  & \qquad t-t' \gg t_r^v
\; . 
\end{array} 
\right.
\end{eqnarray}
A simple calculation proves that in the short time-differences limit 
this is the result for Newton dynamics.

\vspace{0.2cm}

\noindent
\textcolor{orange}{\bf Exercise \thesection.\theexercise} show the property mentioned above.

\addtocounter{exercise}{1}

\vspace{0.2cm}

The correlation between the position and the velocity reads
\begin{eqnarray}
&&
\langle (x(t)-\langle x(t)\rangle) (v(t')-\langle v(t')\rangle) \rangle =
\frac{2k_BT}{m} 
\left[
\frac{m}{\gamma_0} - \left(1+\frac{m}{\gamma_0}\right) e^{-\frac{\gamma_0}{m} t'} 
\right] 
\nonumber\\
&& \qquad\qquad 
\to \frac{2k_BT}{\gamma_0}
\end{eqnarray}
and it is only a function of $t'$.  One notices that in the asymptotic limit
in which both sides of the equation saturate
\begin{equation}
\fbox{$
2k_B T \ R_{xx}(t,t') = C^{c}_{xv}(t,t')
\;\;\;\;\; \mbox{for} \;\;  t-t' \gg t_r^v \;\; \mbox{and} \;\; t' \gg t_r^v \; ,  
$}
\end{equation}
with a factor of 2 different from the relation in Eq.~(\ref{eq:FDT-vv}). 

In conclusion, the position is also a Gaussian variable but it is 
explicitly  out of equilibrium. Its 
average and variance grow linearly in time, the latter as in normal diffusion, 
and the fluctuation-dissipation relation has an additional factor of 
1/2 (or 2, depending on on which side of the equality one writes it) 
with respect to the form expected in equilibrium.

A measure for the time dependent fluctuating position and 
velocity can be written down, taking advantage of the fact that 
both variables are Gaussian:
\begin{equation}
P(v,x) \propto 
\exp\left[ 
- \frac{1}{2} \ \int dt \int dt' \
\delta y^t(t) A(t,t') \delta y(t')
\right]
\end{equation}
with the $2\times 2$ matrix $A$ being the inverse of the matrix of correlations, 
${A^{-1}}_{ij}(t,t') = \langle \delta y_i(t) \delta y_j(t')\rangle$ with $i,j=1,2$,  
$\delta y^t(t) = (\delta v(t) \;  \delta x(t) )$ and $\delta v(t) = v(t) -\langle v(t)\rangle$
(similarly for $x$). The correlations are given above so the dynamic pdf can be easily 
constructed. There will be elements in the matrix that remain time-dependent for all 
times.

\vspace{0.2cm}

\noindent\textcolor{orange}{\bf Exercise \thesection.\theexercise} Confront
\begin{equation}
\langle v^m(t) x^n(t) x^k(t') \rangle \qquad \mbox{and} \qquad 
\langle v^m(t) x^n(t) k x^{k-1}(t') v(t')\rangle
\; ,
\end{equation}
conclude.

\addtocounter{exercise}{1}

\pagebreak

\noindent{\textcolor{red}{\it The energy}} 

\vspace{0.25cm}

The averaged kinetic energy can be computed using $\langle v^2(t)\rangle = \sigma_v^2(t) + \langle v(t)\rangle^2$
and the results already derived. It reaches, in the $t\gg t_r^v$ limit,  a constant value: 
$\langle K(t) \rangle \to k_BT/2 + F/(2\gamma_0)$. 
The averaged potential energy diverges in the long-time limit if $F\neq 0$
since the potential is unbounded 
in the $x\to\infty$ limit: 
$\langle V(t)\rangle =  - F \langle x(t) \rangle 
\simeq -F^2/\gamma_0 t$ for $t\gg t_r^v$. In the particular case $F=0$
the total energy is just kinetic and it approaches the constant expected from equipartition 
asymptotically $\langle K(t) \rangle \to k_BT/2$.

It is also interesting to investigate the sign of $dE/dt$ on the mean, 
$\langle dE/dt\rangle = -\gamma_0 \langle v^2\rangle + \langle v\xi\rangle$. The first term tends to
$-\gamma_0 k_BT/m - F$. 
The 
second term also yields a non-trivial contribution $\langle v\xi\rangle \to m^{-1} \int_0^t dt' e^{-\gamma_0(t-t')/m} \langle \xi(t)\xi(t')\rangle = 
\gamma_0k_BT/m$. Adding these two together one finds $\langle dE/dt\rangle \to -F$ asymptotically,  for $t\gg t_r^v$.

\vspace{0.25cm}
\noindent{\textcolor{red}{\it Two kinds of variables}}
\vspace{0.25cm}

 This example shows that even in this very simple problem the velocity
and position variables have distinct behavior: the former is in a
sense trivial, after the transient $t_r^v$ and for longer times, all
one-time functions of $v-F/\gamma_0$ saturate to their equilibrium-like
values and the correlations are stationary. Instead, the latter
remains non-trivial and evolving out of equilibrium. One can loosely
ascribe the different behavior to the fact that the velocity feels a
confining kinetic energy $K=mv^2/2$ while the position feels an unbounded
potential $V=-Fx$ in the case in which a force is applied, or a flat
potential $V=0$ if $F$ is switched off. In none of these cases the
potential is able to take the particle's position to equilibrium with the bath.
The particle slides on the slope and its excursions forward and backward
from the mean get larger and larger as time increases.

\vspace{0.25cm}
\noindent{\textcolor{red}{\it Over-damped (Smoluchowski) limit}}
\vspace{0.25cm}

Quite generally, the classical problems we are interested in 
are such that the friction coefficient $\gamma_0$ is large and 
the inertia term can be neglected, in other words, all times are much 
longer than the characteristic time $t_r^v$. 

\vspace{0.25cm}
\noindent{\textcolor{red}{\it Ergodicity}}
\vspace{0.25cm} 

The ergodic hypothesis states that, in equilibrium, one can exchange
ensemble averages by time averages and obtain the same results. Out of
equilibrium this hypothesis is not expected to hold and one can
already see how dangerous it is to take time-averages in these cases
by focusing on the simple velocity variable. Ensemble and time
averages coincide only if the time-averaging is done over a time-window
that lies after $t_r^v$ but it does not if the integration
time-interval goes below $t_r^v$. Moreover, in the case of the position variable, 
there is no finite $t_r^x$.

Tests of equilibration have to be done very carefully in experiments and 
simulations. One can be simply mislead by, for instance, looking just at
the velocities statistics.

\pagebreak

\noindent{\textcolor{red}{\it Effect of a colored bath: anomalous diffusion}}
\vspace{0.25cm} 

The \textcolor{blue}{anomalous diffusion} ($F=0$) of a particle governed  by the 
generalized Langevin equation, 
Eq.~(\ref{eq:langevin_generalized}), with colored noise characterized by power-law 
correlations as the ones given in Eq.~(\ref{eq:Gamma}), a problem also known as 
\textcolor{blue}{fractional 
Brownian motion}, was studied in detail by N. Pottier~\cite{Pottier}.
The particle's velocity equilibrates with the environment although it does at a much 
slower rate than in the Ohmic case: its average and mean-square displacement 
decay as a power law - instead of exponentially - 
to their asymptotic values (still satisfying the regression theorem). The particle's mean square 
displacement  is determined
by the exponent of the noise-noise correlation, 
\begin{equation}
\Gamma(t) \simeq t^{-\alpha}
\qquad \mbox{and}
\langle x^2(t) \rangle \simeq t^\alpha
\; , 
\end{equation}
the dynamics is \textcolor{blue}{subdiffusive} for $\alpha<1$, \textcolor{blue}{diffusive} 
for $\alpha=1$ and \textcolor{blue}{superdiffusive} for $\alpha>1$. A time-dependent diffusion coefficient verifies $D_x(t) \equiv 1/2 \ d\langle x^2(t)\rangle/dt \propto t^{\alpha-1}$: 
it is finite and given by Eq.~(\ref{eq:diffusion-coeff}) for normal diffusion, 
it diverges for superdiffusion and it vanishes for subdiffusion.  
The ratio between the linear response and the time-derivative
of the correlation ratio reads $TR_{xx}(t,t')/\partial_{t'} C_{xx}(t,t')=D_x(t-t')/[D_x(t-t')+D_x(t')]$. It 
approaches $1/2$ for normal diffusion and the two-time dependent function
$1/[1+(t'/(t-t'))^{\alpha-1}]$ in other cases.

\vspace{0.2cm}

\noindent\textcolor{orange}{\bf Exercise \thesection.\theexercise} 
Work out these results. 

\addtocounter{exercise}{1}

\vspace{0.25cm}
\noindent\textcolor{red}{\it Perrin's experiment}
\vspace{0.25cm}

Jean-Baptiste Perrin used these results to measure the Avogadro number experimentally and, more importantly, give evidence for the 
discrete character of matter. The idea is already described in Lucretius's poem {\it De rerum natura, On the nature of things}.
The reasoning goes as follows. 
Take a spherical tracer particle with radius $a$ and immerse it in a liquid with viscosity $\eta$. These 
two quantities can be measured. Assume that the liquid behaves as a white noise. Stokes law states that the friction coefficient for this 
particle is 
\begin{equation}
\gamma_0 = 6\pi\eta a
\; . 
\end{equation}
The Boltzmann constant $k_B$ is given by the gas constant $R$, that is also known, divided by the Avogadro number since
$k_B=nR/N=R/N_A$ with $n$ the number of moles and $N$ the number of atoms in a gas. Therefore
\begin{equation}
\sigma_x^2(t) \simeq 2D_x t = \frac{R}{3\pi\eta a} \frac{T}{N_A} t 
\end{equation}
and, by measuring the tracer's diffusion one can extract $N_A$.

\begin{figure}
\centerline{
\includegraphics[scale=0.25,angle=90]{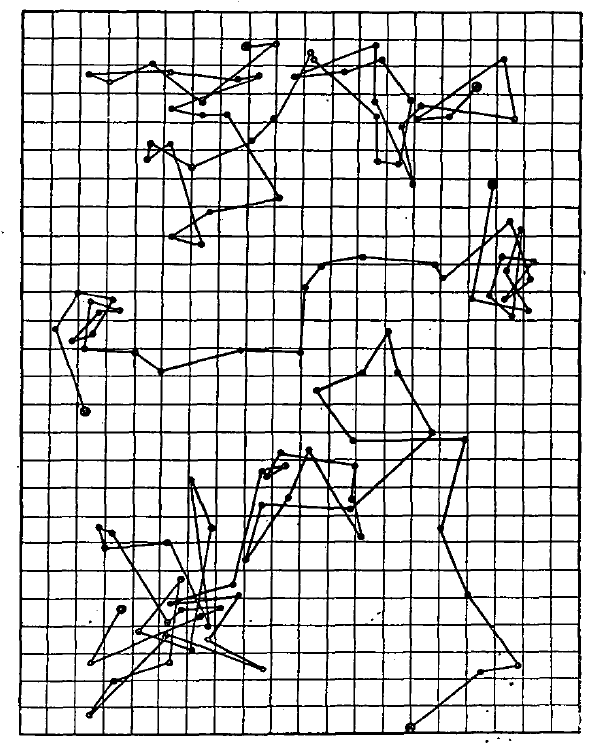}
}
\caption{\small Colloidal particle trajectories. The radius of the particles is $0.53 \, \mu$m, the points are successive positions taken at intervals of 
30\, s. The grid has spacing $3.2 \, \mu$m. Image taken from Perrin's original article.}
\end{figure}

\textcolor{red}{
\subsubsection{A harmonic potential and white additive noise}
\label{sec:harmonic oscillator}
}

\vspace{0.25cm}

\noindent
\textcolor{red}{\it Full analysis}

\vspace{0.25cm}

The Hamiltonian of a one-dimensional
harmonic oscillator of mass $m$ and  spring constant
$k$ is
\begin{equation}
H_{\rm syst}=\frac{p^2}{2 m} + \frac{k x^2}{2} 
\label{harmonic_hamiltonian}
\; .
\end{equation}
The Langevin equation of motion reads
\begin{equation}
m \ddot x(t) = -\gamma_0 \dot x(t) -  k x(t) + h(t) + \xi(t)
\; .
\label{Langevin_m}
\end{equation}
with $h(t)$ a time-dependent deterministic force.
By setting $k=0$ one recovers the motion of a Brownian particle,
see Sect.~\ref{sec:brownian}.
$\xi$ is the white noise with zero mean and correlations 
$\langle \xi(t) \xi(t') \rangle=  2 k_B T \gamma_0 \delta(t-t')$.

\vspace{0.25cm}
\noindent
\textcolor{red}{\it Qualitative analysis: time-scales}
\vspace{0.25cm}

From an order of magnitude analysis of the the three terms in the 
homogeneous equation one can extract the time-scales that rule the 
dynamics of this problem. 
The easiest way to determine these time scales is to first Fourier 
transform the homogeneous equation that determines the Green 
function
\begin{equation}
-m \omega^2 \tilde x(\omega) - i \gamma_0 \omega \tilde x(\omega) + k \tilde x(\omega)
\; . 
\end{equation}
For the sake of completeness, we present this analysis for a coloured noise with power-law 
correlations, Eq.~(\ref{eq:spectral-density-bath}), in the limit in which the cut-off has been sent to infinity, for which the equation above generalises to 
\begin{equation}
-m \omega^2 \tilde x(\omega) - i \gamma_0 \omega^\alpha \tilde{\omega}^{\alpha-1} \tilde x(\omega) + k \tilde x(\omega)
\; . 
\end{equation}

In the absence of dissipation ($\gamma_0=0$) the particle 
oscillates within the harmonic well, and the parameter dependence of the 
frequency and period of oscillation 
is determined with an order of magnitude argument 
\begin{equation}
m \omega_{\rm osc}^2 \approx k 
\qquad \Rightarrow \qquad 
\omega^2_{\rm osc} \approx \frac{k}{m}
\quad\Rightarrow \qquad 
\fbox{$\displaystyle{t_{\rm osc} \approx \left(\frac{m}{k}\right)^{1/2}}$}
\end{equation}

Under the action of the dissipative force new time-scales appear. First, 
one can expect that the competition between inertia and dissipation gives rise to the 
velocity relaxation time, $t_r^v$. Therefore,
\begin{equation}
m {\omega_r^v}^2 \approx \gamma_0 {\omega_r^v}^\alpha \tilde{\omega}^{\alpha-1}
\qquad
\Rightarrow
\qquad
\omega_r^v \approx \left(\frac{\gamma_0}{m} \ \tilde \omega^{1-\alpha} \right)^{1/(2-\alpha)}
\end{equation}
and
\begin{equation}
\fbox{$
t_r^v \approx \displaystyle{\left(\frac{m}{\gamma_0} \ \tilde \omega^{\alpha-1} \right)^{1/(2-\alpha)}}
$}
\end{equation}
In the particular case $\alpha=1$ the bath becomes Ohmic and $t_r^v \approx m /\gamma_0$ 
as in Eq.~(\ref{eq:velocity-relaxation-time}).

Second, the comparison between dissipation and harmonic force yields
\begin{equation}
\gamma_0 {\omega_r^x}^\alpha \tilde{\omega}^{\alpha-1} \approx k 
\qquad
\Rightarrow 
\qquad 
{\omega_r^x} \approx \left( \frac{k}{\gamma_0} \tilde\omega^{1-\alpha} \right)^{1/\alpha}
\end{equation}
and 
\begin{equation}
\fbox{
$
{t_r^x} \approx 
\displaystyle{
\left( \frac{\gamma_0}{k} \ \tilde\omega^{\alpha -1} \right)^{1/\alpha}
}
$
}
\end{equation}
that for $\alpha=1$ becomes $t_r^x \approx \gamma_0/k$ a time-scale that we will 
see appearing in the exact calculation below.

Now, we can compare these time-scales and conclude about the possible types of 
particle motion. If $t_{\rm osc} < t_r^x$ the particle continues to oscillate during its
dissipative evolution while,  on the contrary, if $t_{\rm osc} > t_r^x$,  
the oscillations are damped and the relaxation of observables 
is monotonic. One has
\begin{eqnarray}
\begin{array}{rcll}
 t_{\rm osc} &<& t_r^x \qquad \mbox{Underdamped motion} &\qquad {\tilde\omega}^{2(1-\alpha)} k^{2-\alpha} m^\alpha > \gamma_0^2
\; , 
\\
t_{\rm osc} &>& t_r^x \qquad \mbox{Overdamped motion} &\qquad {\tilde\omega}^{2(1-\alpha)} k^{2-\alpha} m^\alpha < \gamma_0^2
\; .
\end{array}
\end{eqnarray}
The crossover occurs at parameters such that $\gamma_0^2 \approx km$ for $\alpha=1$, see below.
 
Finally, the comparison between the oscillation time, $t_{\rm osc}$, and the velocity relaxation time, $t_r^v$, yields
\begin{eqnarray}
\begin{array}{rcll}
 t_{\rm osc} &<& t_r^v \qquad \mbox{Underdamped motion} &\qquad {\tilde\omega}^{2(1-\alpha)} k^{2-\alpha} m^\alpha > \gamma_0^2
\; , 
\\
t_{\rm osc} &>& t_r^v \qquad \mbox{Overdamped motion} &\qquad {\tilde\omega}^{2(1-\alpha)} k^{2-\alpha} m^\alpha < \gamma_0^2
\; ,
\end{array}
\end{eqnarray}
exactly the same conditions as for the position, as it should. The difference in behaviour between position and velocity is 
decided by the comparison between the relaxation time of the velocity and position. If $t_r^v \ll t_r^x$ the 
velocity equilibrates with the environment well before the position. 
 
\vspace{0.25cm}
\noindent
\textcolor{red}{\it Quantitative analysis in the white noise case}
\vspace{0.25cm}

The full differential equation (\ref{Langevin_m}) can be easily
solved by first evaluating the Green function $G(t)$ from
\begin{equation}
m \ddot G(t) + \gamma_0 \dot G(t) + k G(t) = \delta(t)
\; ,
\label{eq:equation-Green}
\end{equation}
that, after Fourier transforming, implies 
\begin{equation}
\tilde G(\omega) = 1 /(-m \omega^2 - i\gamma_0\omega + k ) 
\; .
\end{equation}
The right-hand-side has two poles:
\begin{eqnarray}
\omega_{\pm} = -\frac{i\gamma_0}{2m} \pm 
\sqrt{\frac{k}{m} - \frac{\gamma_0^2}{4m^2}}
\; ,
\label{eq:modes-harmonic}
\end{eqnarray}
that are complex or imaginary depending on the relative values of
the parameters:
\begin{eqnarray}
 4km -\gamma_0^2 >0 \;\;\;\; &&\omega_{\pm} \;\; 
\mbox{complex (under-damped case)} \; ,
\label{eq:complex}
\\
4km -\gamma_0^2 \leq 0 \;\;\;\; &&
\omega_{\pm} \;\;\mbox{imaginary (over-damped)} 
\label{eq:imaginary}
\; .
\end{eqnarray}
It is important
to note that in both cases the poles are located 
in the lower half complex plane. 

Using Cauchy's formula to transform back in time
one finds that, for $t>0$, the Green function reads
\begin{eqnarray}
G(t) &=& \left\{
\begin{array}{lcl}
\displaystyle{\frac{1}{m \omega_R} \, \sin\omega_R t 
\; e^{-|\omega_I| t}}
\;\; && \;\; \mbox{if} \;\; 
\omega_\pm = \pm \omega_R - i |\omega_I|
\nonumber\\
\displaystyle{\frac{i}{m(\omega_+-\omega_-)} \, \left(
e^{-|\omega^{(+)}_I| t} - e^{-|\omega^{(-)}_I| t}
\right)}
\;\; && \;\; \mbox{if} \;\; 
\omega_\pm = - i |\omega^{(+,-)}_I|
\end{array}
\right.
\end{eqnarray}
and it vanishes identically for $t<0$. Two other important properties
of $G(t)$ are $G(0)=0$ and $m\dot G(0)=1$ that follow from integrating 
(\ref{eq:equation-Green}) 
between $t=-\delta$ and $t=\delta$ and taking $\delta\to 0$. One also checks
$2|\omega_I| \dot G(0) + \ddot G(0) =0$ in the under-damped case.

In the under-damped case the time-dependent position of the particle is given by
\begin{eqnarray}
&& 
x(t) = 
e^{-|\omega_I| t} \left\{ [\dot x(0) + x(0) | \omega_I| ] \ \frac{\sin \omega_R t}{\omega_R} + x(0) \cos\omega_R t
\right\}
\nonumber\\
&& 
\qquad
+
 \int_0^\infty dt' \; G(t-t') \, [\xi(t') + h(t')]
\label{eq:x-sol-harm}
\label{xoscclas}
\end{eqnarray}
and this can be rewritten as
\begin{eqnarray}
&& 
x(t) = 
[\dot x(0) + x(0) |\omega_I|] m G(t) + x(0) [m \dot G(t) + |\omega_I| m G(t)] 
\nonumber\\
&& 
\qquad
+
 \int_0^\infty dt' \; G(t-t') \, [\xi(t') + h(t')]
\label{eq:x-sol-harm}
\label{xoscclas}
\end{eqnarray}

The first two terms on the {\sc rhs} represent the effect of 
the initial conditions. Note that $G(t)$ is proportional to a
Heaviside theta function and hence the integration over time 
has an effective upper limit at $t'=t$. One can find corresponding expression for the
over-damped case.

Let us first discuss the asymptotic values of one-time quantities.
The simplest cases are the averaged position and momentum themselves. 
In the absence of an external field, the 
potential is symmetric with respect to $x\to -x$ and $p\to -p$.
Since the noise $\xi$ has zero average,
after a characteristic-time needed to forget the 
initial conditions,
the average of both $x$ and $p$ vanish if $k\neq 0$. 
This is consistent with the 
result expected in equilibrium, $\langle x\rangle_{\sc eq} = 
\langle p\rangle_{\sc eq}=0$, though it is not sufficient to prove that the
particle equilibrates with its environment.
The way in which this zero 
limit is approached depends strongly on the value of $4km-\gamma_0^2$
and we shall discuss it later. 

When $k=0$ the result is different.  In the absence of external
forces, while the average momentum vanishes, the average coordinate
approaches a non-zero value for $t\gg t_c^v$, $\langle x(t)\rangle \to
x(0)+p(0)/\gamma_0$: the initial condition is remembered forever by the
particle's motion. It is a first indication of the non-equilibration
of the coordinate for a flat potential.

Independently of the parameters $k$, $\gamma_0$ and $T$ and as long
as $m\neq 0$, after a tedious but straighforward calculation
one finds that 
\begin{displaymath}
\lim_{t \gg t_c^v}
C_{pp}(t,t) = \lim_{t \gg t_c^v} \langle p(t) p(t) \rangle = 
mk_B T = \langle p^2 \rangle_{\sc eq} 
\; ,
\;\;\;\;\;\;\;\;\;\;\;
t_c^v \equiv \frac{m}{\gamma_0}
\end{displaymath}
where the last term indicates the static average.
The same kind of calculation can be pursued to 
show that the average of any function of the momentum approaches 
its equilibrium limit asymptotically. This is good evidence for
establishing the equilibration of the momentum. [Note that even if 
one of the characteristic times that determine the relaxation of the
Green function diverges when $k=0$, the velocity-velocity correlation
is well-behaved since it only involves $\dot G(t)$.]

The observables that are functions of the  position 
depend on the value of $k$. As long as $k >0$ there is a confining
harmonic potential for the position and all equal-time functions
of it approach an asymptotic limit that coincides with the
one dictated by the equilibrium distribution. For instance,
\begin{displaymath}
\lim_{t \gg t_c^v} C_{xx}(t,t) = 
\lim_{t\gg t_c^v  } \langle x(t) x(t) \rangle = 
\frac{k_B T}{k} =
\langle x^2 \rangle_{\sc eq}
\; .   
\end{displaymath}
Instead, if $k=0$ there is no confining potential and the particle 
diffuses to infinity. 
If $k<0$ the potential pushes the 
particle away from the origin towards $\pm \infty$ depending on the 
sign of the initial position. In none of these cases
one can define a normalisable measure over the full
infinite space and the position of the particles does not equilibrate with 
its environment. We discuss these two cases in detail below
focusing on the study of the temporal evolution of 
correlation functions that depend on two times.
We analyse the auto-correlation 
\begin{displaymath}
C_{xx}(t,t')=
\mbox{Effect of initial cond}+
2 k_B T \gamma_0  \int_0^\infty ds \; G(t-s) G(t'-s)
\; ,
\end{displaymath} 
and the linear response of the
position of the particle at time $t$ after a kick to this same variable
has been applied at a previous time $t'$. From 
eqn~(\ref{eq:x-sol-harm}), this is given by the 
Green function itself:
\begin{equation}
R_{xx}(t,t') \equiv \delta \langle x(t) \rangle/\delta h(t')|_{h=0}= G(t-t')
\label{eq:resp_osc}
\end{equation}
We distinguish the relaxation with different damping
arising from different values of the parameters.

\vspace{0.25cm}
\noindent
\textcolor{red}
{\it Relaxation in  the under-damped limit}
\vspace{0.25cm}

\begin{figure}
\centerline{
\includegraphics[scale=0.4]{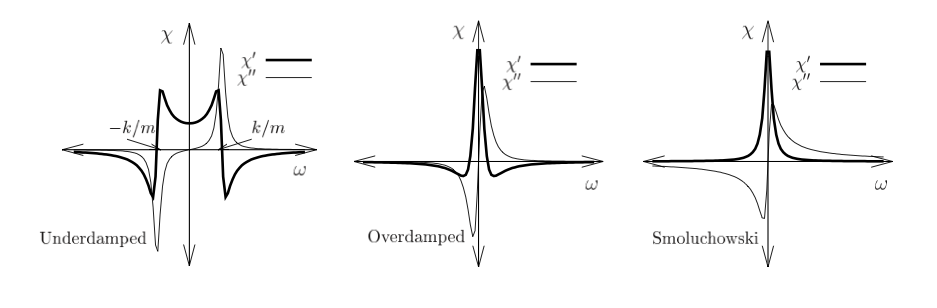}
}
\caption{\small Green functions of the damped harmonic oscillator in different limits.}
\label{fig:green-function}
\end{figure}

When $\omega_R\neq 0$, see eqn~(\ref{eq:complex}),  
the self correlation and linear response {\it oscillate} with frequency
$\omega_R=\sqrt{k/m-\gamma_0^2/(4m^2)}$ and {\it decay exponentially}
with a characteristic time $t_c=|\omega_I|^{-1}=2m/\gamma_0$.  
They are displayed with dashed lines in Fig.~\ref{corr_resp_osc_fig}. 
The Fourier representation of the response function
is shown in Fig.~\ref{fig:green-function}-left where we plot 
$\chi'_{xx}$ and
$\chi''_{xx}$ as functions of $\omega$. We observe that 
$\chi'$ changes sign at $\omega=\pm k/m$ and 
$\chi''(\omega)$ has peaks at $\omega=\pm \sqrt{k^2/m^2-\gamma_0^2/4}$
with half-width at half maximum equal to $\gamma_0/2$. If $\gamma_0\to 0$ 
these peaks approach the frequencies $\pm k/m$ of the undamped 
oscillator.

\vspace{0.25cm}
\noindent
\textcolor{red}
{\it Relaxation in the over-damped limit}
\vspace{0.25cm}

If, instead, we take the case in eqn~(\ref{eq:imaginary}) 
 for which $\omega_R=0$ (and $k\neq 0$) 
the self correlation and linear response have 
{\it pure exponential} decays with two
time constants:
\begin{eqnarray}
\begin{array}{ccc}
t_{\sc fast} = \omega_-^{-1} &=& 
\displaystyle{\frac{2m}{\gamma_0 + \sqrt{\gamma_0^2 - 4km}}}
\;\; \to \;\; \frac{m}{\gamma_0} \equiv t_c^v 
\\
t_{\sc slow} = \omega_+^{-1} &=& 
\displaystyle{\frac{2m}{\gamma_0 - \sqrt{\gamma_0^2 - 4km}}}
\;\; \to \;\; \frac{\gamma_0}{k}  \equiv t_c^x
\end{array}
\;\;
\mbox{when}  \;\; 4km \ll \gamma_0^2
\end{eqnarray}
When $km \ll \gamma_0^2$ the fast decay time, which is the caracteristic
time for relaxation of the velocity correlations, is much shorter than
the slow one, $t_{\sc fast}\ll t_{\sc slow}$. For long observation
times compared to $t_{\sc slow}$ one can neglect the fast mode. This
is equivalent to neglecting the inertial term in the original Langevin
equation and using the Smoluchowski limit 
to construct the properties of the
coordinate.

The real and imaginary parts of the Fourier transform of the 
linear response are usually called $\chi'$ and $\chi''$. 
$\chi'$ is peaked at the origin.
In the extreme over-damped limit in which one can neglect inertia 
$\chi''/\omega$ is a Lorentzian centered at the origin with 
width $t_{\sc slow}^{-1}$.


\vspace{0.25cm}
\noindent
\textcolor{red}
{\it Relaxation in the Smoluchowski limit}
\vspace{0.25cm}

In this purely viscous case, where $m=0$, there is only one characteristic time left, 
$t^x_c = \gamma_0/k$. 
The response decays exponentially, $R(t) = \gamma_0^{-1}e^{-t/t^x_c}$ and 
the susceptibility is then given by 
\begin{displaymath}
\tilde \chi(\omega)=\frac{1}{-i\gamma_0 \omega+k} = 
\frac{k}{k^2+\gamma_0^2\omega^2}
+ i \frac{\gamma_0 \omega}{k^2+\gamma_0^2\omega^2} 
\; .
\end{displaymath}
Its real part is positive for all values of $\omega$ and the imaginary part is 
usually said to take a {\it Debye} form. See the right panel in 
Fig.~\ref{fig:green-function}.

\vspace{0.25cm}
\noindent
\textcolor{red} 
{\it Diffusion in the random walk limit}
\vspace{0.25cm}

When $k\to 0$ the coordinate $x$ does not have a confining potential 
and a normalized equilibrium distribution cannot be defined
for this degree of freedom. In this case there is  no reason to expect 
that any equilibrium property will apply to this variable.
Indeed, when $k\to 0$ the characteristic time $t_{\sc slow}$ diverges:
there is no relaxation and a Brownian particle diffuses. 
The Green function approaches, exponentially in $t-t'$, a finite limit:
\begin{equation} 
G(t-t') \sim \frac1\gamma_0\; 
\left( 1-e^{-\frac{\gamma_0}{m} (t-t')} \right)
\; .
\end{equation}

For any fixed time-difference, the correlation 
function diverges linearly with 
the shorter time.
If $t' \leq t$, for $t' \gg t_c^v$ and $t-t'$ fixed, choosing the
simplest initial condition $x(0)=p(0)=0$, we have
\begin{eqnarray*}
\lim_{t'\gg   t_c^v, \;\; t-t' \, \mbox{\tiny{fixed}}} 
C_{xx}(t,t') =  - \frac{2mk_BT}{\gamma_0^2} 
\left( 1 - \frac{1}{2} e^{- \frac{\gamma_0}{m} |t-t'|} \right)
+ \frac{2k_B T}{\gamma_0} \min(t,t')
\end{eqnarray*}
 In particular, at equal long times $t=t' \gg   m/\gamma_0$,
$C_{xx}(t,t)\sim 2k_B T/\gamma_0 \; t$. This demonstrates the breakdown of 
stationarity and hence the fact that the system is far from equilibrium.
For $\min(t,t')$ fixed, $C_xx(t,t')$
decays exponentially with the time-difference towards
the constant $2k_B T/\gamma_0 (\min(t,t')-m/\gamma_0)$.

The displacement $\Delta_{xx}$ instead is a simpler function of $t-t'$,
and for long time-differences it becomes the usual diffusion law.

\vspace{0.25cm}

\noindent
\textcolor{red}{
{\it  Over-damped limit}
}

\vspace{0.25cm}

Another relevant example is the relaxation of a particle in a harmonic
potential, with its minimum at $x^*\neq 0$:
\begin{equation}
V(x) = \frac{k}{2} (x-x^*)^2
\; ,
\end{equation} 
in contact with  noise that we take to be white as the simpler starting case.
The potential confines the particle and one can then expect the coordinate 
to reach an equilibrium distribution.

This problem can be solved exactly keeping inertia for all values of
$\gamma_0$ but the calculation is slightly tedious. The behavior of the
particle velocity has already been clarified in the constant force
case.  We now focus on the over-damped limit, 
\begin{equation}
\gamma_0 \dot x =-k (x-x^*)+\xi
\; , 
\end{equation}
with $k$ the spring constant of the harmonic well, that can be readily solved, 
\begin{equation}
x(t) = x_0 \ e^{-\frac{k}{\gamma_0} t} + \gamma_0^{-1}  
\int_0^t dt' \ e^{-\frac{k}{\gamma_0} (t-t')} \ [\xi(t')+kx^*] 
\; , 
\qquad  x_0 =x(0)
\; .
\label{eq:x-traj-osc}
\end{equation}
This problem becomes formally identical to the velocity dependence in the 
previous example. 

\vspace{0.25cm}
\noindent
\textcolor{red}{\it Convergence of one-time quantities}
\vspace{0.25cm}

The averaged position is
\begin{equation}
\fbox{$
\langle x(t) - x^*\rangle= (x_0 - x^*) e^{-\frac{k}{\gamma_0}t} \to 0 \qquad\qquad 
t_r^x \gg \gamma_0/k
\;\;\;$ (Convergence)
}
\end{equation}
Of course, one-time quantities should approach a constant
asymptotically if the system equilibrates with its environment.

\begin{figure}[h]
\centerline{
\includegraphics[scale=0.24]{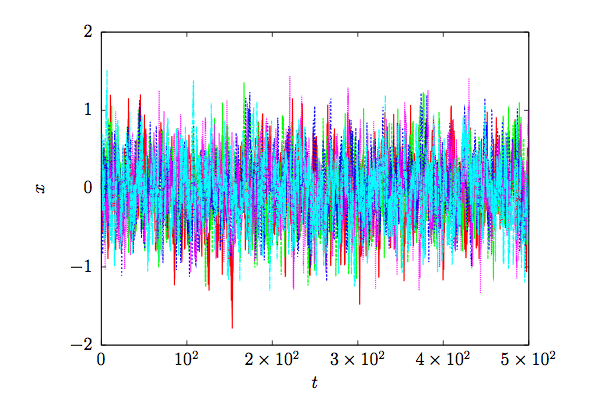}
\hspace{-0.25cm}
\includegraphics[scale=0.23]{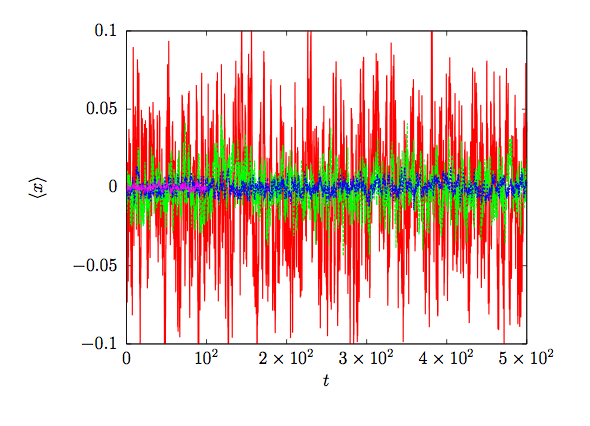}
\hspace{-0.25cm}
\includegraphics[scale=0.16]{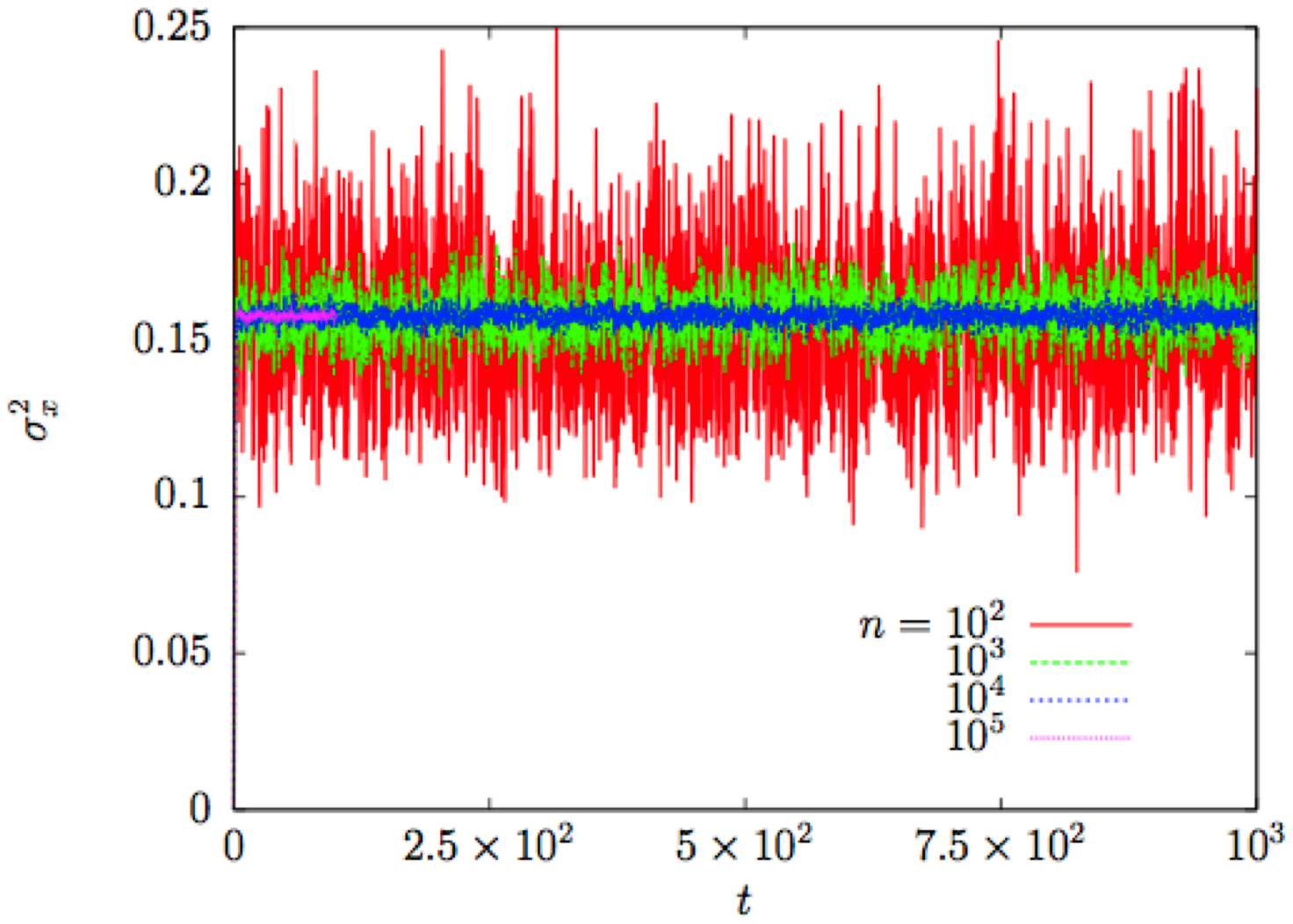}
}
\caption{\small Left panel: five runs of the Langevin equation in the over-damped limit with a quadratic external potential (oscillator)
and a Gaussian white noise
at temperature $T$. Central panel: the average $\langle x\rangle$ computed with 
$n=10^2, \ 10^3, \ 10^4, \ 10^5$ runs. Right panel: the variance $\sigma^2_x=\langle x^2 \rangle - \langle x\rangle^2= k_BT/k$.
}
\end{figure}


\vspace{0.25cm}
\noindent
\textcolor{red}{\it Two-time quantities}
\vspace{0.25cm}

The two-time connected correlation (where one extracts, basically, the 
asymptotic position $x^*$) reads
\begin{equation}
\langle \delta x(t) \delta x(t') \rangle = 
k_B T \ k^{-1} \ 
e^{-\frac{k}{\gamma_0} (t+t') } \left[ e^{2\frac{k}{\gamma_0} \min(t,t')}-1\right]
\; . 
\label{eq:Dirichlet}
\end{equation}
Again, the \textcolor{blue}{Dirichlet correlator} ($\delta x(t)=x(t)-\langle x(t)\rangle$).  For at least one
of the two times going well beyond the position relaxation time
$t_r^x=\gamma_0/k$ the memory of the initial condition is lost and the
connected correlation becomes \textcolor{blue}{ stationary}:
\begin{equation}
C_c(t,t') = \langle \delta x(t) \delta x(t') \rangle \to
k_B T \ k^{-1} \ 
e^{-\frac{k}{\gamma_0} |t-t'| } 
\qquad \min(t,t') \gg t_r^x
\; . 
\end{equation}
For time-differences that are longer than $t_d^x=\gamma_0/k$ the correlation 
decays to $1/e$ and one finds $t_d^x=t_r^x$. Interestingly enough, the relaxation 
and decay times diverge when $k\to 0$ and the potential becomes flat.

Note that when the time-difference $t-t'$ diverges the average of 
the product factorizes, in particular, for the correlation one gets 
\begin{equation}
\langle x(t) x(t') \rangle \to 
\langle x(t) \rangle \langle x(t') \rangle \to
x^* \langle x(t')\rangle 
\label{eq:factorization}
\end{equation}
for any $t'$, even finite.  We will see this factorization property
at work later in more complicated cases.

\vspace{0.2cm}
\noindent{\textcolor{red}{\it Fluctuation-dissipation theorem (FDT)}}
\vspace{0.2cm}

One can also compute the linear response to an infinitesimal perturbation that couples linearly to 
the position changing the energy of the system as $H \to H-f x$ at a given time~$t'$:
\begin{equation}
R(t,t') = \left. \frac{\delta \langle x(t)\rangle_f}{\delta f(t')} \right|_{f=0}
\; . 
\end{equation}
The explicit calculation yields 
\begin{eqnarray}
&& R(t,t') = 
\gamma_0^{-1} \ e^{-k\gamma_0^{-1} (t-t')} \ \theta(t-t') 
\nonumber\\
&&
\fbox{$
R(t,t') = 
\frac{1}{k_BT} \ \dfrac{\partial C_c(t,t')}{\partial t'} \ \theta(t-t')
\;\;\;$ (FDT)}
\label{eq:FDT}
\end{eqnarray}
The last equality holds for times that are longer than $t_r^x$.  It
expresses the \textcolor{blue}{  fluctuation-dissipation theorem 
({\rm fdt})}, a
model-independent relation between the two-time linear response and
correlation function. Similar - though more complicated - relations
for higher-order responses and correlations also exist in
equilibrium. There are many ways to prove the {\rm fdt} for stochastic
processes. We will discuss one of them in
Sect.~\ref{sec:Langevin-eq-gen} that is especially interesting since 
it applies easily to problems with correlated noise.

It is instructive to examine the relation between the linear response and the 
correlation function in the limit of a flat potential ($k\to 0$). The linear 
response is just $\gamma_0^{-1} \theta(t-t')$. The Dirichlet correlator approaches the diffusive limit:
\begin{equation}
\langle \delta x(t) \delta x(t') \rangle =
 2 \gamma_0^{-1} k_BT \  \min(t,t') \qquad \mbox{for} \qquad k\to 0
\end{equation}
and its derivative reads $
\partial_{t'} \langle \delta x(t) \delta x(t') \rangle = 
2 \gamma_0^{-1} k_BT \ \theta(t-t') 
$. Thus, 
\begin{eqnarray}
&& R(t,t') = \frac{1}{2k_BT} \ 
\frac{\partial}{\partial t'} \langle \delta x(t) \delta x(t')\rangle \ \theta(t-t')
\nonumber\\
&& \fbox{$
R(t,t') = \dfrac{1}{2k_BT} \ \partial_{t'} C_c(t,t') \ \theta(t-t')
\;   \;\;\;$ (FDR for diffusion)}
\end{eqnarray}
 A factor $1/2$ is now present in the relation between $R$ and $C_c$. It
is another signature of the fact that the coordinate is not in
equilibrium with the environment in the absence of a confining potential.

\vspace{0.25cm}

\noindent{\textcolor{orange}{\bf Exercise \thesection.\theexercise}}
Evaluate the two members of the FDT, Eq.~(\ref{eq:FDT}), 
in the case of the tilted potential
$V(x)=-Fx$. Conclude.

\addtocounter{exercise}{1}

\vspace{0.25cm}
\noindent{\textcolor{orange}{\bf Exercise \thesection.\theexercise}}
Compute $\langle x^n(t) x(t')\rangle$ and $\delta \langle x^n(t)\rangle/\delta h(t') |_{h=0}$
and compare. Discuss.

\addtocounter{exercise}{1}

\vspace{0.25cm}
\noindent{\textcolor{orange}{\bf Exercise \thesection.\theexercise}}
Take the diffusive problem in the over-damped limit, $\dot x = \xi$, where the friction coefficient has been 
absorbed with a redefinition of time. Compute the linear response of the $n$-th moment of the coordinate
position $\langle x^n(t)\rangle_h$ under a perturbation that modifies the energy according to 
$V \mapsto V- h x$. Compute the time-derivative of the correlation $\langle x^n(t) x(t') \rangle$. Compare
the two and conclude about the pre factor that relates them. Does it depend on $n$? Is it equal to $(k_BT)^{-1}$?

\addtocounter{exercise}{1}

\vspace{0.25cm}

\noindent
\textcolor{red}{
\noindent{\it Reciprocity or Onsager relations}
}
\vspace{0.2cm}

Let us compare the two correlations $\langle x^3(t) x(t')\rangle$ 
and $\langle x^3(t') x(t)\rangle$ within the harmonic example. One finds
$\langle x^3(t) x(t') \rangle = $
\newline
$3 \langle x^2(t)\rangle \langle x(t)x(t')\rangle$ and 
$\langle x^3(t') x(t) \rangle = 3 \langle x^2(t')\rangle \langle x(t')x(t)\rangle$. 
Given that $\langle x^2(t)\rangle = \langle x^2(t')\rangle \to  \langle x^2\rangle_{eq}$ and 
the fact that the two-time self-correlation is symmetric, 
\begin{equation}
\langle x^3(t) x(t') \rangle = \langle x^3(t') x(t) \rangle 
\; . 
\end{equation}
With a similar argument one shows that for any functions $A$ and $B$ of $x$:
\begin{eqnarray}
&&
\langle A(t) B(t') \rangle = \langle A(t') B(t) \rangle 
\nonumber\\
&& 
\fbox{$
C_{AB}(t,t') = C_{AB}(t',t) 
\; \;\;\;
$ (Reciprocity)
} 
\end{eqnarray}
This equation is known as \textcolor{blue}{Onsager relation} and applies 
to $A$ and $B$ that are even under time-reversal (e.g. they 
depend on the coordinates but not on the velocities or they have an 
even number of verlocities).

All these results remain unaltered if one adds a linear potential 
$-Fx$ and works with connected correlation functions.

\noindent
\textcolor{red}{\subsubsection{A harmonic potential and white multiplicative noise}}
\label{subsubsec:harmonic-multiplicative}

Take the equation
\begin{equation}
\frac{dx(t)}{dt} = - k x(t) +x(t) \xi(t)
\end{equation}
with zero average Gaussian white noise. Use  
the \textcolor{blue}{Stratonovich} convention in which usual rules of calculus apply.
In order to solve this stochastic differential equation, first absorb the first term in the rhs with the redefinition $y(t) = e^{k t} x(t)$ and write
\begin{equation}
\frac{dy}{dt} = y(t) \xi(t)
\end{equation}
with the solution
\begin{equation}
y(t) = y(0) \ e^{\int_0^t dt' \ \xi(t')} 
\qquad
\Rightarrow 
\qquad 
x(t)  = x(0) \ e^{-kt + \int_0^t dt' \ \xi(t')}
\end{equation}
For a Gaussian noise the average yields
\begin{equation}
\langle x(t) \rangle = x(0) \ e^{-kt} \ e^{\frac{1}{2} \langle (\int_0^t dt' \ \xi(t') )^2\rangle }
= x(0) \ e^{-kt} e^{k_BT t} = x(0) e^{-(k-k_BT) t}
\end{equation}
interestingly enough, the average particle position vanishes or diverges depending upon $k>k_BT$ or $k < k_BT$.
It is clear from this example that in the latter case the particle does not equilibrate with the potential $V(x)=x^2/2$ as one 
could have expected. We will see how the departing equation has to be modified, by an additional drift term, to ensure equilibration to a 
quadratic potential, after discussing the Fokker-Planck equation for the 
stochastic process with multiplicative noise in Sec.~\ref{subsubsec:FP-multiplicative}. The equation ensuring this fact is
\begin{equation}
\frac{dx(t)}{dt} = -V'_{\rm eff} + x(t) \xi(t) = k_BT x(t) - k x^3(t) + x(t) \xi(t)
\end{equation}
but this one is no longer  solvable analytically. Note the double well structure induced by the noise in the 
effective potential
\begin{equation}
V_{\rm eff} = -k_BT  \frac{x^2}{2}  + \frac{x^4}{4} 
\end{equation}
that gives rise to \textcolor{blue}{drifted force}
\begin{equation}
f_{\rm drifted} = - g^2 V' + 2k_BT(1-\alpha) gg' = - k x^3 + k_BT x
\; .
\end{equation}
We will justify the generic form of $f_{\rm drifted}$ in Sec.~\ref{subsubsec:FP-multiplicative}.
In interacting many-body problems as the ones described by field equations this phenomenon can induce a phase transition~\cite{Sancho}.

One can still compute the linear response by using the solution under a perturbation linearly coupled to the 
particle's position ($-k x \mapsto -k x +h$)
\begin{eqnarray}
x(t) = x_0 \ e^{-k t + \int_0^t dt' \ \xi(t')} + \int_0^t dt' \ e^{-k (t-t') + \int_{t'}^t dt'' \ \xi(t'') } \ h(t')
\end{eqnarray}
from where 
\begin{eqnarray}
R(t,t') 
&=& 
e^{-k (t-t')} \langle e^{\int_{t'}^t dt'' \ \xi(t'')} \rangle 
=
e^{-k (t-t')}  \ e^{\frac{1}{2} \langle (\int_{t'}^t dt'' \ \xi(t'') )^2 \rangle } 
\nonumber\\
&=&
e^{-k (t-t')} \ e^{k_BT (t-t')} 
= 
e^{-(k-k_BT) (t-t')}
 \end{eqnarray}
 where we ignored the $\theta(t-t')$ factor assuming $t\geq t'$.  The linear response is a stationary function, in the sense that 
 it only depends upon $t-t'$. 
 What about the position-position correlation and mean-share displacement? They read
 \begin{eqnarray} 
C(t,t') &=& \langle x(t) x(t') \rangle 
= 
x_0^2 \ e^{-k(t+t')} \ e^{k_BT [ \max(t,t')]}
\nonumber\\
\Delta(t,t') &=& \langle (x(t) - x(t'))^2\rangle 
\end{eqnarray}
and they both behave very weirdly.


\noindent
\textcolor{red}{\subsubsection{Colored noise}
}

\vspace{0.2cm}
\noindent
\textcolor{red}{
\noindent{\it Colored noise with exponential correlation}
}

\vspace{0.25cm}

\noindent\textcolor{orange}{\bf Exercise \thesection.\theexercise}  Solve the stochastic dynamics of a particle in a 
harmonic potential with a exponentially decaying memory kernel $\Gamma(t-t') = \gamma_0 e^{-|t-t'|/\tau_D}$. 
Hint: use Laplace transform
techniques.

\vspace{0.25cm}

\addtocounter{exercise}{1}

\begin{figure}
\centerline{
\includegraphics[scale=0.35]{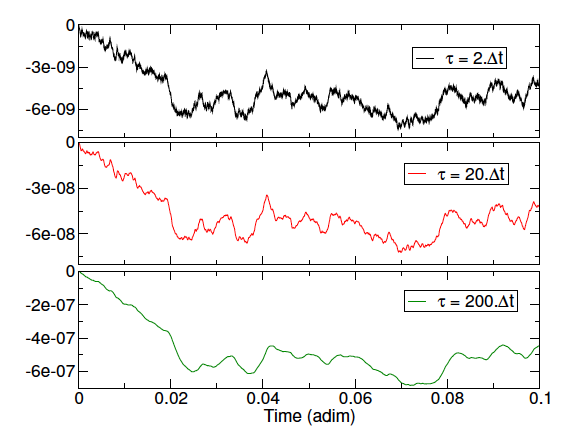}
}
\caption{\small Random walk under the effect of a noise with exponential decaying correlations controlled by the parameter $\tau_D$ given as 
labels in the figure, see Eqs.~(\ref{eq:exp-corr-noise}). The time-step for the integration is $\Delta t = 10^{-4}$, temperature fixed by $2k_BT= 0.5$, and the damping parameter is $\lambda=10^{-2}$.
Figure taken from~\cite{Tranchida16}.}
\label{fig:tranchida}
\end{figure}

Figure~\ref{fig:tranchida} shows the solution to the Langevin equations:
\begin{eqnarray}
\dot x &=& - \lambda x + \eta \; ,  \\
\dot \eta &=& -\eta/\tau + \xi \; , 
\label{eq:exp-corr-noise}
\end{eqnarray}
with $\langle \xi(t) \rangle =0$ and $\langle \xi(t) \xi(t') \rangle = 2k_BT/\tau_D \ \delta(t-t')$. This white noise induces an exponentially decaying correlation function of the 
noise $\eta$ that appears  in the first equation. (Note, however, that this first equation lacks the memory kernel that should be present in the left-hand-side to ensure the approach to 
equilibrium of such an equation under a potential.) The three panels in the equation show the trajectories $x(t)$ for the same realization of the white noise $\xi$. The smoothing effect of 
correlation is quite clear.

\vspace{0.2cm}
\noindent
\textcolor{red}{
\noindent{\it Colored noise with power law correlation}
}

\vspace{0.2cm}

Let us now take a power-law correlated noise. The Langevin equation can be solved by using the Laplace transform.
Correlation and linear responses can be computed. As the system should equilibrate -- there is confining
potential -- the FDT holds. However, the decay of these two functions (and more complex ones involving more
times) are not trivial in the sense that their temporal dependence is not exponential. Instead, the position correlation
function and its linear response are given by the Mittag-Leffer function
\begin{eqnarray}
&& 
C_{xx}(t,t') = \frac{1}{k} E_{\alpha,1}\left(\frac{k |t-t'|^\alpha}{\overline\gamma_0}\right) 
\; , 
\\
&& 
R_{xx}(t,t') = \frac{1}{\overline\gamma_0} E_{\alpha,\alpha} \left(\frac{k |t-t'|^\alpha}{\overline\gamma_0}\right) \theta(t-t')
\; , 
\end{eqnarray}
where $\overline\gamma_0$ is a constant that is proportional to $\gamma_0$ and all other 
pre-factors in $\Gamma(t-t')$. For the Ohmic $\alpha=1$ case the Mittag-Leffer function becomes an 
exponential, as expected. For $\alpha \neq 1$ the decay is algebraic, $E_{\alpha,1}(x) \simeq x^{-1}$ 
that implies $C_{xx}(t-t') \simeq |t-t'|^{-\alpha}$. 
The ratio between linear response and time derivative of the correlation function is
\begin{equation}
\frac{k_BTR_{xx}(t-t')}{\partial_{t'} C_{xx}(t-t')} 
=
1+\left( \frac{t}{t'}-1\right)^{1-\alpha} 
\frac{E_{\alpha,1}(-k t^\alpha/\overline\gamma_0)  E_{\alpha,\alpha}(-k{t'}^{\alpha}/\overline\gamma_0)}
        {E_{\alpha,\alpha}(-k (t-t')^\alpha/\overline\gamma_0)}
\end{equation}
In the long time limit, $t\geq t' \gg 1$, 
the second term vanishes as long as $k>0$ and one recovers the equilibrium result.

\begin{figure}[h]
\begin{center}
\raisebox{4cm}{
\includegraphics[scale=0.2,angle=-90]{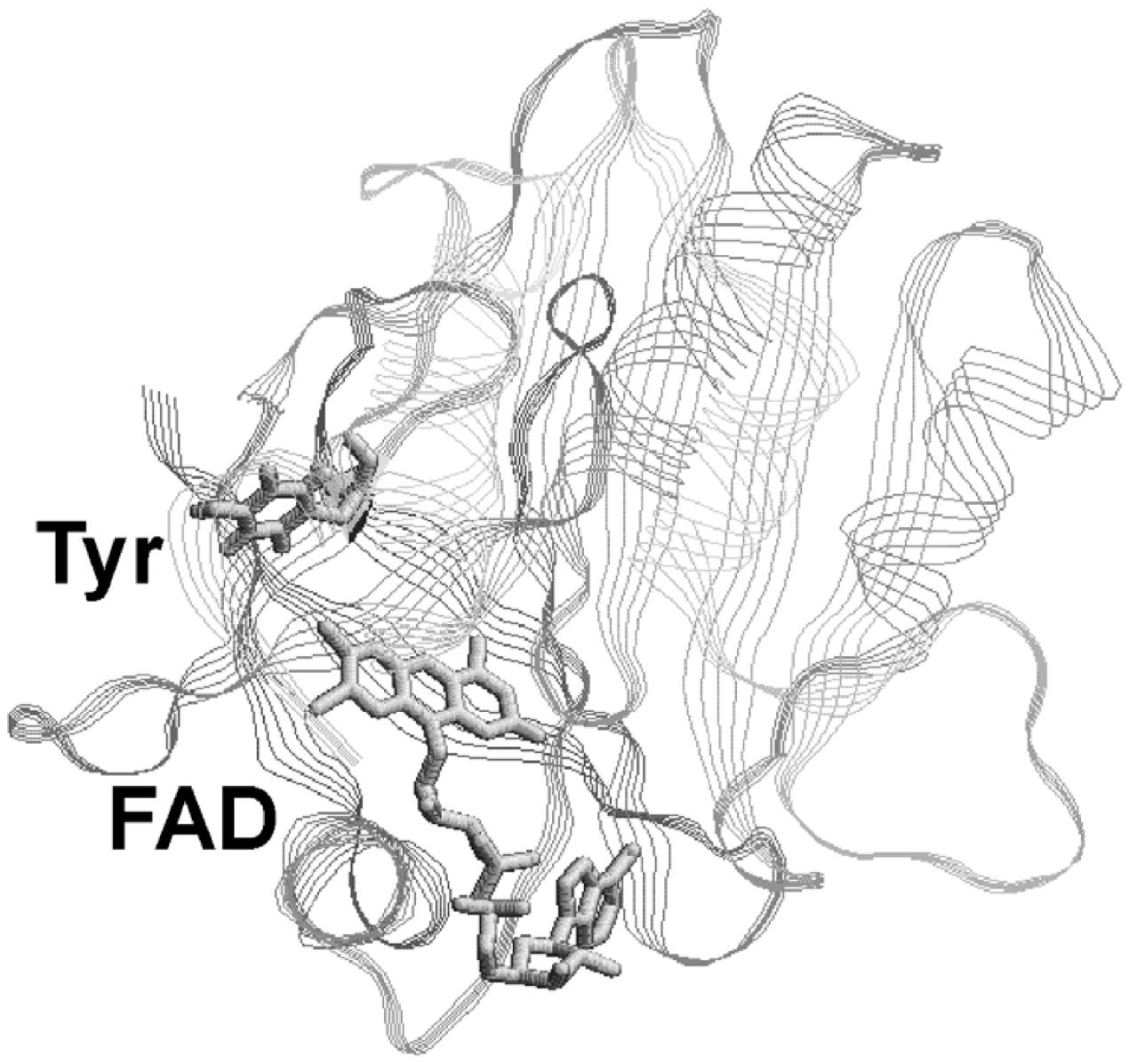}
}
\includegraphics[scale=0.2]{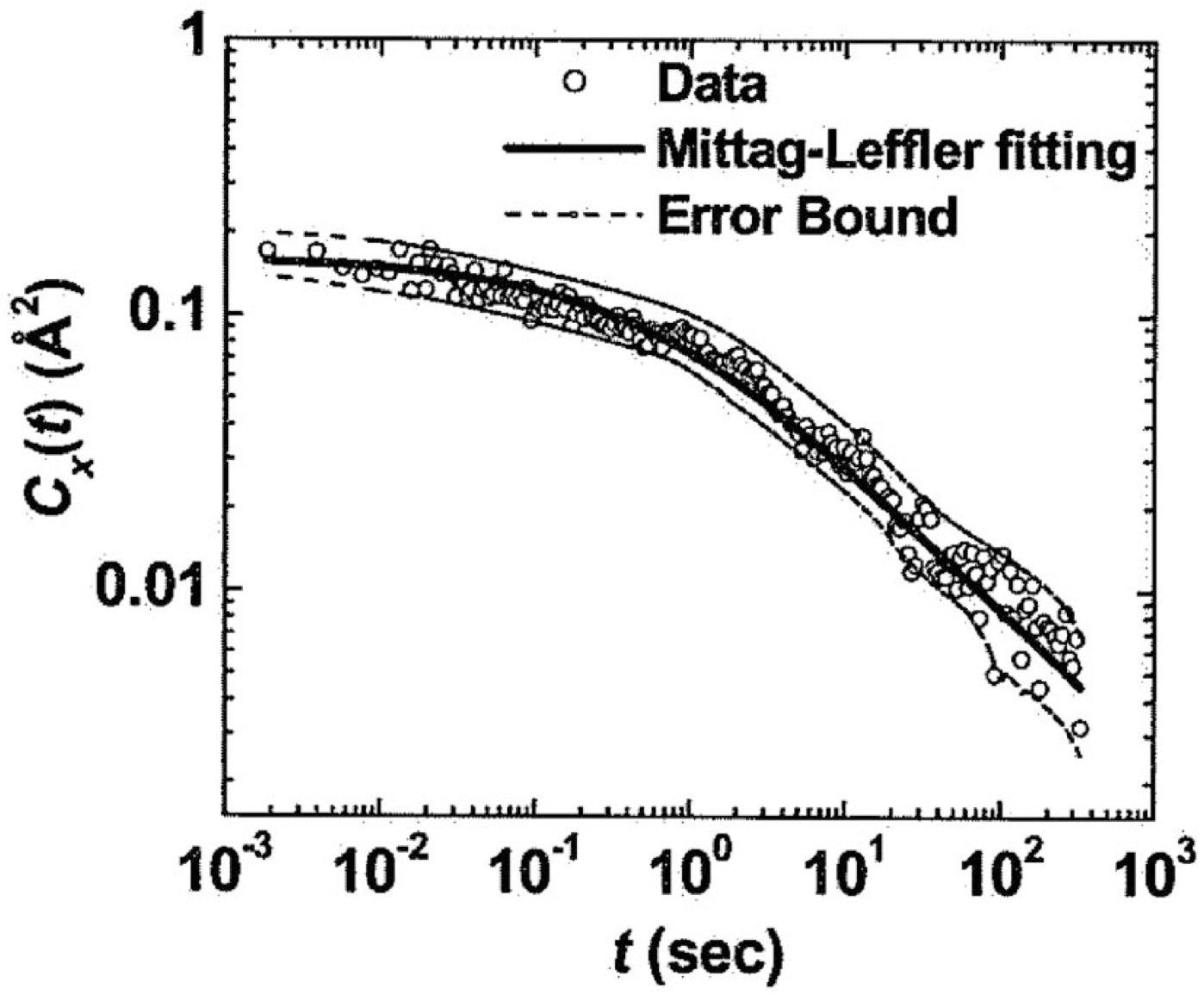}
\caption{\small Sketch of the experiment. Correlation function measured in~\cite{Yang-etal03} and~\cite{Min-etal05}, $\alpha=0.51\pm 0.07$.}
\end{center}
\end{figure}

\pagebreak

\textcolor{red}{
\subsubsection{A two-dimensional example}}

Take a point-like particle with mass $m$ moving in a two dimensional space.
The position of this particle is  $\vec r = (x,y) = x \hat e_x + y \hat e_y$ in a 
Cartesian coordinate system. The particle feels a potential 
$V(x,y)= k x^2/2$ and it is in contact with a generic environment in thermal equilibrium 
at temperature $T$. 

Using what we have already computed for one dimensional problems we can guess the asymptotic 
behaviour of the phase space variables $(\vec p, \vec r)$. For simplicity,
we will use  a white bath with friction coefficient $\gamma_0$. 
The momentum (or velocity) should equilibrate to its Maxwellian form, $\propto \exp(-\beta mv^2/2)$, 
after a characteristic time $\tau^v_r=m/\gamma_0$.
The position $\vec r = (x,y)$ will have different 
behaviour in the $x$ (confined) and $y$ (flat) directions. The $x$ component should reach
equilibrium after a characteristic time $\tau^x_r = \gamma_0/k$. This means that it will 
reach a pdf $\propto \exp(-\beta kx^2/2)$.  
The $y$ component of the position, instead, should undergo normal diffusion and 
it will not equilibrate. 

The expectations exposed in the previous paragraph can be shown analytically.
Take the over-damped (Smoluchowski) limit in which the inertia term in
the dynamic equation is neglected. In this limit the Langevin equation becomes
\begin{eqnarray*}
&&
\gamma_0 \dot x(t) = -k x(t) + \xi_x(t)
\; , 
\\
&&
\gamma_0 \dot y(t) =  \xi_y(t)
\; . 
\end{eqnarray*}
The solutions are 
\begin{eqnarray*}
&& x(t) = x(0) e^{-k t /\gamma_0} + \gamma_0^{-1} \int_0^t dt' \ e^{-k (t-t')/\gamma_0} \ \xi_x(t')
\; . 
\\
&&
y(t) = y(0) + \gamma_0^{-1} \int_0^t dt' \ \xi_y(t')
\; . 
\end{eqnarray*}
The four correlations are given by 
\begin{eqnarray*}
&& C_{xx}(t,t') = \langle x(t) x(t')\rangle = x^2(0) e^{-k(t+t')/\gamma_0} + 
k_BT\gamma_0^{-1} \ [ e^{-k|t-t'|/\gamma_0} - e^{-k (t+t')/\gamma_0} ]
\; , 
\\
&& C_{xy}(t,t') = \langle x(t) y(t')\rangle = x(0) y(0) e^{-kt/\gamma_0} 
\; , 
\\
&& C_{yx}(t,t') = \langle y(t) x(t')\rangle =x(0) y(0) e^{-kt'/\gamma_0}
\; , 
\\
&& C_{yy}(t,t') = \langle y(t) y(t')\rangle = y^2(0) + 2k_BT\gamma_0^{-1} \min(t,t')
\; ,
\end{eqnarray*}
where we used $\langle \xi_x(t)\xi_x(t') \rangle = \langle \xi_y(t)\xi_y(t') \rangle = 2k_BT\gamma_0 \delta(t-t')$,
and the fact that different noise components are uncorrelated, $\langle \xi_x(t)\xi_y(t') \rangle = 0$.
As already announced, in the long times limit, $t \gg \gamma_0/k$ and $t'\gg \gamma_0/k$, one finds stationarity for 
the $xx$ correlation, 
$C_{xx}(t,t') \to k_BT \gamma_0^{-1} e^{-k |t-t'|/\gamma_0}$, decorrelation of the crossed
functions, $C_{xy}(t,t') \to 0$ and $C_{yx}(t,t') \to 0$, and diffusion along the $y$ direction,
$C_{yy}(t,t') \to 2k_BT\gamma_0^{-1} \min(t,t')$.

Apply now a small perturbation to the particle that modifies the potential $V$ according to 
$V\to V - \vec h \cdot \vec r$. The solutions under the perturbation are
\begin{eqnarray*}
&&
\langle x\rangle_{\vec h} =  
x(0) e^{-k t /\gamma_0} + \gamma_0^{-1} \int_0^t dt' \ e^{-k (t-t')/\gamma_0} \ [ \xi_x(t') + h_x(t')] 
\; , 
\\
&& 
\langle y\rangle_{\vec h} = y(0) + \gamma_0^{-1} \int_0^t dt' \ [\xi_y(t') + h_y(t')]
\; ,
\end{eqnarray*}
and these imply
\begin{eqnarray*}
&& R_{xx}(t,t') = \delta \langle x(t)\rangle_{\vec h}/\delta h_x(t'){\large |}_{\vec h=\vec 0}
=
\gamma_0^{-1} \ e^{-k(t-t')/\gamma_0} \ \theta(t-t')
\; , 
\\
&& R_{yy}(t,t') = 
\delta \langle y(t)\rangle_{\vec h}/\delta h_y(t'){\large |}_{\vec h=\vec 0} = 
\gamma_0^{-1} \ \theta(t-t')
\; , 
\\
&& R_{xy}(t,t') = 
\delta \langle x(t)\rangle_{\vec h}/\delta h_y(t'){\large |}_{\vec h=\vec 0}
=
R_{yx}(t,t') = 
\delta \langle y(t)\rangle_{\vec h}/\delta h_x(t'){\large |}_{\vec h=\vec 0}
=
0
\; . 
\end{eqnarray*}
The comparison to the time-derivatives of the associated correlation functions yields
\begin{eqnarray*}
&& 
k_BT R_{xx}(t,t') = \partial_{t'} C_{xx}(t,t') \theta(t-t') 
\qquad\qquad \mbox{and FDT holds}
\; , 
\\
&&
k_BT R_{yy}(t,t') = \frac{1}{2} \partial_{t'} C_{yy}(t,t') \theta(t-t')
\qquad\;\;\;\;\; \mbox{there is a factor of 1/2}
\; , 
\\
&&
k_BT R_{xy}(t,t') = \partial_{t'} C_{xy}(t,t')\theta(t-t') = 0 
\; , 
\\
&&
k_BT R_{yx}(t,t') = 0 \;\; \mbox{and}\;\; \partial_{t'} C_{yx}(t,t')\theta(t-t') \to 0 \;\; \mbox{for} \; t' \gg \gamma_0/k
\; . 
\end{eqnarray*}


\textcolor{red}{
\subsubsection{Perturbation theory}
\label{subsubsec:pert-theory}
}

For the moment we only treated cases in which the potential was, at most, quadratic, and the 
Langevin equation was, therefore, linear in the variable. Quite generally one faces non-linear 
stochastic differential equations that cannot be solved exactly. 

In some fortunate cases, perturbation theory can be easily formulated in this context. Take, for instance,
the case of a quartic potential $V(x) = k x^2/2 + \lambda x^4/2$ with $k>0$ and $\lambda>0$ and 
let us focus on the over-damped dynamics of a particle that starts from the position $x_0$ initially.
The Langevin equation for $\lambda=0$ has already been solved. Let us then take the trajectory (\ref{eq:x-traj-osc})  
as the zero-th order of a systematic expansion in powers of the coupling constant $\lambda$:
\begin{equation}
x(t) = \sum_{n=0} x_n(t) \lambda^n
\end{equation}
with 
\begin{equation}
x(0) = x_0 = \sum_{n=0} x_n(0) \lambda^n
\; .  
\end{equation}
Quite naturally, we choose 
\begin{equation}
x_0(0) = x_0 \qquad\qquad\mbox{and} \qquad\qquad
x_{n>0}(0) =  0
\; . 
\end{equation} 
Order by order in $\lambda$ we then have
\begin{eqnarray}
O(\lambda^0): & \quad &
\gamma_0 \dot x_0(t) = - k x_0(t) + \xi(t) 
\\
O(\lambda^1): &\quad &
 \gamma_0 \dot x_1(t) = -k x_1(t) - x_0^3(t) 
\\
O(\lambda^2): & \quad &
\gamma_0 \dot x_2(t) = - k x_2(t) - 3 x_0^2(t) x_1(t) 
\\
O(\lambda^3): & \quad &
\gamma_0 \dot x_3(t) = - k x_3(t) - 3 x_2(t) x_0^2(t) - 3 x_1^2(t) x_0(t)
\end{eqnarray}
etc.  The structure of these equations is the same, with a linear operator $\gamma_0 d_t + k$ acting on the 
unknown functions at each order and a source term that is known (as a functional of $\xi$) from the previous orders.
Their solutions are
\begin{equation}
x_n(t) = x_n(0) e^{-kt/\gamma_0} + \int_0^t dt' \ e^{-k(t-t')/\gamma_0} \ \mbox{\small source}(t')
\end{equation}
Note that the power expansion in $\lambda$ transforms into a power expansion in $\xi$. The averages can be easily computed by using 
the factorization properties of the noise averages for Gaussian statistics (Wick's theorem).

\begin{figure}[h]
\centerline{
\includegraphics[scale=0.25]{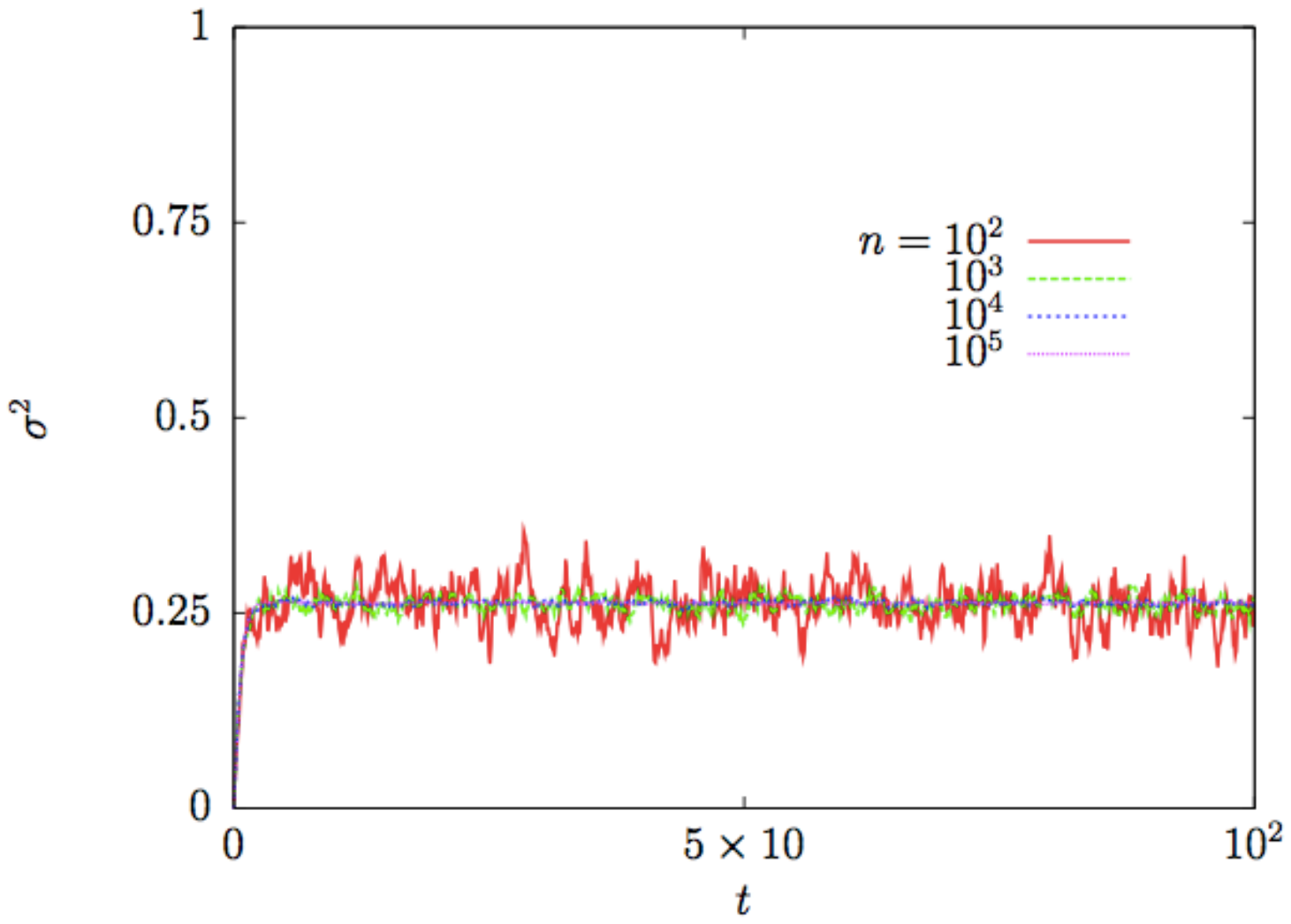}
\includegraphics[scale=0.25]{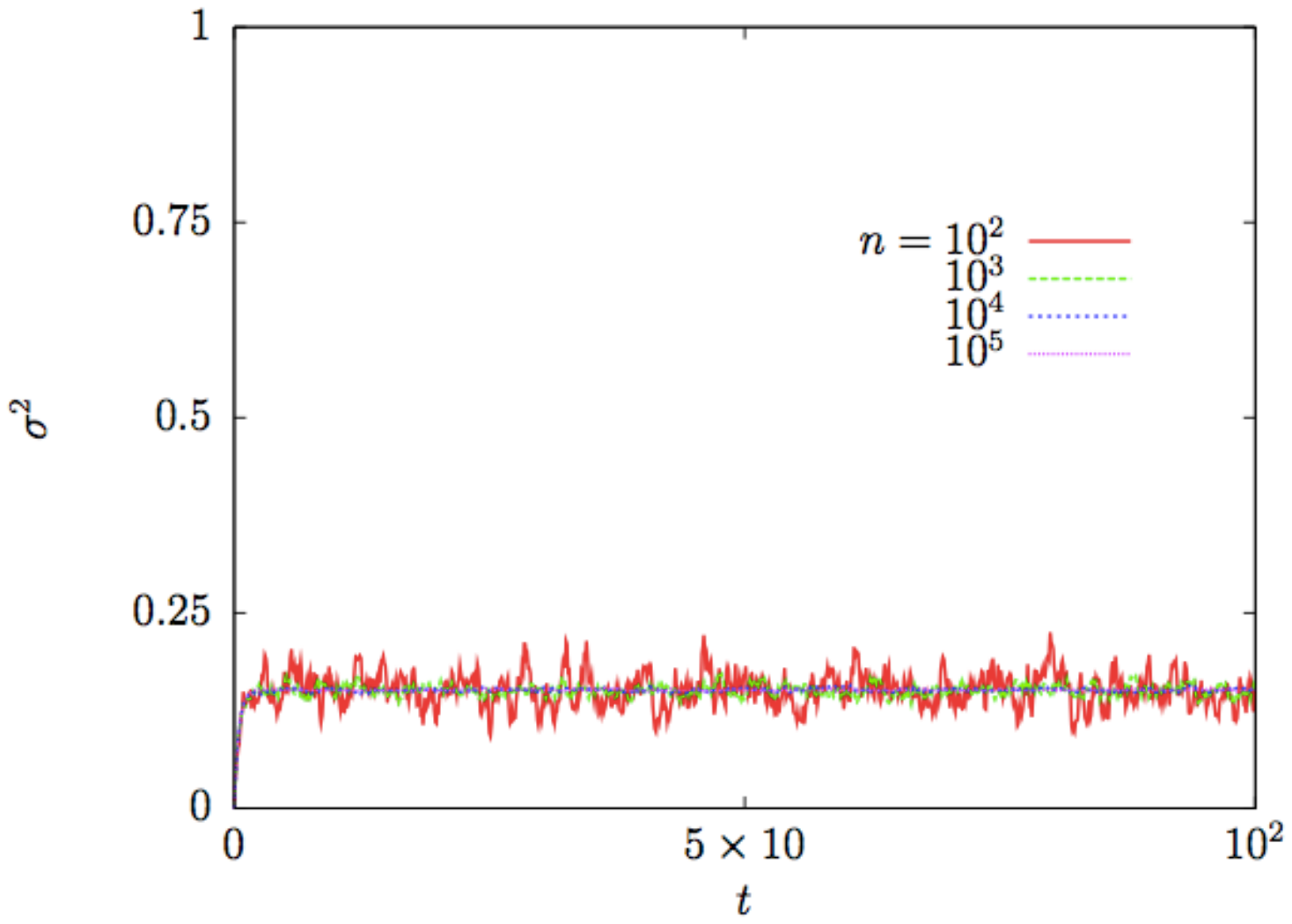}
}
\caption{\small Dynamics in a quartic potential $V(x)=k x^2/2+ \lambda x^4/4$ with $k=0$ and $\lambda=1$ (left) and 
$k=1$ and $\lambda=0.1$ (right). In both cases $\gamma_0=1$ and $k_BT=0.15$. Different curves are for different number of 
samples as explained in the key.}
\label{fig:sigma2-quartic}
\end{figure}

\vspace{0.2cm}
\noindent
\textcolor{orange}{\bf Exercise \thesection.\theexercise} 
Compute the first terms in the expansion above. Compare the outcome for $\sigma^2$ to the 
numerical result shown in Fig.~\ref{fig:sigma2-quartic} generated with $\gamma_0=1$ and $k_BT=0.15$ for a pure quartic 
potential with $\lambda=1$ and for a potential with $k=-1.$ and $\lambda=0.1$. 

\addtocounter{exercise}{1}

\vspace{0.2cm}

With this perturbative method one cannot, however, access non-perturbative processes as the ones leading to the thermal 
activation over barriers discussed below.

\textcolor{red}{
\subsubsection{Thermally activated processes}
\label{subsubsec:Arrhenius}
}
 
The phenomenological \textcolor{blue}{ Arrhenius
  law}~\cite{Arrhenius}
  yields the typical time needed
to escape from a potential well as an exponential of the ratio between
the height of the barrier and the thermal energy scale $k_BT$, (with
prefactors that can be calculated explicitly, see below). This
exponential is of crucial importance for understanding slow (glassy)
phenomena, since a mere barrier of $30 \, k_B T$ is enough to transform a
microscopic time of $10^{-12}$ s into a macroscopic time scale. See
Fig.~\ref{fig:kolton-coulomb-arrhenius}-right for a numerical study of
the Coulomb glass that demonstrates the existence of an Arrhenius
time-scale in this problem. In the glassy literature such systems are
called \textcolor{blue}{ strong} glass formers as opposed to
\textcolor{blue}{ weak} ones in which the characteristic time-scale
depends on temperature in a different way.

In 1940 Kramers estimated the \textcolor{blue}{  escape rate} 
from a potential well
as the one shown in Fig.~\ref{fig:kolton-coulomb-arrhenius}-center 
due to thermal
fluctuations that give sufficient energy to the particle to allow it
to surpass the barrier~\cite{Kramers}.  After this seminal paper this problem has
been studied in great detail given that it is of
paramount importance in many areas of physics and chemistry~\cite{Hanggi-arrhenius}. An
example is the problem of the dissociation of a molecule where $x$
represents an effective one-dimensional \textcolor{blue}{  reaction coordinate} and
the potential energy barrier is, actually, a \textcolor{blue}{  free-energy barrier}.

Kramers assumed that the reaction coordinate is coupled to an
equilibrated environment with no memory and used the probability
formalism in which the particle motion is described in terms of the
time-dependent probability density $P(x,v,t)$ (that for such a stochastic process
follows the Kramers partial differential equation).

If the thermal energy is at least of the order of the barrier height, 
$k_B T \sim \Delta V$, the reaction coordinate, $x$,  moves freely from the 
vicinity of one well to the vicinity of the other. 

The treatment we discuss applies to the opposite weak noise limit in
which the thermal energy is much smaller than the barrier height, $k_B
T \ll \Delta V$, the random force acts as a small
perturbation, and the particle current over the top of the 
barrier is very small. Most of the time $x$ relaxes towards
the minimum of the potential well where it is located. Eventually, the
random force drives it over the barrier and it escapes to
infinity if the potential has the form in
Fig.~\ref{fig:kolton-coulomb-arrhenius}-center, or it remains in the
neighbourhood of the second well, see
Fig.~\ref{fig:kolton-coulomb-arrhenius}-left.

\begin{figure}[h]
\centerline{
\includegraphics[scale=0.2]{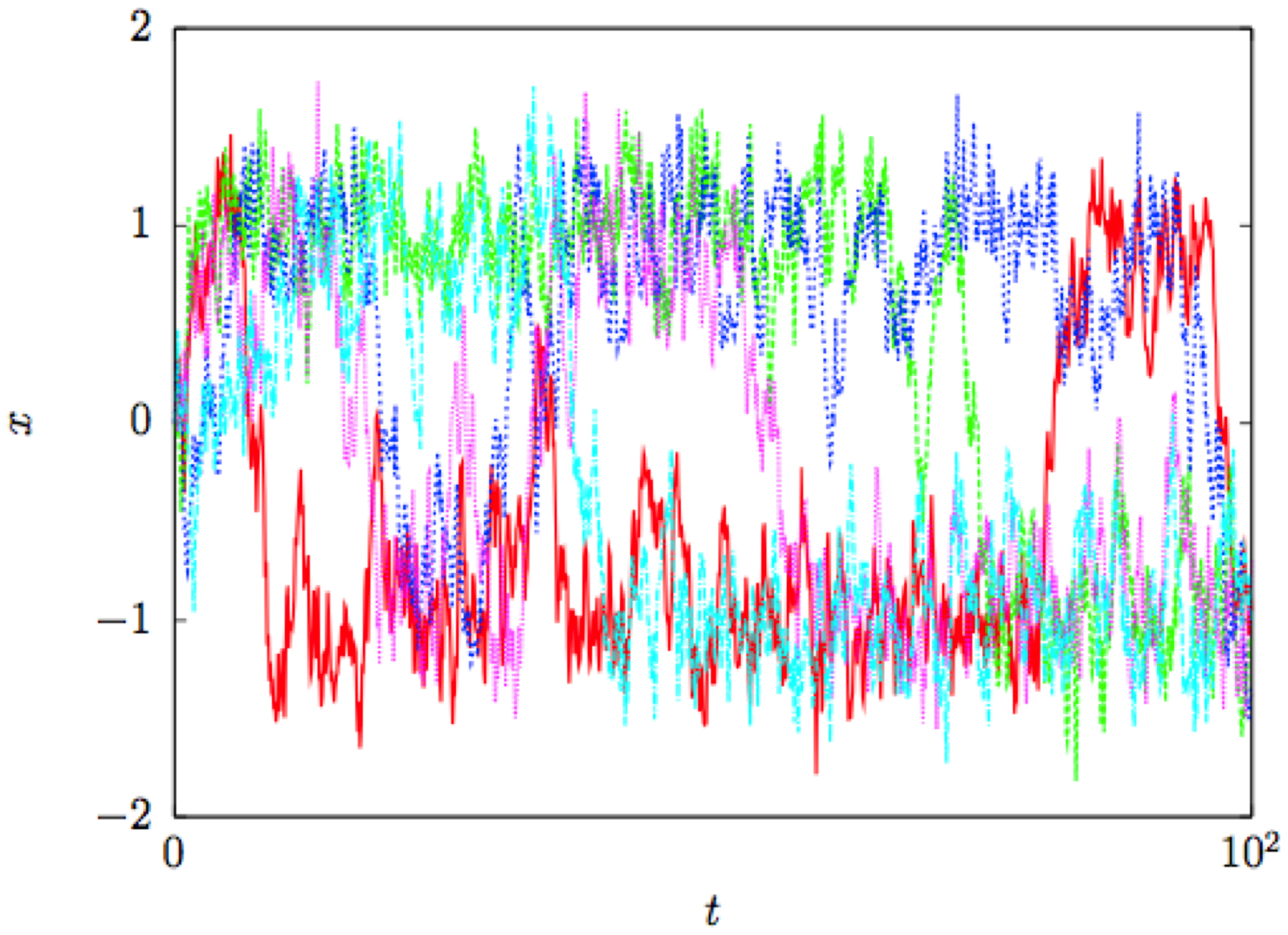}
\hspace{-0.8cm}
\includegraphics[scale=0.2]{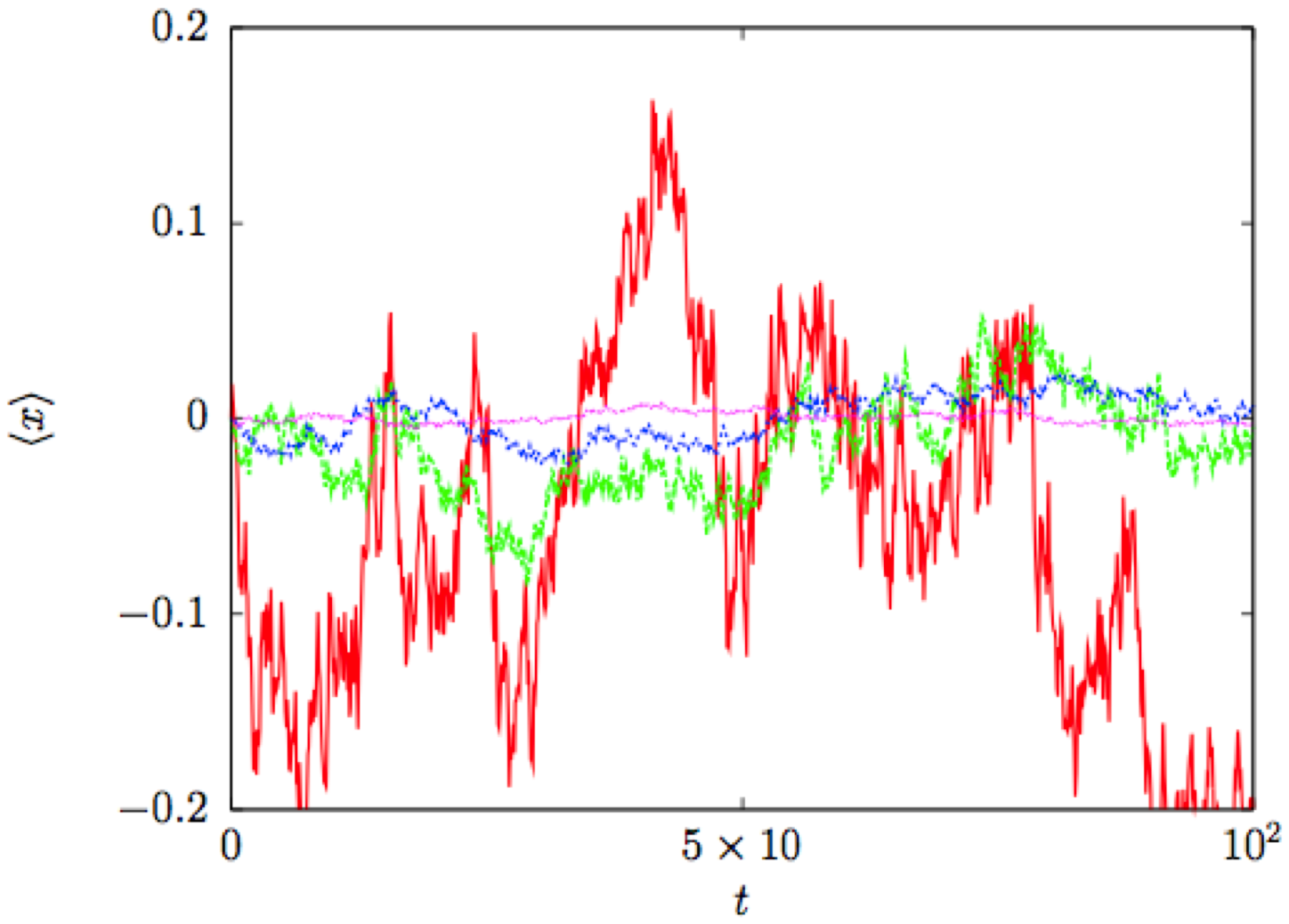}
\hspace{-0.8cm}
\includegraphics[scale=0.2]{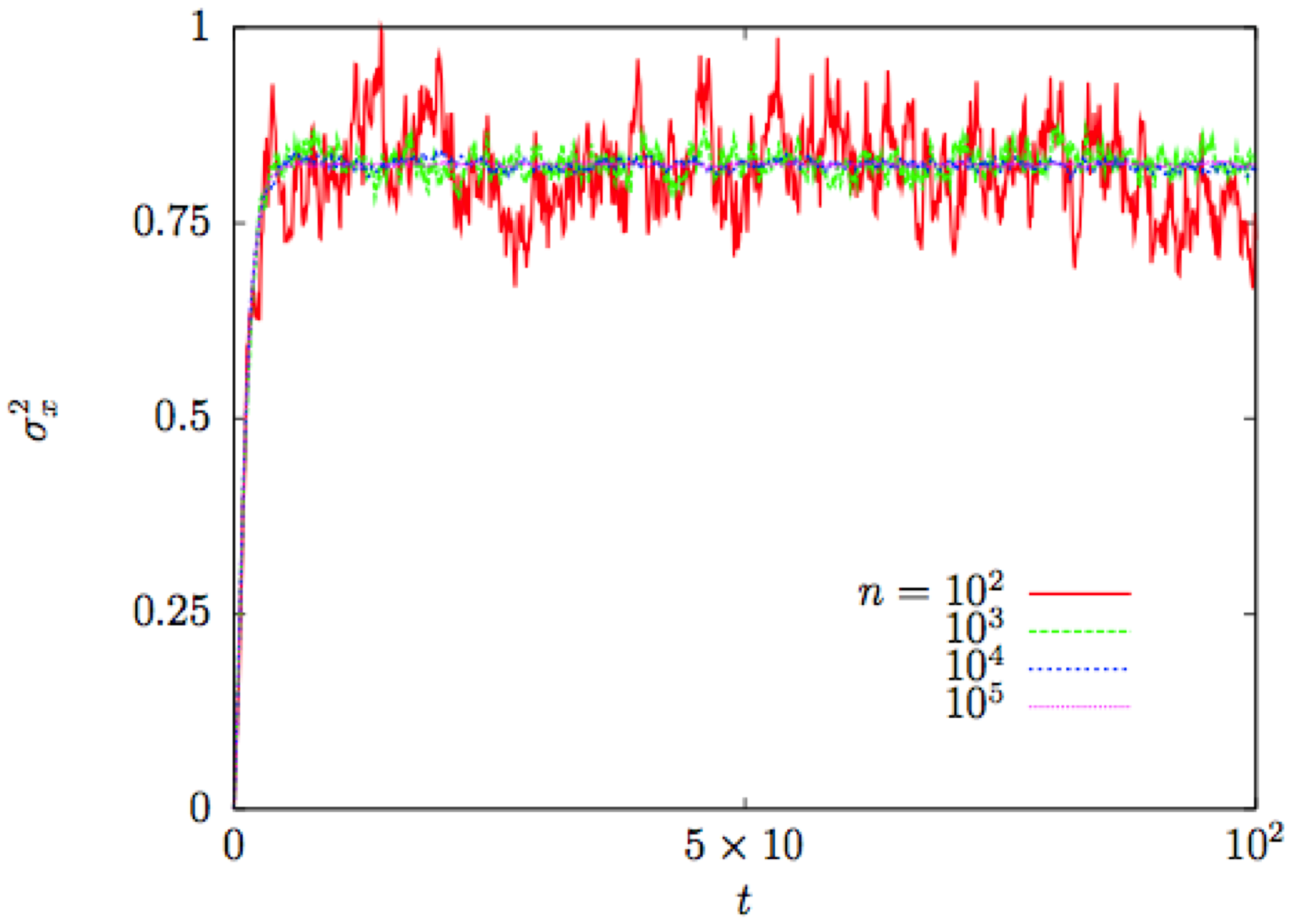}
}
\caption{\small Left panel: five runs of the Langevin equation in the over-damped limit with a double well external potential (oscillator)
and a Gaussian white noise
at temperature $T$. Central panel: the average $\langle x\rangle$ computed with 
$n=10^2, \ 10^3, \ 10^4, \ 10^5$ runs. Right panel: the variance $\sigma^2_x=\langle x^2 \rangle - \langle x\rangle^2$.
}
\end{figure}

The treatment is simplified if a constant current can be imposed by injecting particles within the
metastable well and removing them somewhere to the right of it.  In
these conditions Kramers proposed a very crude approximation whereby
$P$ takes the stationary canonical form
\begin{equation}
P_{\rm st}(x,v) = {\mathcal N} e^{-\beta \frac{v^2}{2} -\beta V(x)}
\; .
\label{eq:Pstat-kramers}
\end{equation}  
($m=1$ for simplicity here.)
If there is a sink to the right of the maximum, the 
normalization constant ${\mathcal N}$ is fixed by further 
assuming that $P_{\rm st}(x,v) \sim 0$ for $x\geq \tilde x > x_{\rm max}$.
The resulting integral over the coordinate
can be computed with a saddle-point approximation
justified in the large $\beta$ limit.
After expanding the 
potential about the minimum and keeping the quadratic 
fluctuations one finds
\begin{displaymath}
{\mathcal N}^{-1} = \frac{2\pi}{\beta \sqrt{V''(x_{\rm min})}} \; 
e^{-\beta V(x_{\rm min})}                                    
\; .
\end{displaymath}

The escape rate, $r$, over the top of the barrier can now be readily
computed by calculating the outward flow across the top of the barrier:
\begin{eqnarray}
\fbox{
$r \equiv \dfrac1{t_A} \equiv \bigintsss_{\; 0}^{\, \infty} dv \; v P(x_{\rm max},v) 
=
\dfrac{\sqrt{V''(x_{\rm min})}}{2\pi} \; 
e^{-\beta (V(x_{\rm max})-V(x_{\rm min}))}
$}
\;\;\;\;\;\;
\label{eq:escape-rate1}
\end{eqnarray}
Note that we here assumed that no particle comes back 
from the right of the barrier. This assumption is justified 
if the potential quickly decreases on the right side of the barrier. 

The crudeness of the approximation (\ref{eq:Pstat-kramers}) 
can be grasped by noting that the equilibrium form is justified only 
near the bottom of the well. Kramers estimated an 
improved $P_{\rm st}(x,v)$ that leads to 
\begin{equation}
r = \frac{\left(\frac{\gamma^2}{4}+V''(x_{\rm max})\right)^{1/2} -
\frac{\gamma}{2}}
{\sqrt{V''(x_{\rm max})}} \; 
\frac{\sqrt{V''(x_{\rm min})}}{2\pi} \; 
e^{-\beta (V(x_{\rm max})-V(x_{\rm min}))}
\; .
\label{eq:escape-rate2} 
\end{equation}
This expression approaches (\ref{eq:escape-rate1}) when $\gamma \ll
V''(x_{\rm max})$, {\it i.e.} close to the under-damped limit, and 
\begin{equation}
r = \frac{\sqrt{V''(x_{\rm max}) V''(x_{\rm min})}}{2\pi\gamma}
 \, 
e^{-\beta (V(x_{\rm max})-V(x_{\rm min}))}
\label{eq:escape-rate3} 
\end{equation}
when $\gamma \gg V''(x_{\rm max})$, {\it i.e.} in the over-damped limit (see Sect.~\ref{sec:smoluchowski} for the 
definition of these limits).

The inverse of (\ref{eq:escape-rate2}), $t_A$, is called the \textcolor{blue}{ 
  Arrhenius time} needed for \textcolor{blue}{  thermal activation} over a barrier
$\Delta V\equiv V(x_{\rm max})-V(x_{\rm min})$. The prefactor that
characterises the well and barrier in the harmonic approximation is
the \textcolor{blue}{  attempt frequency} with which the particles tend to jump over
the barrier. In short,
\begin{equation}
\fbox{$
t_A \simeq {\tau} \; e^{\beta|\Delta V|}
\;\;\;\; 
$ (Arrhenius time)}
\end{equation}

The one-dimensional reaction coordinate can be more or less easily
identified in problems such as the dissociation of a molecule. In
contrast, such a single variable is much harder to visualize in an
interacting problem with many degrees of freedom.  The Kramers problem
in higher dimensions is highly non-trivial and, in the
infinite-dimensional \textcolor{blue}{  phase-space}, is completely out of reach.

The Arrhenius time can be derived within the path-integral formalism~\cite{dilute-gas-instanton}. 

\begin{figure}
\begin{center}
\includegraphics[scale=0.8]{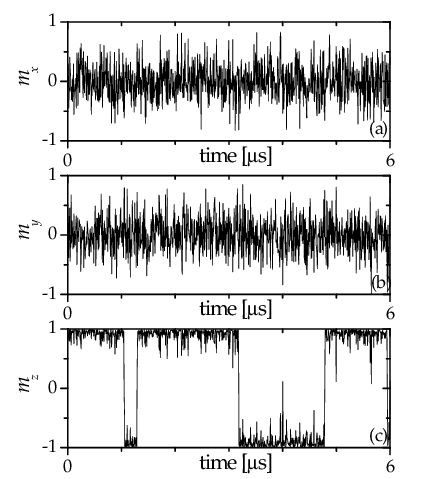}
\caption{\small Magnetization reversal (an activated process) in the LLGB equation 
(picture taken from~\cite{Roma}.}
\end{center}
\end{figure}

\textcolor{red}{
\subsubsection{Driven systems}
}

In the introduction we mentioned that systems can be externally maintained out of equilibrium 
We list here two solvable examples, in the form of exercises, that illustrate this point.

\vspace{0.2cm}

\noindent
\textcolor{orange}{\bf Exercise \thesection.\theexercise}
Study the Langevin equation for a single 
particle moving in $d=1$ under no external potential, in a case in which the 
friction kernel is $\gamma_1(t-t')$ and the noise-noise correlation in $\Gamma_2(t-t')$. 

\addtocounter{exercise}{1}

\vspace{0.2cm}

\noindent
\textcolor{orange}{\bf Exercise \thesection.\theexercise}
Take a harmonic oscillator, in its over-damped limit to make the calculations simpler, and 
couple it to two external reservoirs, at different temperatures, $T_1$ and $T_2$, and 
different memory kernels, for instance, a delta function (white noise) and an exponential 
decay (Ornstein-Uhlenbeck process).  The Langevin relaxation of the particle can be solved
exactly and it is quite interesting. The particle inherits the two time-scales ($\tau_1 \to 0$ and $\tau_2$ finite)
from the baths as can be seen from the decay, in two steps, of the position correlation function or
linear response. The temperatures of the environments appear in the fluctuation dissipation 
relation between these two functions in the corresponding time regimes~\cite{Cuku-Japan,Ilg-Barrat}.
Moreover, these temperatures appear also in the fluctuation theorem~\cite{Zamponi}.

\addtocounter{exercise}{1}

\vspace{0.2cm}

\noindent
\textcolor{orange}{\bf Exercise \thesection.\theexercise}
Take now a symmetric two-dimensional harmonic oscillator $V(x, y) = k(x^2 + y^2)/2$ 
and apply the non-potential force $\vec f(x,y) = a (y,-x)$ on it, with $a$ a parameter. This force makes a particle turn within the potential 
well. Describe the trajectories and compute mean-square displacement, correlation function and linear response.
One can check, by direct calculation, that the fluctuation-dissipation theorem does not hold.

\addtocounter{exercise}{1}

\textcolor{red}{
\section{The Kramers/Fokker-Planck approach}
}
\setcounter{equation}{0}
\setcounter{exercise}{1}

The Kramers  
or Fokker-Planck 
approach is useful to prove that a given Langevin equation with \textcolor{blue}{white noise} takes the system to 
equilibrium at the working temperature. It is a deterministic partial differential equation on the probability distribution 
for the stochastic variable at time $t$ to take a given value, say $y$, that can be closed as 
such for problems with white (additive or multiplicative) noise. The stochastic variables can be both velocity and position
and then one speaks about the \textcolor{blue}{Kramers equation}
or just the position variables, in the Smoluchowski $t\gg t_I$ limit
and one speaks about the \textcolor{blue}{Fokker-Planck equation}. here, for the sake of simplicity, we focus on the latter
case.

\textcolor{red}{
\subsection{First derivation for the Smoluchowski case with white noise}
}

Consider the probability density of finding the particle close to $y$ at time $t$ and call it 
$P(y,t+\Delta t) $. We start from the identity for \textcolor{blue}{Markov processes},
\begin{equation}
P(y,t+\Delta t) = \int dx_0 \ P(y, t+\Delta t|x_0, t) \ P(x_0,t)
\; , 
\label{eq:recursion-P}
\end{equation}
where  $P(y, t+\Delta t|x_0, t)$ is the \textcolor{blue}{conditional probability} of finding $y$ at
time $t+\Delta t$ provided the system was in the state $x_0$ at the previous time $t$
(note that $x_0$ is not necessarily the initial value here). The probability density of this 
last event is $P(x_0, t)$. The integral
runs over all accessible values of $x_0$. This equation holds for
any value of the time increment $\Delta t$ but we will later focus on infinitesimal ones.
It is also called the \textcolor{blue}{Chapman-Kolmogorov} equation.

To make contact with the stochastic process in the Langevin description,
it is convenient to define the conditional probability in the following
way:
\begin{equation}
P(y, t+\Delta t|x_0, t)
=
\langle \delta(y-x(t+\Delta t)) \rangle
\label{eq:def-Px}
\end{equation}
where the mean value is taken over the noise $\xi$ weighted with its 
probability distribution $P[\xi]$, and
$x(t+\Delta t)$  is determined
by the Langevin  equation with the `initial condition' 
$x(t)=x_0$. Expanding Eq.~(\ref{eq:def-Px})
 in powers of $\Delta x \equiv  x(t+\Delta t) - x(t)=y-x_0$
we immediately obtain
 \begin{equation}
P(y, t+\Delta t|x_0, t)
=
\delta(y-x_0) - d_y \delta(y-x_0) \langle \Delta x\rangle
+
\frac{1}{2} d^2_y \delta(y-x_0) \langle (\Delta x)^2\rangle + \dots
\label{eq:cond-prob-x}
\end{equation}
where the ellipsis indicate terms involving higher order
moments of $\Delta x$.
The idea is to compute the averages  $\langle \Delta x\rangle$ and 
$\langle (\Delta x)^2\rangle$ to 
leading order in $\Delta t$ and then take the limit
$\Delta t\to 0$. To do this, we need to use the Langevin equation of motion 
and it is at this point that its form (additive or multiplicative noise) will 
play a role.

We recall the discretization of the stochastic different equations with white noise discussed 
in Sec.~\ref{subsubsec:discretization}:
\begin{equation}
\gamma_0 (x_{n+1} - x_n ) = f(\overline x_n) \Delta t + g(\overline x_n) \xi_n \Delta t
\end{equation}
where we reintroduced the friction coefficient $\gamma_0$. The overline variables are defined as
$\overline x_n = \alpha x_{n+1} + (1-\alpha) x_n$, and $\langle \xi_n \Delta t \rangle =0$
and $\langle (\xi_n \Delta t)^2 \rangle = 2k_BT\gamma_0  \Delta t$. 

\textcolor{red}{
\subsubsection{Additive white noise}
}

We will here present an evaluation of $\Delta x$ obtained from the integration of the 
Langevin equation over the interval $[t,t+\Delta t]$. It reads
\begin{equation}
\Delta x \equiv x(t+\Delta t) - x(t)  = x_{n+1} - x_n  = 
- \frac{1}{\gamma_0} V'(\overline x_n) + \frac{1}{\gamma_0}   \xi_n \Delta t
\end{equation}
and $x_n$ being set to the `reference value'  $x_0$.
From here the averages are readily computed:
\begin{eqnarray}
\langle \Delta x\rangle &=& -\frac{\Delta t}{\gamma_0} V'(x_0)
\\
\langle (\Delta x)^2 \rangle  &=& \frac{2k_BT\Delta t}{\gamma_0} + O(\Delta t^2)
\end{eqnarray}
where, in the second line, we identified the contribution from the deterministic force 
as being $O(\Delta t^2)$ and we used the fact that $x(t)=x_n $ will be fixed to $x_0$ to set to zero the contribution from the 
cross product. Interestingly enough, the mean value as well as the second moment  are
of order  $\Delta t$. Higher momenta of the distribution such as
$\langle (\Delta x)^3 \rangle$ and so on and so forth are of higher order in
$\Delta t$ and do not contribute to the expansion
for sufficiently small $\Delta t$. It is important to note that these results depend
on $x_0$. Replacing now  these averages in (\ref{eq:cond-prob-x}), next in 
(\ref{eq:recursion-P}), 
\begin{eqnarray}
P(y,t+\Delta t) &= & P(y,t) + \frac{\Delta t}{\gamma_0} \partial_y \int dx_0 \ V'(x_0) \ \delta(y-x_0) \ P(x_0,t)
\nonumber\\
&&
+ \frac{2k_BT\Delta t}{2\gamma_0} \partial^2_y \int dx_0 \  \delta(y-x_0) \ P(x_0,t)
\; , 
\end{eqnarray}
performing the integrals over $x_0$, and taking the $\Delta t\to 0$ limit
\begin{equation}
\gamma_0 \partial_t P(y,t) =   \partial_y [ V'(y) \ P(y,t) ]
+ k_BT \partial^2_y  P(y,t)
\label{eq:FP-additive}
\end{equation}
This is the Fokker-Planck (or Smoluchowski) equation for a one variable Langevin process with 
white additive noise. 
The limit $\gamma_0$ can then be safely taken to recover the dissipation-less limit.

\pagebreak

\noindent
\textcolor{red}
{\it Stationary solution}

\vspace{0.2cm}

We look now for a solution that is time-independent, $P_{\rm st}(y)$, and normalizable. We have 
\begin{equation}
0 = \partial_y [ V'(y) \ P_{\rm st}(y) ]
+ k_BT \partial^2_y  P_{\rm st}(y)
\; . 
\end{equation}
A first integration over $y$  implies
\begin{equation}
\mbox{cst} =   V'(y) \ P_{\rm st}(y) + k_BT  \ \partial_y  P_{\rm st}(y)
\; . 
\end{equation}
To ensure the normalization of the pdf it is natural to impose $\lim_{y\to\infty} P_{\rm st}(y) = 0$ and  
$\lim_{y\to\infty}\partial_y P_{\rm st}(y) = 0$. Therefore, the constant must vanish and we find
\begin{equation}
\frac{\partial_y P_{\rm st}(y)}{P_{\rm st}(y)} = - \frac{ V'(y)}{k_BT} \qquad\Rightarrow \qquad 
P_{\rm st}(y) \propto e^{- V(y)/(k_BT)} = e^{-\beta V(y)}
\end{equation}

\vspace{0.2cm}

\noindent
\textcolor{red}
{\it Approach to the stationary solution}

\vspace{0.2cm}

The question remains as to whether the dynamics of the system takes it to this 
stationary solution asymptotically or not. An elegant way to prove this fact is to 
consider the `dynamic free-energy functional'
\begin{equation}
{\cal F}[P] = \int dy \ P(y,t) \ [ k_BT \ln P(y,t) + V(y)] 
\end{equation}
where $P$ is a generic solution of the Fokker-Planck equation. The time derivative 
of ${\cal F}$ reads
\begin{equation}
d_t {\cal F}[P] = \int dy \ \partial_t P(y,t) \ \left[ 
k_BT \ln P(y,t) + V(y) + k_BT
\right]
\end{equation}
Using now the FP equation to replace $\partial_t P(y,t)$
\begin{eqnarray}
\gamma_0 d_t {\cal F}[P] &=& 
\int dy \ 
\left\{
\partial_y [ V'(y) P(y,t) ] + k_BT  \partial^2_y P(y,t)
\right\}
\nonumber\\
&&
\qquad\quad\quad \times \left[ 
k_BT \ln P(y,t) + V(y) + k_BT
\right]
\end{eqnarray}
We now integrate by parts and drop the border terms as $P$ and $\partial_y P$ are 
expected to vanish at infinity to obtain
\begin{eqnarray}
\gamma_0 d_t {\cal F}[P] &=& 
- 
\int dy \ 
\left[
  V'(y) P(y,t) + k_BT \partial_y P(y,t)
\right]
\nonumber\\
&&
\qquad\qquad
\times
\partial_y \left[ 
k_BT \ln P(y,t) + V(y) +k_BT
\right]
\nonumber\\
&=& 
-
\int dy \ 
\left[
  V'(y) P(y,t) + k_BT \partial_y P(y,t)
\right]^2 \frac{1}{P(y,t)} \leq 0
\nonumber
\end{eqnarray}
Moreover, one sees that the numerator in the integrand vanishes identically 
only for $P_{\rm eq} \propto e^{-\beta V}$. For the Boltzmann equilibrium distribution function, therefore, 
$d_t {\cal F}[P_{\rm eq}] =0$. 

As $V$ is bounded from below for a potential that may lead to equilibrium, ${\cal F}$ is also bounded from below.
In the course of time, for any $P \neq P_{\rm eq}$,  its derivative is always negative. Therefore, 
${\cal F}$  has to approach its asymptotic value where $d_t {\cal F}$ must 
vanish. As we also showed that $d_t {\cal F}[P_{\rm eq}]=0$ then 
\begin{equation}
\lim_{t\to\infty} P(y,t) = P_{\rm eq}(y)
\; . 
\end{equation}

\vspace{0.2cm}
\noindent
\textcolor{red}{\it Connection to the Schr\"odinger equation}
\vspace{0.2cm}

The FP equation looks very similar to the Schr\"odinger equation for imaginary time, apart from a term 
proportional to $V'(y) \partial_y P(y,t)$. One can, however, eliminate it by introducing the function 
\begin{equation}
P(y,t) = \psi_0(y) \rho(y,t)
\qquad
\mbox{with}
\qquad \psi_0(y) = \mbox{ct} \ e^{-\frac{\beta}{2} V(y)}
\end{equation}
with $\beta = (k_BT)^{-1}$. 
After a simple calculation one finds
\begin{equation}
\gamma_0 \partial_t \rho(y,t) = \left[ k_BT \partial^2_y - U_{\rm FP}(y) \right] \rho(y,t)
\end{equation}
with 
\begin{equation}
U_{\rm FP}(y) = -\frac{1}{2} V''(y) +\frac{\beta}{4} (V'(y))^2
\end{equation}
where ${\rm FP}$ stands for Fokker-Planck. This is a Schr\"odinger equation in imaginary time, with 
the linear Schr\"odinger operator 
\begin{equation}
H_{\rm FP}(y) = k_BT \partial^2_y - U_{\rm FP}(y)
\end{equation}
that is a symmetric operator on the space of real functions
($\int dx \ (H_{\rm FP} \Phi_1(x) ) \Phi_2(x) = \int dx \  \Phi_1(x) (H_{\rm FP} \Phi_2(x)) $).
A number of properties follow:

\vspace{0.2cm}

\begin{packed_enum}
\item[--] The eigenvalues of $H_{\rm FP}$ are real.

\item[--]
If $U_{\rm FP}$ grows rapidly to infinity for $y \to \pm\infty$ the spectrum of $H_{\rm FP}$ is discrete.

\item[--]
It is easy to check that $\psi_0(y)$ is an eigenvector of $H_{\rm FP}$ with zero eigenvalue, $H_{\rm FP}(y) \psi_0(y) = E_0 \psi_0(y) = 0$, 
implying $E_0=0$.

\item[--]
$\psi_0(y)$ is non-negative (cst is taken to be positive). Hence, it must be the ground state of $H_{\rm FP}$. All other 
eigenvalues $E_n$ are strictly positive, $E_n >0$ for $n>0$. 

\item[--] 
The eigenvectors of $H_{\rm FP}$ associated to different eigenvalues are orthogonal.

\item[--]
The solution is
\begin{equation}
\rho(y,t) = \sum_{n=0}^\infty c_n \psi_n(y) e^{-E_nt}
\end{equation}
(where we absorbed the $\gamma_0$ in a redefinition of time, for simplicity)
with $H_{\rm FP} \psi_n(y) = E_n \psi_n(y)$ and $c_n = \int dy \ \psi_n(y) \rho(y, 0)$. 

\item[--]
When $t\to\infty$ all terms vanish exponentially apart from the one associated to $n=0$. Thus, 
\begin{equation}
\lim_{t\to\infty} \rho(y,t) = c_0 \psi_0(y) = \psi_0(y)
\end{equation}
since $c_0 = \int dy \ \psi_0(y) \rho(y,0) = \int \ dy \ P(y,0) =1$. 

\item[--]
The property above implies 
\begin{equation}
\lim_{t\to\infty} P(y,t) = \psi_0^2(y) = \mbox{cst}^2 e^{-\beta V(y)}
\end{equation}

\item[--]
One can easily show that the probability is normalized at all times
\begin{eqnarray}
&& \int dy \ 
P(y,t)= 
\int dy \ \psi_0(y) \rho(y,t) = \int dy \ \psi_0(y) \sum_n c_n \psi_n(y) e^{-E_nt}
\nonumber\\
&& \qquad
=
\sum_n c_n e^{-E_n t} \int dy \ \psi_0(y) \psi_n(y) = 
\sum_n c_n e^{-E_n t}\delta_{n0}  = c_0 =1
\end{eqnarray}

\item[--]
Finally, 
\begin{equation}
\lim_{t\to\infty} P(y,t) = 
\frac{
e^{-\beta V(y)}}{\int dx \ e^{-\beta V(x)}}
\end{equation}
and this is another way of proving the approach to Boltzmann equilibrium.

\end{packed_enum}

\vspace{0.2cm}
\noindent
\textcolor{red}{\it Relaxation time}
\vspace{0.2cm}

The longest relaxation time is then the inverse of the energy of the first excited state
\begin{equation}
\tau_{\rm eq} = E_1^{-1}
\; . 
\end{equation}
This time can, however, diverge. In particular, if it scales with the size of the system.

\textcolor{red}{
\subsubsection{Multiplicative white noise}
}
\label{subsubsec:FP-multiplicative}

In this calculation we will be more careful with the discrete time analysis. We rely heavily on the 
fact that $\langle \xi_n \xi_m \rangle = 2D/\Delta t \ \delta_{nm}$ implies 
$\xi_n \simeq {\cal O}(\Delta t^{-1/2})$ and $ \xi_n \Delta t \simeq {\cal O}(\Delta t^{1/2})$. 
We work with the generic equation $\gamma_0 d_t x = f(x) + g(x) \xi$.

As discussed in Sec.~\ref{subsubsec:discretization} the discretized equation reads
\begin{equation}
\gamma_0 \Delta x \equiv \gamma_0( x_{n+1} - x_n)  =  f(x_n)  \Delta t +  g(x_n)  \xi_n \Delta t + g'(x_n) \alpha \Delta x  \xi_n \Delta t 
\; . 
\end{equation}
We replace $\Delta x$ in the last term by this very same equation to get 
\begin{eqnarray}
\gamma_0 \Delta x &=&  f(x_n)  \Delta t +  g(x_n)  \xi_n \Delta t 
\nonumber\\
&&
+ g'(x_n) \alpha   \xi_n \Delta t \gamma_0^{-1}
[
 f(x_n)  \Delta t +  g(x_n)  \xi_n \Delta t  + g'(x_n) \alpha  \Delta x  \xi_n \Delta t
]
\; . 
\end{eqnarray}
Keeping now all terms that will contribute to the average up to ${\cal O}(\Delta t)$
\begin{eqnarray}
\gamma_0 \Delta x =  f(x_n)  \Delta t +  g(x_n)  \xi_n \Delta t 
+ \alpha g(x_n) g'(x_n) \gamma_0^{-1}  ( \xi_n \Delta t )^2
\end{eqnarray}
If we fix $x_n$ to take the value $x(t)=x_0$ in the expansion for 
$P(y, t+\Delta t|x_0,t)$, $x_n$ is not correlated with the noise $\xi_n$.
Therefore, under  the noise average  the third term vanishes.
Using $\langle  (\xi_n \Delta t)^2\rangle = 2D\Delta t$, 
\begin{equation}
\gamma_0 \langle \Delta x\rangle = 
f(x_n) \Delta t + 2D \gamma_0^{-1} \alpha g(x_n) g'(x_n) \Delta t
\; . 
\end{equation}

Let us examine $(\Delta x)^2$. Keeping terms that will contribute to the 
average up to ${\cal O}(\Delta t)$ we have 
\begin{eqnarray}
&& \gamma_0^2  \langle \Delta x^2 \rangle \simeq \langle [g(x_n)   \xi_n \Delta t ]^2 \rangle 
= 2 k_BT\gamma_0 g^2(x_n) \ \Delta t
\end{eqnarray}

Once again, the mean value as well as the two point correlation are
of order  $\Delta t$. These results depend
on $x_n = x_0$. Replacing now in (\ref{eq:cond-prob-x}), next in 
(\ref{eq:recursion-P}), 
\begin{eqnarray}
&&
\gamma_0 P(y,t+\Delta t) =  \gamma_0 P(y,t) - \Delta t \ \partial_y \int dx_0 \ [f(x_0) + 2 k_BT \alpha g(x_0)  g'(x_0)] 
\ \delta(y-x_0) \  P(x_0,t)
\nonumber\\
&&
\qquad
+ k_BT\Delta t \partial^2_y \int dx_0 \  \delta(y-x_0) \ g^2(x_0) \ P(x_0,t)
\; , 
\end{eqnarray}
and performing the integrals over $x_0$, in the $\Delta t\to 0$ limit
\begin{equation}
\boxed{
\gamma_0 \partial_t P(y,t) =  - \partial_y \left\{ \left[f(y) + 2k_BT \alpha g(y) g'(y) \right]  \ P(y,t) \right\}
+ k_BT \partial^2_y  \Big[g^2(y) P(y,t) \Big] 
}
\label{eq:FP-multiplicative}
\end{equation}
This is the Fokker-Planck (or Smoluchowski) equation for the stochastic process $\gamma_0 d_t x = f(x) + g(x) \xi$
with white noise. For $g(x)=1$ we recover Eq.~(\ref{eq:FP-additive}) for additive noise.

\vspace{0.25cm}

\noindent
\textcolor{red}{
\it Stationary solution
}

\vspace{0.25cm}

The stationary solution to Eq.~(\ref{eq:FP-multiplicative}) with vanishing current, $J=0$, is
\begin{equation}
P_{\rm st}(x) = \frac{N}{g^2(x)} \exp \left[ \frac{1}{k_BT} \int_{x} dx' \ \frac{f(x')+2k_BT\alpha g(x) g'(x))}{g^2(x')} \right]
\label{eq:stat-sol-no-drift}
\end{equation}
with $N$ a normalization constant.
This stationary probability depends upon $\alpha$ and $g(x)$.
In order to get rid of this undesired feature, we chose to work with the drifted force
\begin{equation}
\fbox{
$f(x) = - g^2(x) V'(x) + 2k_BT (1-\alpha)  g(x) g'(x)$ }
\; . 
\label{eq:equil-choice-f}
\end{equation}
The associated FP equation reads 
\begin{eqnarray}
&&
\displaystyle{
\partial_t P(x,t)
} = 
\displaystyle{
- \partial_x[ - g^2(x) V'(x) + 2k_BT g(x) g'(x)) P(x,t)] 
}
\nonumber
\\
&& 
\qquad\qquad\qquad
\displaystyle{
+ k_BT \, \partial_x^2 [ g^2(x) P(x,t)]
}
\nonumber\\
&&
\qquad\qquad
= 
\displaystyle{
 \partial_x \{ [( g^2(x) V'(x) - 2k_BT g(x) g'(x)) P(x,t)] 
}
\nonumber
\\
&& 
\qquad\qquad\qquad
\displaystyle{
\;\;\;\; \;\;\;+ k_BT \, \partial_x [ g^2(x) P(x,t)]
}
\}
\nonumber\\
&& 
\qquad\qquad
=
\displaystyle{
 \partial_x \{ g^2(x) [ V'(x)  P(x,t) + k_BT \, \partial_x P(x,t) ]
\}
}
\; . 
\label{eq:FP-correct}
\end{eqnarray}
It  is still independent of $\alpha$ though it depends on $g(x)$. However, its asymptotic solution with vanishing current 
 does not and it reads
\begin{equation}
P_{\rm st}(x) = P_{\rm GB}(x) = N \exp \left[ - \beta \ V(x) \right]
\end{equation}
independently of $\alpha$ and $g$, the desired result. Note that the effect of the extra term is to correct the prefactor
in the measure, not what goes in the exponential, that would be the same
$-\beta V$ even without the additional $2k_BT gg'$ term in the drift.

Therefore, meaning physical applications in the sense that the stochastic dynamics 
tends to equilibrium at the Boltzmann measure, need the drifted Langevin equation
\begin{equation}
\boxed{ d_t x(t) = -g^2 V'(x) + 2k_BT(1-\alpha) g(x) g'(x) +
g(x) \xi(t)  
 }
 \label{eq:x-eom-drifted}
\end{equation}
Note that with this force, there is a drift in the Langevin equation 
even in the Stratonovich convention. The extra term {\it is not} the one needed to build the generalized derivative appearing in the 
chain rule~(\ref{eq:x-chain0}), since the factor $2k_BT(1-\alpha)$ in the drift is different from the factor 
$k_BT(1-2\alpha)$ in the chain-rule
The Fokker-Planck equation takes a simple form given in Eq.~(\ref{eq:FP-correct}). 

In Sec.~\ref{subsubsec:harmonic-multiplicative} we saw an explicit realization of this phenomenon, with the study of a 
particle in harmonic potential under a white noise multiplied by the particle's position.

The Langevin equation (\ref{eq:x-eom-drifted}) is equivalent to 
\begin{equation}
\fbox{$g^{-2}(x) \gamma_0 d_t x(t) = -  V'(x) + 2k_BT  (1-\alpha) g'(x)  g^{-1}(x) + g^{-1}(x) \xi(x) $ }
\label{eq:eq-a-la-Aron-Biroli-LFC}
\end{equation}
in the sense that the 
term responsible for dissipation (lhs) is proportional to $g^{-2}$ while the noise is 
accompanied by just one factor $g^{-1}$ and there is no $g$ factor in the 
deterministic force along the gradient descent direction.  The second term in the rhs
is the drift. This form has the same structure as the form of the Langevin equations 
derived from the non-linear coupling of the particle to the oscillator coordinates in 
the bath modelling.

\textcolor{red}{
\subsection{Second derivation for the Smoluchowski case with white noise}
}

The following is an even simpler - and in my view a little bit sloppy - derivation of the Smoluchowski equation.
Choose a test function $g(y)$ of $y$, where $y$ is a value that the stochastic process ruled by the Langevin without the inertial term may take at time 
$t$. At  time $t$ the expected value  of $g(y)$ is 
\begin{equation}
\langle g(y) \rangle = \int dy \, P(y,t) \, g(y)
\label{eq:average-g}
\end{equation}
and we search the equation that determines the time evolution of $ P(y,t) $.

Evaluate $g$ at $x(t)$ as determined from the Langevin equation, 
and consider the noise averaged infinitesimal variation of this $g$ due to an infinitesimal variation of the stochastic 
trajectory $\Delta x$:
\begin{equation}
\langle \Delta g \rangle = \left\langle \frac{\partial g}{\partial x} \Delta x \right\rangle + 
\frac{1}{2} \left\langle  \frac{\partial^2 g}{\partial x^2} (\Delta x)^2 \right\rangle + 
\langle {\mathcal O}(\Delta x)^3) \rangle 
\end{equation}
We keep the order $(\Delta x)^2$ and drop the higher order terms.
The Langevin equation dictates
\begin{equation}
\gamma_0 \Delta x = f(x) \Delta t + \xi(t) \Delta t
\end{equation}
and then
\begin{eqnarray}
\langle \Delta g \rangle 
&=& 
\frac{1}{\gamma_0} \left\langle \frac{\partial g}{\partial x}  \left( f(x) \Delta t + \xi(t) \Delta t \right) \right\rangle + 
  \frac{1}{2\gamma_0^2}  \left\langle  \frac{\partial^2 g}{\partial x^2} \left( f(x) \Delta t + \xi(t) \Delta t\right)^2  \right\rangle  
\nonumber\\
&=&
\frac{1}{\gamma_0} \left\langle \frac{\partial g}{\partial x}  f(x)  \right\rangle \Delta t + 
\frac{k_BT}{\gamma_0}   \left\langle  \frac{\partial^2 g}{\partial x^2}    \right\rangle \Delta t
\label{eq:Delta-g}
\end{eqnarray}
where we used $\left\langle h(x)  \; \xi(t) \right\rangle = \langle h(x)\rangle \langle \xi(t)\rangle=  0$ and 
$\left\langle h(x)  \; \xi^2(t) \right\rangle = \left\langle h(x)  \rangle \langle \xi^2(t) \right\rangle $. (Ito convention), the relation $\langle \xi^2(t) \rangle \sim 2\gamma_0 k_BT  (\Delta t)^{-1}$, and we kept ${\mathcal O}(\Delta t)$ only in the right-hand-side.

Now, on the one hand, consider $\Delta g/\Delta t$ from (\ref{eq:Delta-g}) but applied to the non-stochastic
variable $y$, and average it over the noise with the 
expression Eq.~(\ref{eq:average-g}):
\begin{eqnarray}
\left\langle \frac{\Delta g}{\Delta t}\right\rangle 
&=&
\int dy \; P(y,t) \, \left[ \frac{1}{\gamma_0}  \frac{\partial g}{\partial y}  f(y) + 
\frac{k_BT}{\gamma_0}     \frac{\partial^2 g}{\partial y^2}     \right]
\nonumber\\
&=& 
\int dy \; \, \left\{ - \frac{1}{\gamma_0}  \,  \frac{\partial [ f(y)  P(y,t)] }{\partial y}+ 
\frac{k_BT}{\gamma_0}    \frac{\partial^2 P(y,t)}{\partial y^2}   \right\} g(y)
\end{eqnarray}
up tp ${\mathcal O}(\Delta t)$ corrections.
On the other hand, since the only time dependence in the right-hand-side of Eq.~(\ref{eq:average-g}) 
is in $P$, 
\begin{equation}
\lim_{\Delta t \to 0} \left\langle  \frac{\Delta g}{\Delta  t} \right\rangle 
=
\int dy \; \frac{\partial P(y,t)}{\partial t} \, g(y)
\end{equation}
Since these two expressions have to be identical for all $g(y)$, then 
\begin{equation}
\gamma_0 \frac{\partial P(y,t)}{\partial t} = 
-   \frac{\partial [ f(y)  P(y,t)] }{\partial y}+ 
k_BT  \, \frac{\partial^2 P(y,t)}{\partial y^2}  
\; , 
\end{equation}
the Smoluchowski equation.

\pagebreak

\textcolor{red}{
\subsection{Kramers equation for additive white noise}
}

In the previous section we worked in the Smoluchowski limit. A partial 
differential equation for $P( x,p,t)$ can also be derived. In its more usual form, 
derived for additive white noise, this is the Kramers equation
\begin{eqnarray}
&& 
\frac{\partial P(x, p; t)}{\partial t}
=
\frac{\partial }{\partial p}
\left\{ 
 \left[ 
- \frac{\partial V}{\partial x}  + \gamma_0  \frac{\partial H}{\partial p}
\right] P(x, p; t) 
\right\}
-
\frac{\partial }{\partial x} 
\left[ 
\frac{\partial H}{\partial p} P(x, p; t)
\right]
\nonumber\\
&&
\qquad\qquad\qquad
+ \, \gamma_0 k_BT \;   \frac{\partial^2 P(x,  p; t) }{\partial p^2} 
\; .
\end{eqnarray}
In the usual cases in which $H=K+V=p^2/(2m) + V$ and writing it in terms of the velocity instead of the 
momentum:
\begin{eqnarray}
&& 
\frac{\partial P(x, v; t)}{\partial t}
=
\frac{1}{m} \frac{\partial }{\partial v}
\left\{ 
 \left[ 
- \frac{\partial V}{\partial x}  + \gamma_0 v
\right] P(x, v; t) 
\right\}
-
\frac{\partial }{\partial x} 
\left[ 
v P(x, v; t)
\right]
\nonumber\\
&&
\qquad\qquad\qquad
+ \, \frac{\gamma_0 k_BT}{m^2}  \;   \frac{\partial^2 P(x,  v; t) }{\partial v^2} 
\; .
\end{eqnarray}
Proposing $P(x,v,t) = e^{-\beta mv^2/2} P_x(x,t)$ this equation reduces to Eq.~(\ref{eq:FP-additive}).

\vspace{0.25cm}

\noindent
\textcolor{orange}{\bf Exercise \thesection.\theexercise}
Derive the Kramers equation. Show that it admits the stationary solution 
$P_{\rm st} \propto \exp\{-\beta [p^2/(2m) + V(x)]\}$.

\addtocounter{exercise}{1}

\vspace{0.25cm}

\noindent
\textcolor{orange}{\bf Exercise \thesection.\theexercise}
Derive the corresponding Kramers equation for the multiplicative white noise Langevin process.

\vspace{0.25cm}

\textcolor{red}{
\subsection{Colored noise}
}

The coloured noise case is more tricky. As no Markov property can be used there is no closed partial differential 
equation for $P(u,y,t)$ (with inertia) nor $P(y,t)$ (dropping inertia). Some approximations exist~\cite{Fox}.

\textcolor{red}{
\subsection{Master equation}
}

For the moment we have only treated problems with continuous variables and in the rest of the notes we will stick to this kind
of problems as well. However, here, we wish to present the master equation that is, basically, the partner of the Fokker-Planck equation for 
discrete Markov stochastic processes.

The Chapman-Kolmogorox equation (\ref{eq:recursion-P}) applies also to problems with discrete variables with the simplification 
of replacing the integral by a sum over discrete states. The conditional or \textcolor{blue}{transition probability} over an infinitesimal
time interval $dt$  can be written as
\begin{eqnarray}
P(y, t+ \Delta t|x_0,t) = \underbrace{\left( 1-\Delta t \sum_{z} W_{zy} \right)}_{\rm probabilty \; of \; no \; transition} \delta_{zy} + \Delta t \, W_{yx_0}
\label{eq:W}
\nonumber
\end{eqnarray}
The possibility of there being more than one transition in the interval $dt$ has been neglected; a conjecture that is valid in the limit 
$dt\to 0$. The first term represents the probability for the system to stay in the state $x_0$ between $t$ and $dt$ while the second 
term is the probability that the system leaves the state $x_0$ to  go to $y$ in this same interval. Consistently, this 
expression satisfies $\sum_{y} P(y, t+ \Delta t|x_0,t) =1$.

Replacing now (\ref{eq:W}) in (\ref{eq:recursion-P}) and arranging terms in such a way to have a time-derivative in the left-hand-side, 
\begin{equation}
\frac{dP_x}{dt} = \sum_z (W_{xz} P_z - W_{zx} P_x) 
\end{equation}
The first term in the right-hand-side is a gain, in the sense that it increases the probability $P_x$ while the second term in the 
right-hand-side is a loss since it contains the contribution of all processes that take from $x$ to any $z$ and makes $P_x$ 
diminish. This is the \textcolor{blue}{master equation}.

\textcolor{red}{
\subsection{Concluding remarks}
}

{\rm The Langevin equation and its relation to the Fokker-Planck formalism have been described in many textbooks on stochastic processes 
including Risken's~\cite{Risken}, Gardiner's~\cite{Gardiner} and van Kampen's~\cite{vanKampen}. 
Many applications can be found in Coffrey et al.'s~\cite{Langevin-Coffey}. 
Another derivation of the Langevin equation uses collision theory and admits a generalization 
to relativistic cases~\cite{Hanggi}.
The alternative master equation description of stochastic processes, more adapted to deal with 
discrete variables, is also very powerful but we will not use it is these lectures.
}

\textcolor{red}{
\section{Functional formalism}
\label{subsec:functional}
}
\setcounter{equation}{0}
\setcounter{exercise}{1}

Until here we presented the Langevin equation and, in white noise cases, the corresponding Fokker-Planck equation. 
Yet, within the study of Markov but also non-Markov stochastic processes there exists also the
possibility of approaching the problem via the path integral formulation of generating functionals~\cite{books-PI}. 
The use of generating functionals is an elegant and powerful method to derive generic properties of dynamical systems. 
A path-integral is handy for computing moments, probability distribution functions, transition probabilities
and responses. It is also particularly well suited when it comes to perturbation
theory and renormalization group analysis, as one can easily set up a diagrammatic
expansion.

The classical path integral formalism for stochastic
processes governed by the Langevin equation goes under the name of 
Martin-Siggia-Rose~\cite{MSR} but it was first constructed by Janssen~\cite{Janssen76,Janssen92,cirano}.
Many subtleties have been identified in processes with multiplicative white 
noise~\cite{Edwards,Salomon,Langouche79,Langouche79b,Apfeldorf,Arnold2000,Aron16} 
and with the difficulty of performing 
non-linear transformation of variables~\cite{Salomon,Aron14,Lecomte17}. 

A covariant generating functional that allows to perform
non-linear changes of variables and use the conventional rules of calculus 
was constructed in the past~\cite{Graham} and more recently rederived  with an 
explicit higher order discretization scheme for one dimensional processes~\cite{Lecomte19}
and higher dimensional ones~\cite{Arnoulx}.

\textcolor{red}{
\section{Applications}
\label{subsec:applications}
}
\setcounter{equation}{0}
\setcounter{exercise}{1}

In this Section we briefly present a number of recent uses of Langevin equation
in different contexts, including biophysics and quantum matter.

\textcolor{red}{
\subsection{Single variable effective representation of complex systems}
}

The dynamics of interacting degrees of freedom in a complex system can be studied 
in terms of a {\it self-consistent one-body} problem which takes the form of a 
 single variable effective Langevin equation. This equation can be derived from the
 functional representation of the complete problem under some approximations
 that turn out to be mean-field like. They read
\begin{equation}
G_0^{-1}(t) x(t) + \int_0^{t} dt' \; \tilde\Sigma(t,t') x(t')
= \beta_0 \tilde D(t,0) x(0)  + h(t) + \zeta(t) + \xi(t)
\; . 
\label{eq:DMFT}
\end{equation}
The representative variable is $x$ and $G_0^{-1}$ is a differential 
operator. There are two 
uncorrelated noises: the original zero-average white noise $\xi$, and
a new effective noise $\zeta$, also with zero mean and 
\begin{equation}
\langle \zeta(t) \zeta(t') \rangle = \tilde D(t,t') 
\; . 
\label{eq:self-cons-noise}
\end{equation}
The vertex $\tilde D$ acts as the coloured noise correlation and the self-energy $\tilde \Sigma$ as the time-derivative of 
a retarded friction in the non-Markovian Langevin Equation~\ref{eq:DMFT}. 
with 
\begin{equation}
\tilde  C(t,t') = 
x(t) x(t') \; , 
\qquad\qquad\quad
\tilde R(t,t')  =  
\left. \frac{\delta x(t)}{\delta h(t')} \right|_{h=0}
\!\!\!\! 
\end{equation}
and they collect the effects of the other degrees of freedom on the selected one.  
Note that $\tilde C$ and $\tilde R$ are not averaged over the thermal noises but for 
cases in the thermodynamic limit $N\to\infty$ the fluctuations 
should be suppressed, and $\langle \tilde C\rangle  \to C$ and $\langle \tilde R \rangle \to R$, 
where the angular brackets represent an average over both $\xi$ and $\zeta$. 
The set of equations has to be solved self-consistently; the difficulty lies in imposing the 
condition~\ref{eq:self-cons-noise}. 
The single variable equation can also be derived with an extension of the static {\it cavity} method
to the time-dependent problem~\cite{Agoritsas18}.
This approach is similarity to the Dynamical Mean-Field Theory~\cite{Georges96} which has been 
so successfully applied to condensed matter systems, though mostly in equilibrium.

\textcolor{red}{
\subsection{Stochastic thermodynamics}
\label{sec:stochastic-thermodynamics}
}

Thermodynamics is a phenomenological theory of equilibrium macroscopic systems
which deals with heat, work, and temperature, and their relation to energy and entropy. 
It developed from the will to increase the efficiency of early steam engines in the 
XIXth and it was later justified microscopically by the equilibrium Statistical Physics
of Boltzmann and Gibbs.
Linear response theory allows to express transport properties caused by small external fields through equilibrium correlation functions. This applies to weak perturbations away from equilibrium only. 

General laws applicable to non-equilibrium systems were derived in the last 20 years or so. They concern
fluctuations and typically characterize the distribution functions of thermodynamic quantities like exchanged 
heat, applied work or entropy production.

These results are particularly relevant for small systems with appreciable (typically non-Gaussian) fluctuations, 
which are now accessible thanks to improvements in spatio-temporal resolution in nanotechnology. 
Stochastic energetics quantifies individual realizations of a stochastic process on the mesoscopic scale of thermal fluctuations.
They helped to better understand the non-equilibrium dynamics of microscopic systems such as colloidal particles, biopolymers, enzymes, and molecular motors.

Sekimoto introduced the idea of stochastic energetics~\cite{Sekimoto} 
and this was combined 
with the idea that entropy can consistently be assigned to a single fluctuating trajectory to build the 
stochastic thermodynamics~\cite{Sekimoto,Peliti} framework. 

Many exact relations for out of equilibrium processes have been proven. These include the
fluctuation theorem, the Jarzynski relation, the Crooks relation, etc. Possibly, the easiest 
derivations are those that concern Langevin processes. We very briefly introduce some of these.

\vspace{0.2cm}

\noindent
\textcolor{red}{\it Stochastic energetics.}

\vspace{0.2cm}

Let's multiply the generic Langevin equation by $v(t)$.\footnote{Subtle issues about the chain-rule for derivatives of functions of stochastic 
variables were discussed in the text. For equations with inertia and/or coloured noise, these problems do not 
usually pose, see however~\cite{Martin21,Ariel24}.} 
We find
\begin{eqnarray}
\frac{d}{dt} \left[\frac{1}{2} mv^2 + V(x) \right] = v(t) \xi(t) - v(t)  \int_0^t dt' \ \Gamma(t-t') v(t')
\; . 
\end{eqnarray}
Now integrate over time between $t_1$ and $t_2$
\begin{equation}
H_{\rm syst}(t_2) - H_{\rm syst}(t_1) = \int_{t_1}^{t_2} dt \ \xi(t) v(t) - \int_{t_1}^{t_2} dt \ \int_{0}^{t} dt' \ v(t) \Gamma(t-t') v(t') 
\; . 
\end{equation}
As there is no reason to suppose that the rhs be identical to zero, one finds that the energy of the system fluctuates
and is not constant. 
Note also that, although the velocity is a Gaussian random variable, the kinetic energy, being given by its square, is not.

\vspace{0.2cm}

\noindent
\textcolor{red}{\it Non potential forces.}

\vspace{0.2cm}

Time-dependent,
$f(t)$, and constant non-potential forces, $f^{\rm np}$ (in higher dimensions),
as the ones applied to granular matter and in rheological
measurements, respectively, are simply included in the right-hand-side
({\rm rhs}) as part of the deterministic force. 

The energy balance under non-potential forces can be done as above, by multiplying the 
Langevin equation by $v(t)$. One gets an additional term due to the work done by the non-potential 
force over the interval $[t_1,t_2]$:
\begin{eqnarray}
H_{\rm syst}(t_2) - H_{\rm syst}(t_1) 
&=& 
\int_{t_1}^{t_2} dt \ f(t) v(t) + 
\int_{t_1}^{t_2} dt \ \xi(t) v(t) 
\nonumber\\
&&
- \int_{t_1}^{t_2} dt \int_{0}^{t} dt' \ v(t) \Gamma(t-t') v(t') 
\end{eqnarray}
The first term on the rhs is the work done by the time-dependent force. The 
second and third terms can be associated to the heat given or taken from the bath, 
if a fluctuating energy balance relation
\begin{equation}
\Delta H_{\rm syst} = W^f_{t_1\to t_2} + \Delta Q
\label{eq:energy-balance}
\end{equation}
is proposed.  This is the \textcolor{blue}{first law of thermodynamics at the trajectory level}. 

The sign of the dissipated heat is opposite to the sign convention in classical
thermodynamics, where heat transfer from the system to the environment has a negative sign. Here,
\begin{eqnarray}
 \Delta Q = 
 \left\{ 
 \begin{array}{l}
 {\rm positive, \; heat \; goes \; from \; the \; system \; to \; the \; bath} 
 \\
 [5pt]
 {\rm negative,\;  heat \; goes \; from \; the \; bath \; to \; the \; system}
 \end{array}
 \right.
\end{eqnarray} 
Heat is viewed from the perspective of the heat bath, since the
amount of dissipated heat from the system leads to an increase in entropy of the environment. 
For the work the usual sign convention applies, i.e., work applied to the system has a positive sign and work done by
the system has a negative sign.

\comments{
Imagine that the force $f(t)$ derives from a potential which depends on a parameter $\lambda$, 
itself a function of time varying according to some protocol:
\begin{equation}
f(t) = \frac{\partial U(x, \lambda)}{\partial x} 
\; .
\end{equation}
The work is then
\begin{equation}
W^f_{t_1\to t_2} = \int_{t_1}^{t_2} dt \ f(t) v(t) = 
 \int_{t_1}^{t_2} dt \ \frac{\partial U(x, \lambda)}{\partial x} \frac{dx}{dt} 
\end{equation}
}

\vspace{0.25cm}
\noindent
\noindent
\textcolor{orange}{\bf Exercise \thesection.\theexercise}
Calculate all terms in Eq.~(\ref{eq:energy-balance}) for the Langevin dynamics in the 
Smoluchowski over-damped limit of a particle in a
one dimensional harmonic potential and driven by a constant force $F$. Take their averages and verify that the energy 
balance equation is satisfied. Check that in the long-time limit, the deterministic potential energy is increased by the thermal energy provided by the heat bath, whereas the pure mechanical dissipation is reduced correspondingly.

\addtocounter{exercise}{1}

\vspace{0.25cm}

Interest in computing the probability distribution functions of each of these terms
is current in the literature. This is part of the so-called \textcolor{blue}{stochastic thermodynamics}, or 
the idea to extend notions of thermodynamics such as work, heat and entropy, to individual trajectories~\cite{Sekimoto}.
These pdfs are of relevance in the study of small 
systems, especially biological molecules and the like, but also in microfluidics, nanomachines, nano sensing devices,
a many other fields. Exact relation for the probability 
of measuring a positive over the probability of measuring a negative quantity such as 
the work or heat, have been derived and are special cases of the so-called 
\textcolor{blue}{fluctuation theorems}.


\setcounter{equation}{0}
\textcolor{red}{
\subsection{Phenomenological Langevin equations}
\label{subsec:phenomenological-Langevin-eqs}
}

In so far we have discussed a system with position and momentum degrees
of freedom. 
Many fields in physics and other sciences use Langevin-like equations to 
describe the dynamic behavior of a selected set of variables, of different kind, in contact
with an environment. Sometimes, these equations look different from 
the one that we derived above. 

\vspace{0.25cm}

\noindent
\textcolor{red}{
{\it The electric analog.}
}

\vspace{0.25cm}

Take an LRC circuit. The resistance is of the usual Ohmic type, that is to say, the 
potential drop, $V_R$, across it is given by $V_R=IR$ with $I$ the current 
and $R$ the resistance. The potential drop, $V_L$, across the inductor $L$ is
given by $V_L=LdI/dt$. Finally, the potential drop across the capacitor is 
$V_C=-C^{-1} \int I dt$. The balance between these potentials implies 
a Langevin type equation for the current circulating across the circuit:
\begin{equation}
L \frac{d^2 I}{dt^2} + R \frac{dI}{dt} + C^{-1} I = 0 
\; . 
\end{equation}
This analogy justifies the Ohmic name given to a dissipative 
term proportional to the velocity in the general presentation.

\vspace{0.25cm}

\noindent
\textcolor{red}
{\it Classical Ising spins: the soft spin description}

\vspace{0.25cm}

A continuous
Langevin equation for classical spins can also be used if one replaces
the hard Ising constraint, $s_i=\pm 1$, by a soft one implemented with
a potential term of the form $V(s_i) = u (s_i^2-1)^2$ with $u$ a
coupling strength (that one eventually takes to infinity to recover a
hard constraint). The soft spins are continuous unbounded variables,
$s_i\in (-\infty,\infty)$, but the potential energy favors the
configurations with $s_i$ close to $\pm 1$. Even simpler models are
constructed with spherical spins, that are also continuous unbounded
variables globally constrained to satisfy $\sum_{i=1}^N s_i^2=N$.  The
extension to fields is straightforward and we will discuss one
when dealing with the $O(N)$ model.

\vspace{0.25cm}

\noindent
\textcolor{red}
{\it Classical Heisenberg spins}

\vspace{0.25cm}

Another example is the \textcolor{blue}{Landau-Lifshitz-Gilbert-Brown equation} for the stochastic dynamics 
of a magnetic moment with constant magnitude:
\begin{equation}
\dot {{\vec M}} =  
 - \frac{\zeta}{1+\zeta^2 \gamma_0^2}
 \
{\vec M} \wedge
\left( 
 {\vec{H}}_{\rm eff} + {\vec{H}}
 + \frac{\gamma_0 \zeta}{M_s} \ {{\vec M}} \wedge  ({\vec{H}}_{\rm eff} + {\vec{H}})
 \right)
 \; , 
\label{eq:landau}
\end{equation}
in the Landau formulation or
\begin{equation}
\dot {{\vec M}} =  - \zeta {{\vec M}} \wedge \left( {\vec H}_{{\rm eff}}+ {\vec H}  - \frac{\gamma_0}{M_s} \dot{{\vec M}} \right)
 \label{eq:sLLG1}
\end{equation}
in the Gilbert formulation. ${\vec{H}}$ is a Gaussian white noise with zero mean and delta correlations. $\gamma_0$ is friction coefficient.
Noise is multiplicative 
and, as they are written, these equations conserve the modulus of the magnetization only if the Stratonovich calculus is 
used. Otherwise a drift term has to be added~\cite{Aron14}. Note that this is not an irrelevant detail. Numerical codes written with the 
discretised stochastic differential equation in a different from Stratonovich scheme do not conserve the modulus of the magnetisation.

\vspace{0.25cm}

\noindent
\textcolor{orange}{\bf Exercise \thesection.\theexercise}
Show that a connection between the 
two formalisms is possible after the adequate identification of parameters.
 
\addtocounter{exercise}{1}

\vspace{0.25cm}

\noindent
\textcolor{red}
{\it Classical dipoles}

\vspace{0.25cm}

The translational motion of a particle $i$ is characterized by its time-dependent position ${\vec r}_i$ 
and velocity ${\vec v}_i$. 
Imagine that each particle carries an electric dipole orientation ${\vec p}_i$. 
Although constant in modulus, the orientation of the dipole is time-dependent and can be characterized by an angular velocity vector
${\vec \omega}_i$ such that $d {\vec p}_i/d t={\vec \omega}_i\times{\vec p}_i$. It is subjected to an external force 
${\vec F}_i$ and to an external torque ${\vec \Gamma}_i$. For a particle of mass $m$ and inertia tensor 
$I$ (one can have thin rods in mind) one has
\begin{eqnarray}
\label{eqdebase}
m \ \dot{\vec v}_i&=&-\gamma_0{\vec v}_i+{\vec F}_i+{\vec \xi}_i
\; ,
\\
\label{eqdebase1}
I \ \dot {\vec \omega}_i&=&-\zeta_0{\vec \omega}_i+{\vec \Gamma}_i+{\vec \lambda}_i
\; .
\end{eqnarray}
Here ${\vec \xi}_i$ and ${\vec \lambda}_i$ are Gaussian random forces and torques, respectively, introduced 
to account for the thermal exchanges with the surrounding medium. The friction coefficients $\gamma_0$ and $\zeta_0$ govern 
the dissipation into the thermal bath. Just as $\gamma_0$ and ${\vec \xi}_i$ are related by a Stokes-Einstein relation, 
a similar relation between $\zeta_0$ and ${\vec \lambda}_i$ exists. The Gaussian random contributions ${\vec \xi}_i$ 
and ${\vec \lambda}_i$ have $\delta$-correlations in time, the amplitude of which is constrained by the condition 
that for conservative forces the equilibrium distribution should be the standard Boltzmann-Gibbs exponential factor,
$\langle \xi^\mu_i(t) \xi^\nu_j(t') \rangle = 2\gamma_0 k_BT \delta_{ij} \delta^{\mu\nu} \delta(t-t')$, and similarly for 
$\vec \lambda_i(t)$ with $\gamma_0$ replaced by $\zeta_0$. 
In physical conditions under which inertial effects can
be discarded, at low Reynolds numbers, one obtains a set of over-damped Langevin equations
\begin{eqnarray}
\label{Langevin}
\gamma_0 \ \dot {\vec r}_i &=& {\vec F}_i+{\vec \xi}_i
\; ,
\\
\label{Langevin1}
\zeta_0 \ \dot{\vec p}_i &=& {\vec \Gamma}_i\times {\vec p}_i+ {\vec \lambda}_i \times {\vec p}_i
\; ,
\end{eqnarray}
in which the latter equation, which features a multiplicative noise, is to be understood with the Stratonovich, mid-point, discretization scheme,
that ensures the conservation of the modulus of the dipole, as can be seen directly from \eqref{Langevin1}.
The ingredients entering the force ${\vec F}_i$ felt by particle $i$ include an external force field and possible interactions with other colloids. 
Similarly, the torque ${\vec \Gamma}_i$ felt by particle $i$ can include the effect of an external field. 
In  general, the interparticle interaction energy 
$u({\vec r}_i-{\vec r}_j,{\vec p}_i,{\vec p}_j)$ between particles $i$ and $j$ depends on the distance between 
these particles and on the orientation of the dipoles they carry, as is the case in the well-known dipole-dipole interaction 
$u({\vec r}_i-{\vec r}_j,{\vec p}_i,{\vec p}_j)=(4\pi\epsilon_0)^{-1}
[{\vec p}_i\cdot{\vec p}_j/r_{ij}^3-3 ({\vec p}_i\cdot{\vec r}_{ij})({\vec p}_j\cdot{\vec r}_{ij})/ r_{ij}^5 ]$. 
Both the force ${\vec F}_i$ and the torque then
derive from the total potential energy $V=\frac 12 \sum_{i\neq j}u({\vec r}_i-{\vec r}_j,{\vec p}_i,{\vec p}_j)$ according to
\begin{eqnarray}
{\vec F}_i =-\frac{\partial V}{\partial {\vec r}_i} 
\; , \qquad\qquad
 {\vec \Gamma}_i = -{\vec p}_i\times \frac{\partial V}{\partial {\vec p}_i}
 \; . 
\end{eqnarray}
The combination ${\vec \Gamma}_i\wedge{\vec p}_i$ can also be written in the form
\begin{eqnarray}
{\vec \Gamma}_i\wedge {\vec p}_i = {\vec E}_i p_i^2-({\vec p}_i\cdot{\vec E}_i){\vec p}_i 
\; ,
\qquad\qquad
{\vec E}_i = -\frac{\partial V}{\partial {\vec p}_i}
\; . 
\end{eqnarray}

\setcounter{equation}{0}
\textcolor{red}{
\subsection{Active Matter}
}

A field of use of Langevin process which is currently attracting much attention 
is the one of active matter.
The constituents of active matter, be them particles, lines or other,
absorb energy from their environment or internal fuel tanks and
use it to carry out motion. In this new type of soft condensed matter 
energy is partially transformed into mechanical work and partially dissipated 
in the form of heat~\cite{active-matter-reviews,Cates}. The units interact 
directly or through disturbances propagated in the medium.  
In systems of biological interest, conservative forces (and thermal fluctuations) 
are complemented by non-conservative forces. Realizations of active matter in biology 
are thus manifold and exist at different scales. Some of them are: 
bacterial suspensions, the cytoskeleton in living cells, 
or even swarms of different animals.  
Clearly enough, active matter is far from equilibrium and typically kept in a 
non-equilibrium steady state. The difference between active matter and other driven 
systems, such as sheared fluids, vibrated granular matter and driven vortex lattices 
is that the energy input is located on internal units (e.g. motors) and therefore 
homogeneously distributed in the sample. In the other driven systems mentioned above, 
the energy input occurs on the boundaries of the sample. Moreover, the effect of the 
motors can be dictated by the state of the particle and/or its immediate neighborhood 
and it is not necessarily fixed by an external field.

The dynamics of active matter presents a number of interesting features 
that are worth mentioning  here. Active matter displays 
out of equilibrium phase transitions that may be absent in their passive counterparts.
The dynamic states display large scale spatio-temporal dynamical patterns
and depend upon the energy flux and the interactions between their constituents. 
Active matter often exhibits unusual mechanical properties, very large responses to small
perturbations, and very large fluctuations -- not consistent with the central limit theorem. 
Much theoretical effort has been recently devoted to the description of different aspects of these 
systems, such as self-organization of living microorganisms, the identification and analysis of 
states with spatial structure, such as bundles, vortices and
asters, the study of the rheological properties of active particle suspensions
with the aim of grasping which are the mechanical consequences of  biological activity. 
A rather surprisingly result was obtained with a variational solution to the many-body 
master equation of the motorized version of the standard hard
sphere fluid often used to model colloids:
instead of stirring and thus   destabilize ordered structures, the 
motors do, in some circumstances enlarge the range of stability of crystalline
and amorphous structures relative to the ones with purely thermal motion.


\textcolor{red}{
\subsubsection{Active Brownian Particles}
}

The Active Brownian Particles (ABP) model is a mesoscopic particle-based model
that mimics  the self-propelled motion of a particle 
in a dissipative environment~\cite{FiMa12}. It considers 
a collection of interacting objects coupled with the
surrounding medium through stochastic effective interactions. The microscopic mechanisms 
for self-driven motion are not manifestly treated and only an effective self-propulsion is proposed, which results
in a directional persistency in the motion. 

The self-propulsion results from a force that acts on the particle's center of mass 
and points in the direction of an intrinsic body axis (the particle orientation). There are, therefore, positional, 
$\vec r_i$ and 
orientational, ${\bold{n}}_i$, degrees of freedom.  ${\bold{n}}_i$ is a unit vector pointing in a direction fixed along a
preferred axis of the  particle. 
The particles are assumed to be immersed in a 
bath and therefore they are subject to friction and white noise, acting on both their position and the orientation.

In their simplest realization the particles are spherical and
the propulsive force has constant magnitude, $F_{\rm act}$.
Again for simplicity this model is usually considered in two dimensions. 
The equations of motion are then
\begin{equation}
	m \ddot \vec r_i + \gamma_0\dot\vec r_i=
	F_{\rm act} \vec n_i - \vec {\boldsymbol\bigtriangledown}_i\sum_{j(\neq i)} 
	V(r_{ij}) 
	+ \vec \xi_i \; ,\qquad \dot{\theta}_i=\eta_i \; ,  
	\label{eq:ABP-Langevin}
	\end{equation}
	where $\vec {n}_i=  (\cos\theta_i, \sin \theta_i)$.
Here $V$ is a two-body potential that quantifies the interactions between the particles and 
$r_{ij}=|\vec r_i-\vec r_j|$ is the inter-part distance.
The potential is typically short-range and taken to be  hardly repulsive. In relevant applications the parameters 
are such that one can safely take the over-damped limit, since $m/\gamma_0 \ll 1$. 
The units of length and energy are  $\sigma$ and $\varepsilon$, respectively.

The noises $\vec \xi$ and $\eta$ are  Gaussian with
$\langle {\xi}^\mu_{i}(t)\rangle = \langle \eta_{i}(t)\rangle = 0$ 
and 
$\langle {\xi}^\mu_{i}(t) \, {\xi}^\nu_{j}(t') \rangle = 2 \gamma_0 k_B T \delta^{\mu\nu}_{ij} \delta(t-t')$
and 
$\langle \eta_{i}(t) \, \eta_{j}(t') \rangle = 2 D_{\theta} \delta_{ij} \delta(t-t')$. 
Temperature and the rotational diffusion coefficient $D_\theta$ are, in principle, independent parameters.
However, the latter can be related to the temperature of the bath via $D_\theta = 3k_B T/(\gamma_0 \sigma_d^2)$
assuming that rotational diffusion is of thermal origin. 
This follows from the Navier-Stokes equations assuming Stokes
flow in the solvent and stick boundary conditions on the particle surface. 

It is also worth noticing that dynamics is explicitly out of equilibrium due to the
active force term. It also couples the evolution of the velocity $\dot r$  to the angular diffusion. 
In the passive limit  $F_{\rm act}\to 0$, one recovers the motion of a usual Brownian particle
and the position and angular degrees of freedom decouple. 

The control parameters are the P\'eclet number Pe = $F_{\rm act} {\sigma}/(k_BT)$ which compares the 
work done by the active force when translating a particle but its diameter to the typical energy scale of 
the bath, 
and the packing friction $\phi =\pi{\sigma^2}N/(4S)$, with $S$ the surface of the available space.

The particles undergo a persistent Brownian motion, that is, a tendency to continue moving in 
the same direction for a prolonged period of time. This is illustrated  in the 
right panel of Fig.~\ref{fig:ABP}.  The origin of the persistence
is that the active force acquire a memory. More precisely, 
the integration of the second Eq.~(\ref{eq:ABP-Langevin}) yields
\begin{eqnarray}
&& 
\langle \vec F_{\rm act}(t) \cdot \vec F_{\rm act}(t') \rangle =
F_{\rm act}^2
\langle \vec n(t) \, \cdot \, \vec n(t') \rangle = \langle \cos( \theta(t)-\theta(t') \rangle
\nonumber\\
&&
\qquad\qquad
= F_{\rm act}^2 e^{-(t-t')/t_p} \qquad\mbox{with}\qquad  t_p = 1/D_\theta
\end{eqnarray}
Therefore, the active force points in the same direction for a time proportional to~$t_p$. 

The persistence time $t_p= D^{-1}_\theta = \gamma_0 \sigma^2/(3k_BT)$
is related to the persistence length $l_p = F_{\rm act} \tau_p /\gamma_0 = {\rm Pe} \sigma_d/3$
which grows proportionally to Pe. 

\begin{center}
\vspace{0.5cm}
\begin{figure}[h!]
\centerline{
\includegraphics[scale=0.2]{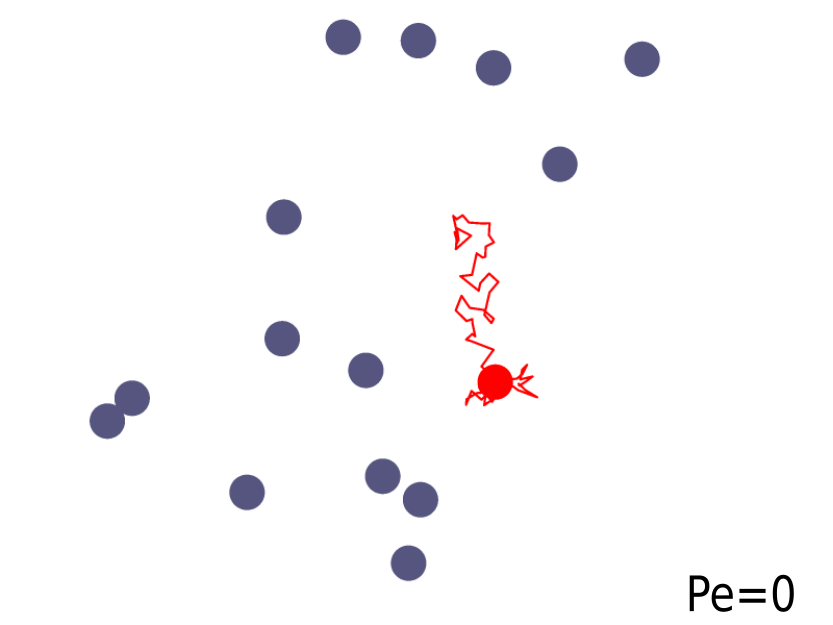}
\hspace{1cm}
\includegraphics[scale=0.2]{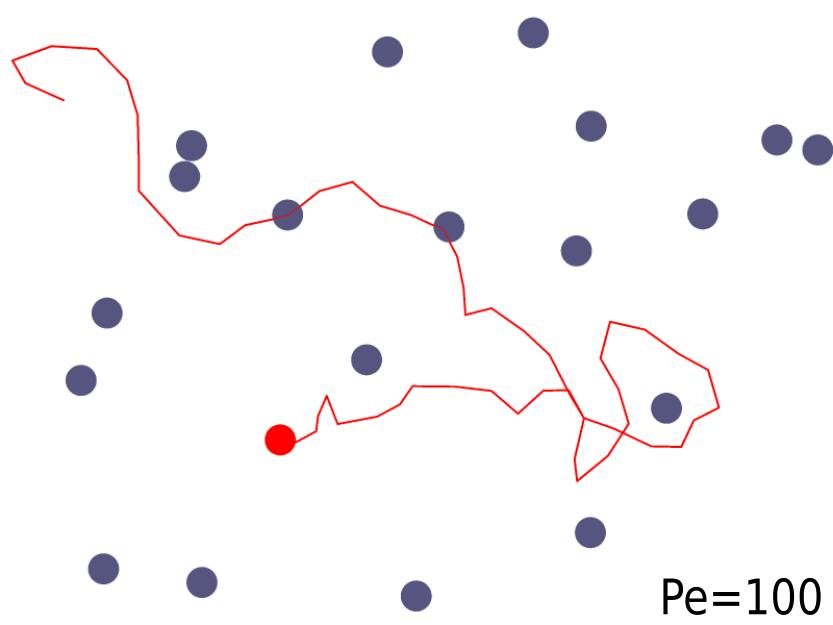}
}
\vspace{0.25cm}
\caption{\small The typical trajectories of a passive Brownian particle (Pe = 0) 
and an active Brownian particle (Pe = 100).}
\label{fig:ABP}
\end{figure}
\end{center}

Information about the translational motion of a single ABP is obtained from the mean-square displacement, 
$\Delta^2(t,t_0) \equiv \langle (\vec r(t) - \vec r(t_0))^2 \rangle$, 
calculated from Eq.~(\ref{eq:ABP-Langevin}).  Dropping the inertial term, an approximation that makes the solution 
easier, 
the ABP position is
\begin{equation}
\vec r(t) = \vec r(t_0) + \frac{1}{\gamma_0} \int_{t_0}^t dt' \left[ \vec F_{\rm act}(t') + \vec \xi(t') \right]
\; . 
\end{equation}
from where we obtain
\begin{eqnarray}
\Delta^2(t,t_0)
\!\! & \!\! = \!\! & \!\! \frac{4 k_BT}{\gamma_0} (t-t_0) + \left( \frac{F_{\rm act}}{\gamma_0} \right)^2
\int_{t_0}^t dt' \int_{t_0}^t dt'' \ e^{-D_\theta(t'+t''-2\min(t',t''))}
\nonumber\\
\!\! & \!\! = \!\! & \!\! 
\frac{4 k_BT}{\gamma_0} (t-t_0) + \left( \frac{F_{\rm act}}{\gamma_0} \right)^2
\! \frac{1}{D_\theta^2} \left[ D_\theta (t-t_0) + e^{-D_\theta (t-t_0)}-1\right]
. 
\end{eqnarray}
The mean-square displacement of the complete equation (with the inertial term) will 
then show four regimes:

\begin{packed_enum}
\item[--]
\textcolor{blue}{Ballistic regime.}
In this regime the velocity approaches the Maxwell distribution 
at the bath temperature $T$, and the mean-square displacement is ballistic, $\Delta^2 \sim (t-t_0)^2$, 
irrespective of there being an applied active force or not. This extremely short 
time regime ends at the time $t_v \sim m/\gamma_0$ that is typically the shortest
time-scale in the problem.  For this reason it can be ignored and the strict over-damped limit taken, 
as we did above.

\item[--]
\textcolor{blue}{Thermal diffusive.}
At time scales longer than $m/\gamma_0$ but still short with respect to the 
time-scale induced by $F_{\rm act}$, that we identify in the next item, 
the particle diffuses as in the absence of the active force, 
\begin{equation}
\Delta^2 \sim D_T (t-t_0)
\quad
\mbox{with} \quad D_T = k_BT/\gamma_0
\quad
\mbox{for} \quad \frac{m}{\gamma_0} \ll t-t_0 \ll \frac{\gamma_0 k_BT}{ F_{\rm act}^2}
\; .
\end{equation}  

\item[--]
\textcolor{blue}{Active ballistic.}
The active force starts acting strongly on the particle motion. 
The Taylor expansion of the 
exponential term between the square brackets two second order, 
valid for $D_\theta (t-t_0) \ll 1$,  yields
$D_\theta^2 (t-t_0)^2/2$ and the cross-over to this new ballistic regime occurs when this term 
becomes of the same order as the one linear in $(t-t_0)$:
\begin{equation}
\frac{k_BT}{\gamma_0} (t-t_0) \approx \left( \frac{F_{\rm act}}{\gamma_0} \right)^2 (t-t_0)^2 
\quad\implies\quad
(t-t_0) \sim \frac{\gamma_0 k_BT}{ F_{\rm act}^2}
\end{equation}
The quadratic term then dominates and 
\begin{equation}
\Delta^2 \sim \frac{F_{\rm act}^2}{\gamma_0^2} (t-t_0)^2
\quad
\mbox{for}
\quad 
\frac{\gamma_0 k_BT}{ F_{\rm act}^2} \ll t-t_0 \ll D^{-1}_\theta 
\; . 
\end{equation}

\textcolor{blue}{Active diffusive.}
For sufficiently long time delays so that the exponential contribution 
vanishes, the  particle starts diffusing again with a new diffusion constant
\begin{equation}
\Delta^2 \sim D_{\rm eff}  (t-t_0) \quad
\mbox{with} \quad
D_{\rm eff} = D_T + \left( \frac{F_{\rm act}}{\gamma_0}\right)^2  \frac{1}{D_\theta}
\quad
\mbox{for} 
\quad
D^{-1}_\theta \ll t-t_0 
\;  . 
\end{equation}
\end{packed_enum}
Of course, if the time scales identified from the equation are not well separated the crossovers 
between the various time-regimes will not be sharp.

The motion of a Janus colloid, which is quite well represented by an Active Brownian Particle, 
in contact with an environment and confined by a harmonic potential has been studied
in great detail experimentally, numerically and analyticially~\cite{Buttinoni22}. 

The collective behavior of large ensembles of these particles leads to a very rich phase diagram, 
with active liquid, hexatic and solid phases, supplemented with a Motility Induced Phase 
Separation at large values of the activity~\cite{DiLeSuCuGoPa18}.

\textcolor{red}{
\subsubsection{Active Uhlenbeck Particles}
}

In the Active Uhlenbeck Particle (AOUP) model, the Gaussian white noise of 
the over-damped Brownian colloids is replaced by a Gaussian colored noise, with no corresponding 
friction memory kernel. This sets the particle explicitly out of equilibrium because it is coupled
to an out of equilibrium bath. 

The position of generic active particles are determined by 
\begin{equation}
\dot{\vec r}_i = -\mu \vec \nabla_i V + \vec v_i
\label{eq:AOUP1}
\end{equation}
where $\vec v_i$ are the self-propulsion velocities, 
$\mu$ is the particle mobility, 
and $V$ is a inter-particle interaction potential. 
The self-propulsion velocities of the AOUP are given by $N$ independent Ornstein-Uhlenbeck processes:
\begin{equation}
\tau \dot{\vec v}_i = -\vec v_i + (2D) \vec \eta_i
\end{equation}
with $\vec \eta_i$  zero-mean Gaussian white noises with delta correlations, 
$\langle \eta_i^\mu \eta_j^\nu\rangle  = \delta_{ij} \delta_{\mu\nu}  \delta(t-t')$.
From direct integration It follows that $\vec v_i$ are a set of zero-mean colored Gaussian noises
\begin{equation}
\vec v_i(t) = \vec v_i(0) e^{-t/\tau} + \int_0^t dt' \; e^{-(t-t')/\tau}  (2D) \vec \eta_i(t')
\label{eq:sol-c-AOUP}
\end{equation}
 with correlations
\begin{equation}
\langle v_i^\mu(t) v_j^\nu(t') \rangle = \frac{D}{\tau} \, \delta_{ij} \delta_{\mu\nu} \, e^{|t-t'|/\tau} 
\end{equation}
Now, $D$ controls the noise amplitude in $\vec v$ and $\tau$ its persistence. 

In the limit $\tau\to 0$,  the self- propulsion velocities become Gaussian white noises. 
For finite $\tau$ instead, the temporal correlations of the $\vec v_i$ are not matched by 
a corresponding memory kernel for the damping in Eq.~(\ref{eq:AOUP1}). 
This is clearly seen by  
introducing (\ref{eq:sol-c-AOUP}) in (\ref{eq:AOUP1}). 
This shows explicitly the 
out of equilibrium nature of these dynamics~\cite{Martin-etal}.

\noindent
\textcolor{red}{
\subsubsection{Active Brownian Dumbbells}
}

Active particles can take different forms and a simple model beyond the spherical ones 
is the one of a dumbbell, that is to say, 
a diatomic molecule formed by two spherical colloids with diameter $\sigma_{\rm d}$ 
and mass $m_{\rm d}$ linked together~\cite{Siebert17,Petrelli18}. 
The atomic  positions are noted ${\vec r}_1$ and ${\vec  r}_2$
in a Cartesian system of coordinates fixed to the laboratory.  
Typically, one assumes that there is an elastic link between the colloids modelled by 
the finite extensible non-linear elastic form
\begin{equation}
{\vec F}_{\rm fene} = - \frac{k {\vec r}}{1- (r^2/r_0^2)} 
\end{equation}
with $k>0$.
The denominator ensures that the spheres cannot go beyond the distance $r_0$
with $r$ the distance between their centres of mass.
An additional repulsive force is added to ensure that the two colloids do not overlap. This 
is the Weeks-Chandler-Anderson (WCA) potential
\begin{eqnarray}
\label{eq:WCA-potential}
V_{\rm wca}( r ) 
&=&
\left\{
\begin{array}{ll}
V_{\rm LJ}( r ) - V_{\rm LJ}(r_c) & \qquad r<r_c
\nonumber\\
0 & \qquad r > r_c
\end{array}
\right.
\end{eqnarray}
with 
\begin{equation}
V_{\rm LJ}(  r  ) = 4\epsilon \left[ \left( \frac{\sigma_{\rm d}}{r} \right)^{12} - \left( \frac{\sigma_{\rm d}}{r} \right)^{6}\right]
\; ,
\end{equation}
where $\epsilon$ is an energy scale and $r_c$ is the minimum of the Lennard-Jones potential, $r_c=2^{1/6} \sigma_{\rm d}$.
The active forces  are polar and act along the main molecular axis  $\hat {\vec  n}$, are constant in modulus, and are the same for the two 
spheres belonging to the same molecule, 
\begin{equation}
{\vec F}_{\rm act} = F_{\rm act} \ \hat {\vec n}
\; . 
\end{equation}
The dynamic equations for one dumbbell  are
\begin{eqnarray}
m_d\ddot{{\vec r}}_{i}(t) &=& -\gamma_0 \dot{{\vec r}}_{i}(t)+
 {\vec F}_{\rm fene}({\vec r}_{i, i+1})
 - 
 \frac{\partial V_{\rm wca}^{i \, i+1}}{\partial  r_{i \, i+1}}
 \frac{ {\vec r}_{i\, i+1} }{ r_{i \, i+1} } 
 + { {\vec F}^{\rm act} }_i
 + {\vec \xi}_{i} 
\; , 
\label{eqdumbattcoll}
\\
m_d\ddot{{\vec  r}}_{i+1}(t) &=&
 -\gamma_0 \dot{{\vec  r}}_{i+1}(t)-
 {\vec  F}_{\rm fene}({\vec  r}_{i,i+1})
-
 \frac{\partial V_{\rm wca}^{i+1,i}}{\partial r_{i+1,i}}
 \frac{{\vec  r}_{i+1,j}}{r_{i+1,i}}+{{\vec  F}^{\rm act}}_i  
  +{\vec \xi}_{i+1}
  \; , 
\end{eqnarray}
with ${\vec  r}_{ij} = {\vec  r}_i - {\vec  r}_j$, $r_{ij} = |{\vec  r}_{ij}|$
 and $V_{\rm wca}^{ij} \equiv
V_{\rm wca}(r_{ij})$ with $V_{\rm wca}$ defined in Eq.~(\ref{eq:WCA-potential}).
Once the active force is attached to a molecule 
a sense of back and forth atoms is attributed to them; ${\vec  F}_{\rm act}$ is  directed 
from the $i$th colloid (tail) to the $i+1$th colloid (head). ${\vec  F}_{\rm act}$ changes direction 
together with the molecule's rotation. 
The coupling to the thermal bath is modelled as usual, with a friction and a noise 
term added to the equation of motion.
$\gamma_0$ is the friction coefficient. 
The noise ${\vec \xi}$ is a  Gaussian random variable with 
\begin{eqnarray}
\langle \xi_{i\mu}(t) \rangle &=& 0 \; , 
\label{eq:noise-ave}\\
\langle \xi_{i\mu}(t) \xi_{j\nu}(t') \rangle &=& 2 \gamma_0 k_BT \delta_{ij} \delta_{ab} \delta(t-t')
\; ,
\label{eq:noise-corr}
\end{eqnarray}
with $k_B$ the Boltzmann constant and $T$ the temperature of the equilibrium environment in which the 
dumbbells move. $a$ and $b$ label the coordinates in $d$ dimensional space. An effective 
rotational motion is generated by the random torque due to the white noise acting independently on the two beads.

\setcounter{equation}{0}
\textcolor{red}{
\subsection{Dynamic constraints}
}

In some applications, one needs to select the trajectories that satisfy 
a given constraint. 
For instance, one may be interested in the dynamics of a particle constrained to move
on a given $N$ dimensional manifold, for example, a sphere with radius $\sqrt{N}$. 
The strict \textcolor{blue}{spherical condition}  reads
\begin{equation}
\phi \equiv \sum\limits_{\mu=1}^N x_\mu^2 = N
\; , 
\qquad\qquad
\phi' \equiv \sum\limits_{\mu=1}^N x_\mu \dot x_\mu = 0
\; , 
\label{eq:constraints-SSK}
\end{equation}
where the first expression is the primary constraint and the second expression ensures that the 
particle does not leave the sphere as time elapses. The index $\mu$ labels the spatial coordinates, 
and runs from 1 to $N$.
The primary constraint can be imposed with a Lagrange multiplier, that turns out to depend on the 
particle's position and momentum and on the noise,  to ensure that the particle remains on the sphere. The equations 
\begin{eqnarray}
&& m\dot x^\mu = p^\mu
\; , 
\\
&&
 \dot p^\mu + \gamma \dot x^\mu =  - z(\vec x, \vec p, \vec \xi) x^\mu - V'(\vec x) +\xi^\mu
\; , 
\end{eqnarray}
with $\xi^\mu$ independent white noises with zero mean, $\langle \xi^\mu(t) \rangle = 0$, and delta correlations
$\langle \xi^\mu(t) \xi^\nu(t') \rangle = 2\gamma k_BT \delta^{\mu\nu} \delta(t-t')$. 
When the potential energy is an anisotropic quadratic potential,
\begin{equation}
V(\vec x) = - \frac{1}{2} \sum\limits_{\mu=1}^N \lambda_\mu x_\mu^2
\label{eq:harmonic-potential-SSK}
\end{equation}
with harmonic constants, $\lambda_\mu$, taken from the Wigner semi-circle law, one 
recovers the Langevin dynamics~\cite{CuDe95a} of the  
Spherical Sherrington-Kirkpatrick (SSK)~\cite{KoThJo76}.

Another application concerns the requirement that  the 
stochastic paths go through a given point at a given time~\cite{ChTo15,MaOr15}. A typical example, 
is the one of \textcolor{blue}{Brownian bridges}, that is, 
Brownian paths that are constrained to return to the origin at some future time.
These processes have wide applications in the context of behavioral ecology and financial stock markets, 
to name a few. Also in this context, a way to achieve the goal is to derive 
effective Langevin equation with an effective force that implicitly accounts for the constraint.
For instance, to generate a Brownian bridge $x_B(t)$ of duration $t_f$ with the bridge constraint 
$x_B(0)=x_B(t_f)$, the effective Langevin equation reads
\begin{align}
  \dot x_B(t) =   - \frac{x_B(t)}{t_f-t} + \sqrt{2\,D}\,\xi(t)
  \,,\label{eq:eff}
\end{align}
where the subscript $B$ refers to ``bridge'' and the first term in the right-hand side is the effective force that accounts for the bridge constraint. Simulating Brownian bridges can then be easily done by discretizing the effective Langevin equation over small time increments. Other effective Langevin equations have been obtained for several constrained processes such as excursions, meanders  and even for some interacting particle systems.

\setcounter{equation}{0}
\textcolor{red}{
\subsection{Inference}
}

Biological systems are so complex that it is too difficult to understand their dynamics from first principles.
Data-driven approaches are then used to quantify their dynamics, and thus try to identify the underlying mechanisms and discover emergent laws.
Microscopy and tracking provide abundant  trajectories from biophysical experiments.
The problem is, however, to infer from these short and noisy experimental data, 
the physical models that reproduces them the best. In other words, the idea is to 
infer the best Langevin equation. Several groups are currently working in devising 
inference methods to achieve this goal. See, for example~\cite{Ronceray}.

\setcounter{equation}{0}
\textcolor{red}{
\subsection{Field equations}
\label{subsec:field-eq}
}

In this Section we extend the Langevin stochastic equations to act on fields. In the 
spirit of Landau, the equations are proposed phenomenologically and are dictated
by symmetry and conservation laws. 

\textcolor{red}{
\subsubsection{The Ginzburg-Landau framework}
}

\vspace{0.2cm}

Equilibrium collective behaviour at second order phase transitions are
largely independent of the microscopic details and, as a
consequence, also of the particular model used to describe it. 
Such \textcolor{blue}{universality} characterizes the physical behaviour 
close to a critical point, where the system undergoes a continuous phase transition.

The onset of collective behaviour is revealed by the
\textcolor{blue}{correlation length} $\xi_{eq}$, the typical distance over which the 
fluctuations of the microscopic variables are correlated.
Far away from a critical point $\xi_{eq}$
is  of the order of the range of the microscopic
interactions, whereas it diverges at the critical point. Accordingly, close
enough to the transition point, $\xi_{eq}$
provides the only relevant length-scale of a critical system.

It is then possible to study the \textcolor{blue}{critical behaviour} in terms of suitable 
field-theoretical models which reflect the internal symmetries
of the underlying microscopic system. The proposed free-energy functionals 
depend only on the \textcolor{blue}{order parameter} (the field) and a few other slow modes, whose actual nature 
are chosen to represent the specific system of interest. For instance, the order parameter can be identified
with the magnetization in magnetic materials, or with the
particle density in fluids. 

By means of field-theoretical techniques it is possible to determine the non-analytic
behaviour observed in various thermodynamic quantities and structure factors upon
approaching the critical point. Such non-analyticities, parametrized by the standard
critical exponents, some associated amplitude ratios and scaling functions turn out to
be universal quantities. The values of the universal quantities and scaling
functions identify the so-called \textcolor{blue}{equilibrium universality class}.

\textcolor{red}{
\subsubsection{Time-dependent Ginzburg-Landau description
}
}

Upon approaching a critical point the typical time scale of the
fluctuations around the equilibrium state diverges as $\xi_{eq}^{z_{eq}}$ (\textcolor{blue}{critical slowing down}), where
$z_{eq}$ is the dynamic critical exponent. This provides the natural
separation between the relevant slow evolution due to the developing collective behaviour and the fast one
related to microscopic processes. This separation makes the field theoretic description of
the dynamics a particularly viable approach.
Indeed it allows one to compute systematically the non-analytic behaviours observed in
dynamical quantities, e.g., in the low-frequency limit of the dynamic structure factor. In
turn the associated universal quantities define the \textcolor{blue}{dynamic universality class}. Each static universality 
class consists of several dynamic sub-universality classes
which differ, e.g., by different conserved quantities, but nonetheless exhibit the same
static universal properties. Moreover, the field theoretic approach can also be used away 
from the critical point to characterise the dynamics within the phases.

In order to treat phase-transitions and the coarsening process
analytically it is preferable to introduce a continuous coarse-grained field,
\begin{equation}
\phi(\vec x, t) \equiv \frac{1}{V} \sum_{i\in V_{\vec x}} s_i(t) 
\; . 
\label{eq:coarse-grained-magn}
\end{equation}
In the magnetic case, this is the fluctuating magnetization density and a particle system it can represent 
a local density (through, for example the identification of an occupation number $n_i=(s_i+1)/2$). 
 
A Landau-Ginzburg free-energy functional
is introduced 
\begin{equation}
F[\phi] = \int d^d x \; \left\{ \frac{c}{2} \, [\vec \nabla \phi(\vec x,t)]^2 + 
V[\phi(\vec x,t)]\right\}
\label{eq:Landau-Ginzburg}
\end{equation}
and the elastic constant $c$ is usually re-absorbed with a series of re-definitions.

With the choice of  potential $V$ one distinguishes between a 
second order and a first order phase transition. In the former case, the 
typical form is the double-well, $\phi^4$, form:
\begin{equation}
V(\phi) = a \phi^4 + b(g) \phi^2
\label{eq:double-well}
\; .
\end{equation}

The first term in Eq.~(\ref{eq:Landau-Ginzburg}) represents the energy
cost to create a domain wall or the elasticity of an interface.  The
second term depends on a parameter, $g$,  
and changes sign from positive at $g>g_c$ to negative at
$g<g_c$. The other parameter $a$ is positive. Above the critical point determined by $b(g_c)=0$ it
has a single minimum at $\phi=0$, at $g_c$ it is flat at $\phi=0$, and
below $g_c$ it has a double well structure with two minima, $\phi=\pm
[-b(g)/(2a)]^{1/2}=\langle \phi\rangle_{eq}(g)$, that correspond to
the equilibrium states in the ordered phase. 

\vspace{0.25cm}
\noindent
\textcolor{orange}{\bf Exercise \thesection.\theexercise} Equation~(\ref{eq:Landau-Ginzburg}) is exact for a fully
connected Ising model where $V(\phi)$ arises from the multiplicity of
spin configurations that contribute to the same $\phi(\vec x)=m$. The
order-parameter dependent free-energy density reads $f(m) =-Jm^2-hm+
k_BT\{(1+m)/2 \ln [(1+m)/2] + (1-m)/2 \ln [(1-m)/2]$ that close to the
critical point where $m\simeq 0$ becomes $f(m) \simeq (k_BT-J)/2 \
m^2-hm+ k_BT/12 \ m^4$ demonstrating the passage from a harmonic form
at $k_BT>k_BT_c=J$, to a quartic well at $T=T_c$, and finally to a
double-well structure at $T<T_c$. Prove these statements.
\vspace{0.25cm}

\addtocounter{exercise}{1}

\vspace{0.25cm}
\noindent
\textcolor{orange}{\bf Exercise \thesection.\theexercise} 
With a six-order potential $V(\phi) = a + b \phi^2 + c \phi^4 + d \phi^6$
one can mimic first order phase transitions.
The sign of $d$, $d>0$, is fixed by the condition that the potential be confining at large values of 
$|\phi |$. The potential has a local minimum at $\phi=0$ for all $b>0$. Next, we choose 
$c<0$ to allow for the existence of two maxima and two minima at
$\phi= \pm [ (- c \pm \sqrt{c^2 - 3bd})/(3d) ]^{1/2}$.  
Derive all these results. 
\vspace{0.25cm}

\addtocounter{exercise}{1}

When discussing dynamics one should write down the stochastic
evolution of the individual components and compute time-dependent averaged
quantities.  This is the
procedure used in numerical simulations. Analytically it is more
convenient to work with a field-theory and an evolution equation of
Langevin-type. This is the motivation for the introduction of
continuous field equations that regulate the time-evolution of the
coarse-grained order parameter. Ideally these equations should be
derived from the microscopic stochastic dynamics but in practice they are
introduced phenomenologically.  In the magnetic case as well as in
many cases of interest, the domain wall and interface dynamics can be
argued to be \textcolor{blue}{over-damped} (i.e. $t\gg t_r^{\dot\phi}$)
and the Smoluchowski limit assumed.

Two very similar approaches are used. Assuming $T$ is only relevant to 
determine the equilibrium coarse-grained field, that is, the form of the potential $V$, 
one uses the phenomenological \textcolor{blue}{zero-temperature 
time-dependent Ginzburg-Landau} equation deterministic equation 
\begin{equation}
\frac{\partial \phi(\vec x,t)}{\partial t} = 
-\frac{\delta F[\phi]}{\delta \phi(\vec x,t)} 
\label{eq:time-dep-GL}
\end{equation}
(the friction coefficient has been absorbed in a redefinition of
time). Initial conditions are usually chosen to be random with short-range 
correlations
\begin{equation}
[\, \phi(\vec x,0) \phi(\vec x',0) \, ]_{ic} = \Delta \delta(\vec x-\vec x')
\end{equation}
thus mimicking the high-temperature configuration
($[\dots ]_{ic}$ represent the average over its probability distribution).
 The numeric solution to this equation with the quartic potential and 
 $b<0$ shows that such a
random initial condition evolves into a field configuration with
patches of ordered region in which the field takes one of the two
values $[-b/(2a)]^{1/2}$ separated by sharp walls.
It ignores 
temperature fluctuations within the domains meaning that 
the field is fully saturated inside them and, consequently, 
one has access to the aging part of the correlations only. 
The phase transition is controlled
by the parameter $b$ in the potential. Importantly enough 
 
Another, similar approach, is to add a thermal noise to the former
\begin{equation}
\frac{\partial \phi(\vec x,t)}{\partial t} = 
-\frac{\delta F[\phi]}{\delta \phi(\vec x,t)} +\xi(\vec x,t)
\; .
\label{eq:time-dep-GL-noise}
\end{equation}
This is the field-theoretical extension of the Langevin equation in
which the potential is replaced by the order-parameter-dependent
functional free-energy in Eq.~(\ref{eq:Landau-Ginzburg}) with a  
potential form with fixed parameters leading to the double well form (independent of 
$T$).   The noise $\xi$ is  taken to be Gaussian
distributed with zero mean and correlations
\begin{equation}
\langle \xi(\vec x,t) \xi(\vec x',t') \rangle  =
2k_B T \delta^d(\vec x-\vec x') \delta(t-t') \; . 
\end{equation}
The friction coefficient has been absorbed in a redefinition of time.
For a quartic potential 
a dynamic phase transition arises at a critical $T_c$; above $T_c$ the 
system freely moves above the two minima and basically ignores the 
double well structure while below $T_c$ this is important. Within the 
growing domains the field $\phi$ fluctuates about its mean
also given by $[-b/(2a)]^{1/2}$ and the fluctuations are determined by $T$.
One can describe the rapid relaxation at times such that the domain 
walls do not move with this approach. 

These equations act on a \textcolor{blue}{scalar} order parameter and 
do not conserve it  neither locally nor globally.
They are called  \textcolor{blue}{ model A} in the classification of Hohenberg-Halperin~\cite{critical-dyn}.

Extensions for cases in which a scalar \textcolor{blue}{order parameter is locally conserved}  (\textcolor{blue}{model B})
are adequate to describe phase separation in particle problems in which the particles cannot be 
eliminated nor created. The equilibrium free-energy density is the same as the one above but the 
stochastic equation must be such that the conservation law is enforced. 
\begin{eqnarray}
\frac{\partial \phi(\vec x, t)}{\partial t} = - \nabla^2 \frac{\delta F[\phi]}{\delta \phi(\vec x, t)} 
\end{eqnarray}
which can be written in the form of a continuity equation with a current $\vec J$:
\begin{equation}
\frac{\partial \phi(\vec x, t)}{\partial t} = \vec \nabla \cdot \vec J (\vec x,t)
\qquad\qquad
 \vec J (\vec x,t) = - \vec \nabla \frac{\delta F[\phi]}{\delta \phi(\vec x,t)}
 \; . 
 \label{eq:continuity}
\end{equation}
The effect of the noise is incorporated by adding a random vector term with i.i.d. components 
white noise characteristics to the current:
\begin{equation}
\vec J(\vec x, t) \mapsto \vec J(\vec x, t)  + \vec \xi(\vec x, t) 
\; .
\end{equation}
This equation is also called the \textcolor{blue}{Cahn-Hilliard equation}.

Cases with vectorial or even tensorial order parameters can be treated
similarly and are also of experimental relevance, notably for vectorial magnets 
or liquid crystals.

These equations are the starting point to study the critical dynamics with renormalization 
group methods~\cite{critical-dyn}. 
Details on their solutions of these equations and the phase ordering kinetics that they describe
can be found in~\cite{review-coarsening,Puri}.

In active matter modelling, an \textcolor{blue}{active model B} was introduced~\cite{Cates}. It 
starts from the continuity equation in (\ref{eq:continuity}) and it modifies the 
current
\begin{equation}
\vec J(\vec x, t) = - \vec \nabla \left\{ \frac{\delta F[\phi]}{\delta \phi(\vec x,t)} + \lambda [\vec \nabla \phi(\vec x, t)]^2 \right\}
+ \vec \xi(\vec x, t)
\; . 
\end{equation}
The current has a term which is the
gradient of the standard non-equilibrium chemical potential/free-energy (\ref{eq:Landau-Ginzburg})
but the last term added is $\neq \delta F[\phi]/\delta \phi$ for any $F[\phi]$.
The model was then modified into the so-called \textcolor{blue}{active model B+}
in which other terms which also break detailed balance are added to the current~\cite{Cates}
\begin{equation}
\vec J \mapsto \vec J + \zeta (\nabla^2 \phi) \vec \nabla \phi - \frac{1}{2} \vec \nabla (\vec \nabla \phi)^2
\end{equation}
which are the allowed terms in an expansion to order $\nabla^3$ and $\phi^2$. Now the terms proportional 
to $\zeta$ and $\lambda$ are of the same order, and the $\zeta$ term vanishes for a flat interface so it 
does not affect the bulk contributions.

\setcounter{equation}{0}
\textcolor{red}{
\subsection{Dean-Kawasaki equation}
}

The Dean-Kawasaki equation is a stochastic field equation obeyed by the density
function for a system of Langevin processes interacting via a pairwise potential. This equation was first proposed by 
Kawasaki on phenomenological grounds~\cite{Kawasaki94} and it was then proven by Dean
in a very elegant paper~\cite{Dean96}. Differently from the phenomenological equations usually used to describe
the dynamics of non-conserved (model A) and conserved (model B) particle systems described in 
Sec.~\ref{subsec:field-eq}, the spatial white noise for this system appears not additively but multiplicatively.
This is a consequence of the fact that the density cannot fluctuate in regions devoid of particles. The
steady state for the density function takes the form of a functional
integral over the coursed grained free energy of the system.

We now briefly review how to introduce a single differential equation with all the information content of the system of 
Langevin equations describing the Brownian particles.
The stochastic ordinary differential equation for each particle are 
\begin{displaymath}
\dot{\vec x}_j = - V'(\vec x_i - \vec x_j)  + \vec \eta_j \; , \qquad \qquad j=1,\ldots, N
\end{displaymath}
The strict over-damped limit in which the inertial term is discarded is usually 
considered in this context.
The $\vec \eta_i$ are independent time-dependent white noises $\eta_j$ with zero mean and 
correlations $\langle \eta^a_i(t) \eta^b_j(t') \rangle = 2 k_BT \delta_{ij} \delta_{ab} \delta(t-t')$. The friction coefficient 
has been absorbed in a redefinition of time.  $V$ is the interaction potential 
between the particles which is assumed to depend only on the inter-particle distance.

The number density field $\varrho(x,t)$ is defined as
\begin{equation} \label{densfield}
\varrho(\vec x,t) = \sum_{j=1}^N \delta(\vec x-\vec x_j(t)) \qquad \mathrm{or} \qquad \varrho_{\vec k}(t) = \frac{1}{V} \sum_{j=1}^N e^{i\vec k\cdot \vec x_j(t)}
\end{equation}

Defining the density field as in (\ref{densfield}), the system of equations can be formally translated into a
partial differential equation governing the evolution of the density field
\begin{eqnarray} \label{dean-eq}
\partial_t \varrho(\vec x,t) &=&  \vec \nabla \cdot \left( \vec \xi\sqrt{\varrho(\vec x,t)} \right)
+ T \nabla^2 \varrho(\vec x,t) 
\nonumber\\
&& + \; \vec \nabla \varrho(\vec x,t) \cdot \int d^dy \, \varrho(\vec y, t) \vec \nabla V(\vec x - \vec y)
\end{eqnarray}
In the process, the $N$ independent noises $\vec \eta_j(t)$ have been turned into a single vector field $\vec \xi(x,t)$ with self-correlation:
\begin{displaymath}
\langle \xi_a(\vec x,t) \xi_b(\vec x^{\, \prime},t^{\prime}) \rangle = 2T \delta_{ab} \delta(t-t^{\, \prime}) \delta(\vec x-\vec x^{\prime})
\end{displaymath}
Since in (\ref{dean-eq}) the noise multiplies the density field, the differential equation becomes a meaningful  string of symbols only once a proper interpretation for the noise term is given. 

It should be stressed that, due to the singular nature of the density field, the natural setting of Dean's equation is a distribution space. In such a setting the equation has been derived exactly, without any coarse-graining assumptions or cut-off scales which are not necessary to obtain the differential equation \cite{Dean96}.

This equation turned out to be extremely useful to analyze passive (as well as active with the necessary extensions) Brownian particle systems. For example, the Poisson statistics of the  density-density correlation functions of
the Brownian gas were derived in~\cite{Velenich}. The formation of periodic clusters in systems of Brownian particle interacting via repulsive soft-core potentials is due the interplay among diffusion, the intracluster  forces,  and  the  forces  between  neighboring  clusters. This surprising phenomenon is captured by the deterministic part  of  the
Dean-Kawasaki equation (neglecting fluctuation effects)~\cite{Delfau16}.
Onsager's results for strong electrolyte conductivity can be derived
in a relatively straightforward fashion using the Kawasaki-Dean equation~\cite{Demery16}.
These are just a few examples where this equation has been successfully applied.

\pagebreak

\textcolor{red}{
\subsection{Quantum}
}

Sometimes open quantum mechanical system are treated in an approximate way with 
Generalized Langevin Equations~\cite{Mazur,dePasquale,FordKac,FordLewis}, in which the quantum Bose-Einstein energy distribution is enforced by tuning the random and friction forces, while the system degrees of freedom remain classical. Although these baths have been formally justified only for harmonic oscillators, they perform well for several systems, while keeping the cost of the simulations comparable to the classical ones~\cite{Fabio}.

In other cases, dissipation and noise are added to 
the deterministic quantum equations (either the Schrödinger equation itself or the equation for the density matrix). 
Often, these approaches break the constraint of unitarity of the evolution, necessary for the conservation of the total probability in the statistical interpretation of quantum mechanics.

A generating functional framework~\cite{weiss} in which the system and the bath are both treated quantum mechanically
and the bath integrated out (since assumed to be simple, made of oscillators~\cite{Feve}, free fermions or free bosons)
is fully consistent. It yields to a reduced dynamic generating functional called
\textcolor{blue}{closed time path} or \textcolor{blue}{Schwinger-Keldysh} which in the classical limit 
approaches the Martin-Siggia-Rose-Janssen one.
 

\comments{ 

xxx

} 

\appendix

\textcolor{red}{
\section{Discretized delta function}
}
\label{app:delta-function}
\setcounter{equation}{0}
\renewcommand{\theequation}{A.\arabic{equation}}

Let us discretize time $t_k =  k\Delta t$ with $k$ an integer and $\Delta t$ the time-step.
A discrete representation of the Dirac delta function is 
\begin{eqnarray}
\delta_{\Delta t}(t_k-t_{k'}) = 
\left\{
\begin{array}{ll}
\dfrac{1}{\Delta t} 
& 
\qquad -\dfrac{\Delta t}{2} < t_k-t_{k'} < \dfrac{\Delta t}{2} 
\\
[10pt]
\ \ 0  & \qquad \ \
\mbox{otherwise}
\end{array}
\right.
\end{eqnarray}
One easily checks that, calling $t=t_k-t_{k'}$
\begin{eqnarray}
&&
\lim_{\Delta t\to 0}\delta_{\Delta t}(t) 
= 
\left\{
\begin{array}{ll}
\infty \qquad\qquad &  \mbox{for} \qquad t\to 0
\\
[5pt]
0 \qquad\qquad & \mbox{otherwise}
\end{array}
\right.
  \end{eqnarray} 
   and 
   \begin{equation}
   \int_{-\infty}^\infty dt \; \delta_{\Delta t}(t) = 1
   \; . 
   \end{equation}
   A short-hand notation for the discrete time delta function uses the Kronecker delta
   $\delta_{kk'} =1$ if $k=k'$ and $0$ otherwise, 
   \begin{equation}
   \delta_{\Delta t}(t_k -t_{k'}) \;\;\; \longleftrightarrow \;\;\; \dfrac{\delta_{kk'}}{\Delta t}
   \; . 
   \end{equation}

\vspace{1cm}

\noindent
\textcolor{red}
{\bf \Large Acknowledgements}

\vspace{0.5cm}

\noindent
I wish to thank F. Ferrari, A. Gonz\'alez, and D. Martinelli for the organization of the ZigZag ICTP School in La Havana, 
R. Ch\'etrite, A. Hedrih, V. M. Ili\v{c}, E. Roldan, and B. Seoane Bartolom\'e  for the organization of the  Information, Noise and Physics of Life 
ICTP school held at Ni\v{s}, C. Calero Borrallo, D. Levis and F. Ritort for the 
organization of the Barcelona school on Non-equilibrium Statistical Physics, and the subsequent directors of 
the Physics of Complex Systems Master in Paris, where I 
taught parts of these lectures.

\end{document}